\pdfoutput=1
\documentclass[11pt]{article}
\usepackage[T1]{fontenc}  % as elsarticle; the bibliography needs it (\k in Jastrz\k{e}bski)
\newif\ifelsart\elsartfalse
\ifelsart\else
  \usepackage[margin=1in]{geometry}
  \usepackage[numbers,sort&compress]{natbib}
  \newsavebox{\absbox}\newsavebox{\kwbox}
  \newcommand{\shimaddr}{}\newcommand{\shimead}{}\newcommand{\shimurl}{}\newcommand{\shimauthor}{}
  \let\authorels\author
  \renewcommand{\author}[2][]{\renewcommand{\shimauthor}{#2}}
  \newenvironment{frontmatter}{}{%
    \authorels{\shimauthor\\[2pt]\small\textit{\shimaddr}\\\small\texttt{\shimead}}%
    \maketitle\noindent\usebox{\absbox}\par\medskip\noindent\usebox{\kwbox}\par\bigskip}
  \renewenvironment{abstract}{\begin{lrbox}{\absbox}\begin{minipage}{\textwidth}\section*{Abstract}\small}{\end{minipage}\end{lrbox}\global\setbox\absbox\box\absbox}
  \newenvironment{keyword}{\begin{lrbox}{\kwbox}\begin{minipage}{\textwidth}\small\textit{Keywords:} }{\end{minipage}\end{lrbox}\global\setbox\kwbox\box\kwbox}
  \newcommand{\sep}{\unskip; }
  
  \newcommand{\address}[2][]{\renewcommand{\shimaddr}{#2}}
  \newcommand{\cortext}[2][]{}
  \newcommand{\ead}[2][]{\def\eadopt{#1}\def\eadurl{url}%
    \ifx\eadopt\eadurl\renewcommand{\shimurl}{#2}\else\renewcommand{\shimead}{#2}\fi}
  \newcommand{\tnotetext}[2][]{}
  \date{}
\fi
\usepackage{amsmath,amssymb,graphicx,booktabs,float,array}
\usepackage{adjustbox}
\usepackage[font=small,labelfont=bf]{caption}
\usepackage{microtype}
\usepackage{xurl}
\usepackage[most]{tcolorbox}
\usepackage{fancyvrb}
\usepackage{hyperref}
\ifelsart
  \usepackage{etoolbox}
  \patchcmd{\pprintMaketitle}{\vskip36pt}{\vskip14pt}{}{\ClassWarning{elsarticle}{title-page patch failed}}
  \patchcmd{\printFirstPageNotes}{\textit{URL:\space}}{\textit{ORCID:\space}}{}{\ClassWarning{elsarticle}{ORCID label patch failed}}
\fi
\hypersetup{hidelinks}
\newcommand{\startsupplement}{%
  \setcounter{section}{0}\setcounter{equation}{0}%
  \setcounter{figure}{0}\setcounter{table}{0}%
  \renewcommand{\thesection}{S\arabic{section}}%
  \renewcommand{\theequation}{S\arabic{equation}}%
  \renewcommand{\thefigure}{S\arabic{figure}}%
  \renewcommand{\thetable}{S\arabic{table}}%
  \renewcommand{\theHsection}{S\arabic{section}}%
  \renewcommand{\theHequation}{S\arabic{equation}}%
  \renewcommand{\theHfigure}{S\arabic{figure}}%
  \renewcommand{\theHtable}{S\arabic{table}}%
  \setkeys{Gin}{width=\linewidth,keepaspectratio}}
\usepackage{placeins}

\graphicspath{{figs/}}
\setkeys{Gin}{keepaspectratio}   % no default width: it would also scale every adjustbox-wrapped table to full width

\newtcolorbox{algobox}{enhanced,colback=white,colframe=black,boxrule=0.7pt,arc=0pt,outer arc=0pt,left=7pt,right=7pt,top=4pt,bottom=4pt,title={Algorithm 1.\enspace Synthesis of a dual-array RopeComb},fonttitle=\bfseries,coltitle=black,colbacktitle=white,titlerule=0.5pt,toptitle=2pt,bottomtitle=2pt}

\newcommand{\Gnet}{G_{\mathrm{net}}}
\newcommand{\Garr}{G_{\mathrm{array}}}
\newcommand{\Gstar}{G^{*}}
\newcommand{\Krope}{K_{\mathrm{rope}}}
\newcommand{\Lrope}{L_{\mathrm{rope}}}
\newcommand{\Lw}{L_{w}}
\newcommand{\dmax}{d_{\max}}
\newcommand{\neff}{n_{\mathrm{eff}}}
\newcommand{\cfg}[2]{\langle #1,#2\rangle}
\newcommand{\ptm}{peak-to-mean}

\date{1 October 2026}

\begin{document}

\begin{frontmatter}

\title{\Large The dual-array RopeComb: a guide-balanced transmission with synthesisable rising mechanical advantage for impedance-matched launch}

\author{Rami N. Mahdi}
\ead{ramimahdi@interestingmachines.com}
\address{Interesting Machines LLC, Sacramento, California, USA}

\begin{abstract}
Efficiently transferring kinetic energy from a slow, high-force driver to a light payload, with a near-uniform output force, requires a transmission whose mechanical advantage rises along a steeply convex profile. Continuous-contour mechanisms (variable-radius drums, cams) become prohibitive at scale through diameter span, inertia and rim burst. The RopeComb bypasses these limits by synthesising profiles ($G = 0 \to 100+$) from discrete constant-radius sheaves deflecting tension members into staggered spans. Each fold's ratio vanishes at first contact, enabling snatch-free engagement at 10~m/s without clutches or dampers. Because the array ratio is closed-form, bounded least squares matches uniform-force targets to within 0.6--1.5\% across mass ratios of 100:1 to 10,000:1 using 8--13 members. To carry large carriage reactions without the overturning moment a single array imposes, two independent arrays flank a central guide and drive a partitioned fixed-ratio stage ($n_1 + n_2 = k$). Sizing array ratios and member counts in the fall-count ratio $n_2 : n_1$ cancels 97.7\% of the single-array reaction and 95\% of its overturning moment in simulated 1,000:1 launches at $k=7$, reducing compliant \ptm{} payload force from 1.40 to 1.13. All results are numerical.
\end{abstract}


\end{frontmatter}

\begin{figure}[!htbp]
\centering
\includegraphics[width=\linewidth]{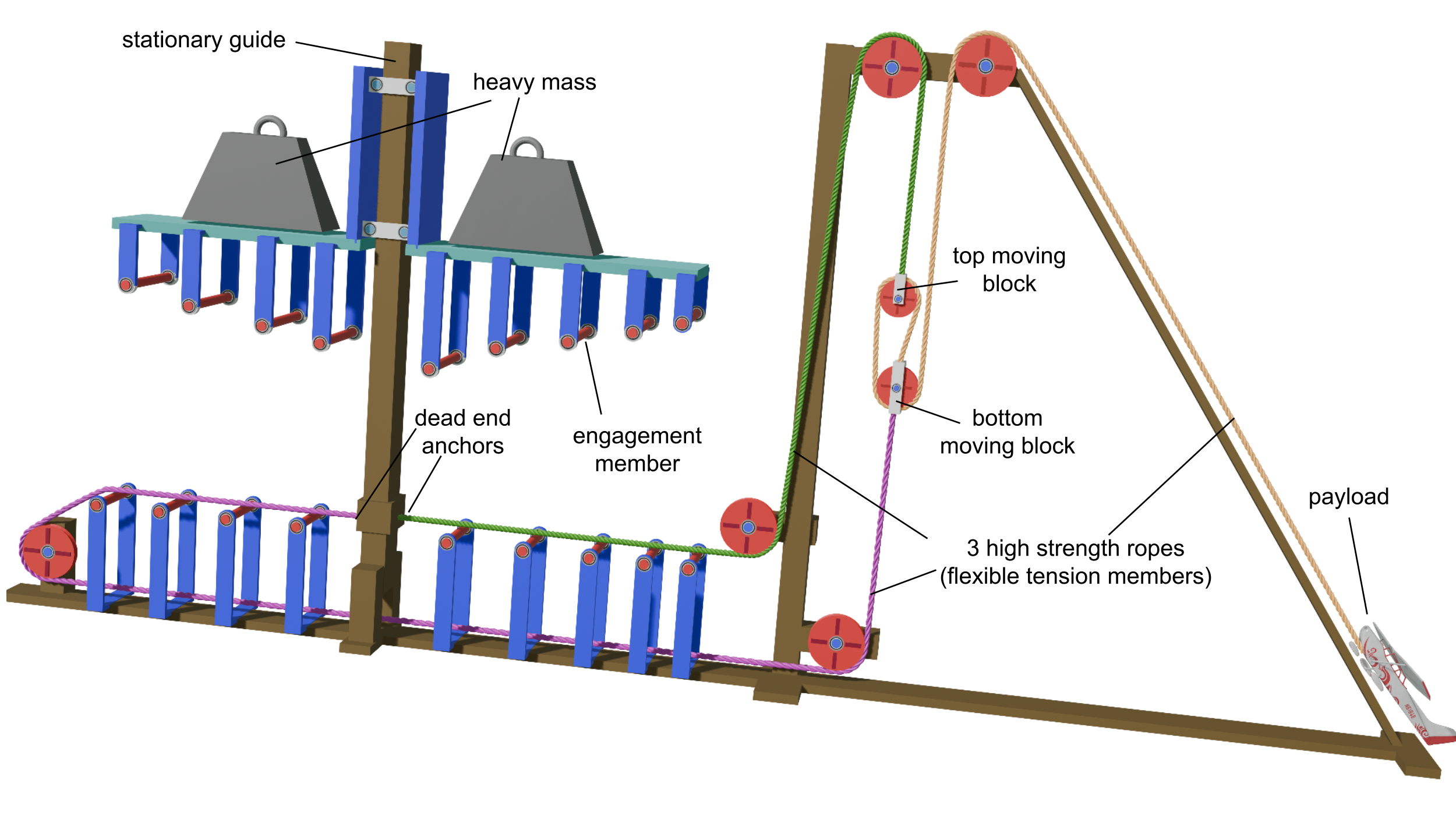}
\caption{The dual-array RopeComb launcher. A carriage carrying the source mass (the heavy mass of the figure) descends the stationary guide; the comb of engagement members on either side deflects its own tension member, dead-ended at an anchor on the guide, into a row of fixed spans, and each line drives its own counter-moving block of the fixed-ratio stage. The third rope is the output member, reeved about the two blocks, which accelerates the payload along the rail. Only the output member and the payload move at launch speed. Geometry is illustrative and not to scale.}
\label{fig:hero}
\end{figure}

% ============================================================================
\section{Introduction}
\label{sec:intro}

\subsection{The impedance problem in mechanical launch}

Assisted launch transfers the work of initial acceleration from a flight vehicle to ground infrastructure, allowing reduced propellant, a greater payload fraction and a lighter airframe \citep{army2010}. Figure~\ref{fig:hero} shows the machine developed here for that purpose. The sources efficient at this scale, a descending mass, a ram, a spring or a flywheel, deliver high force at low velocity, and coupling one directly to a light payload that must reach high velocity gives extreme peak accelerations and poor energy transfer \citep{sibilski2017,kondratiuk2020,doyle1995}. The difficulty is one of impedance. A transmission of fixed ratio $G$ ceases to exert useful force once the payload velocity, referred back through the ratio, exceeds that of the source; efficient transfer requires a ratio that \emph{rises} as the source decelerates. Force uniformity is the figure of merit: for a payload that survives a peak force $F_{\max}$ the energy delivered over a stroke $L$ is $\bar{F}L$, so the stroke required scales with the \ptm{} ratio; the replacement of steam catapults by the electromagnetic system of \citet{doyle1995} was motivated substantially by this, with a reported gain of up to 31\% in airframe life \citep{bushway2001}, and Table~\ref{tab:fielded} collects reported values. Electromagnetic launchers reach nearly uniform acceleration but need electrical storage, power conditioning and cooling beyond the reach of small installations or sites with limited supply \citep{ga2024,navsea2017}; steam and pneumatic pistons impart an initial shock that must be carried by structural mass \citep{qidan2018,sibilski2017}; gravity and elastic systems deliver a strongly non-uniform force fixed by their geometry.

\begin{table}[!htbp]
\centering
\caption{Reported \ptm{} tow force for fielded launch systems.}
\label{tab:fielded}
\begin{adjustbox}{max width=\linewidth}
\begin{tabular}{ll}
\toprule
system & peak / mean force \\
\midrule
pneumatic and elastic catapults & 3--10 \\
gun launch, progressive propellant & $\sim$1.7--2.5 \\
steam catapult & 1.25 nominal, excursions to 2.0 \citep{doyle1995} \\
electromagnetic (EMALS) & 1.05 \citep{doyle1995} \\
\bottomrule
\end{tabular}
\end{adjustbox}
\end{table}

\subsection{Rising mechanical advantage by geometry}

The oldest rising-ratio launchers are the counterweight trebuchet, whose beam and sling present a mechanical advantage that grows through the throw \citep{chevedden1995,denny2005}, the sling alone being a rising-ratio device of some subtlety \citep{denny2025}, and the bow, whose advantage over the arrow varies through the draw with the geometry of limbs and string \citep{kooi1991}. In both the profile is a consequence of the linkage. Where a prescribed nonlinear input--output relation is wanted, the established route is a manufactured contour. Non-circular gears realise a prescribed ratio function directly, with a mature generation theory \citep{litvin2009,mundo2006}; belt and chain continuously variable transmissions vary the ratio under control rather than as a function of displacement \citep{srivastava2009}. In cable systems, non-circular spools synthesise nonlinear springs \citep{schmit2012,kim2014}, variable-radius drums realise prescribed load paths in closed form \citep{seriani2016}, and non-circular pulleys shape joint motion and provide static balancing \citep{shirafuji2017,fedorov2018}. These devices are exact and compact, but each realises its profile as a continuous machined contour: the contour must react the full cable tension over a small radius, at launch loads a bending and bearing problem before a kinematic one; a ratio rising from unity to eighty or more exceeds the radius range one drum or gear pair can carry; and all are configured at robot forces and speeds. The mechanics of cable-and-pulley networks at such loads is itself a subject: rope friction over a pulley departs from the capstan law once bending rigidity is included \citep{jung2008}, a pre-stressed cable over several pulleys has been modelled for structural analysis \citep{spiegelhauer2021}, cable-driven parallel robots supply a kinematic and force framework for multi-cable machines \citep{pott2018}, cable-pulley hysteresis and friction have been characterised experimentally \citep{miyasaka2020}, and rope endurance over sheaves is treated by \citet{feyrer2015} and \citet{mckenna2004}.

What launch needs is a kinematically programmable transmission: a ratio profile specifiable substantially at will, from unity to eighty or more, synthesised from discrete geometric primitives without a machined contour, realised in components most of which move slowly, and able to react forces many times the source's weight. The RopeComb is such a transmission. A carriage driven by the source carries a row of engagement members, the comb, past a row of fixed supports, deflecting a tension member into successive spans. Each fold draws in rope at a rate rising with its depth; each member has two independent geometric parameters, the half-width $R_i$ of its span and the carriage displacement $s_i$ at which it engages; and the array ratio is a closed-form sum in carriage displacement, so matching a target profile is a least-squares fit rather than a search. Throughout, $M$ and $m$ denote the source and payload masses, $v_0$ the source entry velocity, $d$ the source (equivalently carriage) displacement, $y$ the payload displacement, $G = dy/dd$ the instantaneous displacement ratio and $F$ the payload force.

\subsection{Reaction loads, and the embodiment that carries the principle to scale}
\label{sec:reactions}

At mass ratios of 100:1, 1,000:1 and 10,000:1 the designs of Section~\ref{sec:sim} see peak carriage reactions of 23.4~kN, 228~kN and 2.09~MN, 21 to 24 times the weight of the driving mass in every case, and how that reaction is carried into the guide determines the architecture. An array on one side of the guide, the single-array embodiment rendered in Supplementary Section~\ref{si:cases}, reacts the whole load at a lateral offset. Each fold is symmetric about its member, so the transverse components of the rope tension cancel and every member pushes on the carriage along the guide: the reaction is axial, and what the guide bearings feel is its moment about the guide axis. That moment cannot be removed by changing the driving mass placement, because the members engage in sequence along the array and the centre of pressure of the reaction sweeps outward through the stroke, from 0.25~m at first engagement to 1.58~m at the end across the 2.42~m array of the 1,000:1 reference design; a driving mass set at any fixed offset leaves a moment that reverses sign mid-stroke (Section~\ref{sec:moment}). For a carriage guided by two bearings at spacing $h$, with the reaction at offset $e$, the bearings react the moment as a couple, each carrying $F_{\mathrm{carriage}}\,e/h$ sideways while sliding at up to 10~m/s, and with bearing friction coefficient $\mu$ the fraction of the transferred energy lost in the guide is $2\mu e/h$, independent of mass ratio (Supplementary Section~\ref{si:guide}): a single-sided array with $e/h = 0.5$ loses a fraction $\mu$ of its output to guide friction. Two arrays on opposite sides of the guide, each tension member driving its own block of the fixed-ratio stage, oppose their moments. What remains is their difference, which the designs of Section~\ref{sec:moment} hold to between a sixth and a sixteenth of the single array's moment, and the guide loss falls in the same proportion. What governs that difference, the imbalance of the two axial reactions and where along the arrays their members sit, is the question the paper answers. The two arrays are independent, and the answer uses that freedom: the force each applies depends on its geometry, on the stage falls its line actuates and on the tension delivered along its path, so balance is a condition on the ratio of the two geometries rather than on their equality, and whenever the stage ratio is odd, the natural outcome of a counter-moving stage (Section~\ref{sec:mech}), the arrays that balance the guide differ in a small and calculable way. The dual-array embodiment of Figure~\ref{fig:hero} is designed on that condition. It doubles the transmission's geometric freedom at no cost in friction, fast-path inertia or sheave count, two different arrays being no harder to build than two identical ones, and mirror symmetry is the special case the same solve may return when the falls and paths happen to be equal.

\subsection{Contributions and organisation}

The paper makes five contributions. (i) A transmission whose rising ratio profile is synthesised from discrete geometric primitives, each an individually concave fold, without a machined cam or drum: the closed-form array ratio, the target equation in the displacement domain with its closed-form solution and scaling laws, and the demonstration that least squares reaches optima that direct search does not, tracking targets that rise from zero to 25, 79 and 251 to within 0.6 to 1.5\% of full scale. (ii) The dual-array embodiment: partitioned falls with the exact weighting $n_1 G_1 + n_2 G_2$, the accounting of what the second array costs and buys, and the counter-moving stage with $k = 2p+1$. (iii) The balance condition $G_1 : G_2 = n_2 T_2 : n_1 T_1$, its discrete form $n_1 N_1 = n_2 N_2$ on the member counts, the mirror residual $|n_1 - n_2|/k$ at odd $k$, and the rule that the array serving the fewer falls must engage first. (iv) A synthesis with a single balance weight $\lambda$, seeded from a feasible single-array solution, given as Algorithm~1. (v) Dynamic verification against single arrays of matched member count and stage ratio, for reference designs at three mass ratios and for four dual-array designs at 1,000:1, two of equal member counts and two count-balanced. The guide moment of these designs is evaluated, together with the span placement along the array that minimises it. A per-contact friction model carried through the integrators quantifies the two-line saving and the tension term of the balance condition. The balance trade is traced to its end, showing that beyond the member counts near-exact cancellation is bought with array width. Finally, the physical operating limits and the loads a builder must provide for.

Section~\ref{sec:target} derives the target profile and its scaling laws. Section~\ref{sec:mech} sets out the mechanism and the stage. Section~\ref{sec:synth} gives the synthesis for one array. Section~\ref{sec:dual} develops the dual-array embodiment, its balance theory and its synthesis. Section~\ref{sec:sim} reports the simulations, in which the single array ($n_2 = 0$) and the mirror-symmetric pair ($G_1 = G_2$) appear as degenerate cases of the same kinematics and serve as benchmarks. Section~\ref{sec:limits} treats the operating limits and Section~\ref{sec:concl} concludes. Every quantitative result is a simulation output; no apparatus has been built.

\subsection{Nomenclature}
\label{sec:nomen}

The symbols used throughout the paper are collected below; each is also defined where it first appears, and the supplement follows the same notation.

\smallskip
\begin{footnotesize}
\noindent
\begin{tabular}{@{}p{0.115\linewidth}>{\raggedright\arraybackslash}p{0.335\linewidth}p{0.135\linewidth}>{\raggedright\arraybackslash}p{0.315\linewidth}@{}}
$M$, $m$ & source and payload mass & $k$, $p$ & fixed-stage ratio; travelling sheaves, $k = 2p+1$ \\
$v_0$, $v_{\mathrm{end}}$, $v_h$, $v_p$ & source entry and final velocity; source and payload velocity & $n_1$, $n_2$ & output falls about blocks 1 and 2, $n_1 + n_2 = k$ \\
$\nu$, $T$ & $v_{\mathrm{end}}/v_0$; stroke duration & $T_1$, $T_2$ & tension delivered along lines 1 and 2 \\
$d$, $\dmax$, $\delta$ & source displacement, braking distance, $d/\dmax$ & $\Gnet$ & net displacement ratio $n_1 G_1 + n_2 G_2$ \\
$y$, $G$ & payload displacement; $G = dy/dd$ & $\lambda$ & balance weight in the synthesis objective \\
$\Gstar(d)$ & target ratio profile & $\Lw$ & width budget per array \\
$F$ & design (uniform) payload force & $\Krope$, $\Lrope$ & output-member stiffness $EA/\Lrope$; its length \\
$R_i$, $s_i$ & span half-width and engagement offset of member $i$ & $\eta$, $\gamma$, $\Phi$ & transfer efficiency; $2g\dmax/v_0^2$; force number \\
$D_i$ & deflection of member $i$, $\max(0, d - s_i)$ & $c$, $\sigma$, $\rho$, $\rho_\ell$ & critical speed; working stress; density; linear density $\rho A$ \\
$N$, $N_j$ & members per array & $\neff$, $m_s$ & effective fast-path sheave count; sheave mass \\
$x_i$, $\bar{x}_j$, $\mathcal{M}$ & member position along the array; centre of pressure of array $j$; pitch moment on the guide & $\Delta F$ & reaction imbalance $|n_1 T_1 G_1 - n_2 T_2 G_2|$ \\
$\Garr$, $G_j$ & array ratio \eqref{eq:array}; ratio of array $j$ & $\mu$, $e$, $h$, $\varepsilon$ & guide friction coefficient, offset, bearing spacing, peak reaction imbalance as a fraction of the peak reaction \\
\end{tabular}
\end{footnotesize}
\vspace{6pt}

% ============================================================================
\section{Ideal transfer profile in the displacement domain}
\label{sec:target}

\subsection{Uniform payload force and the target equation}

\begin{figure}[!htbp]
\centering
\includegraphics[width=0.7\linewidth]{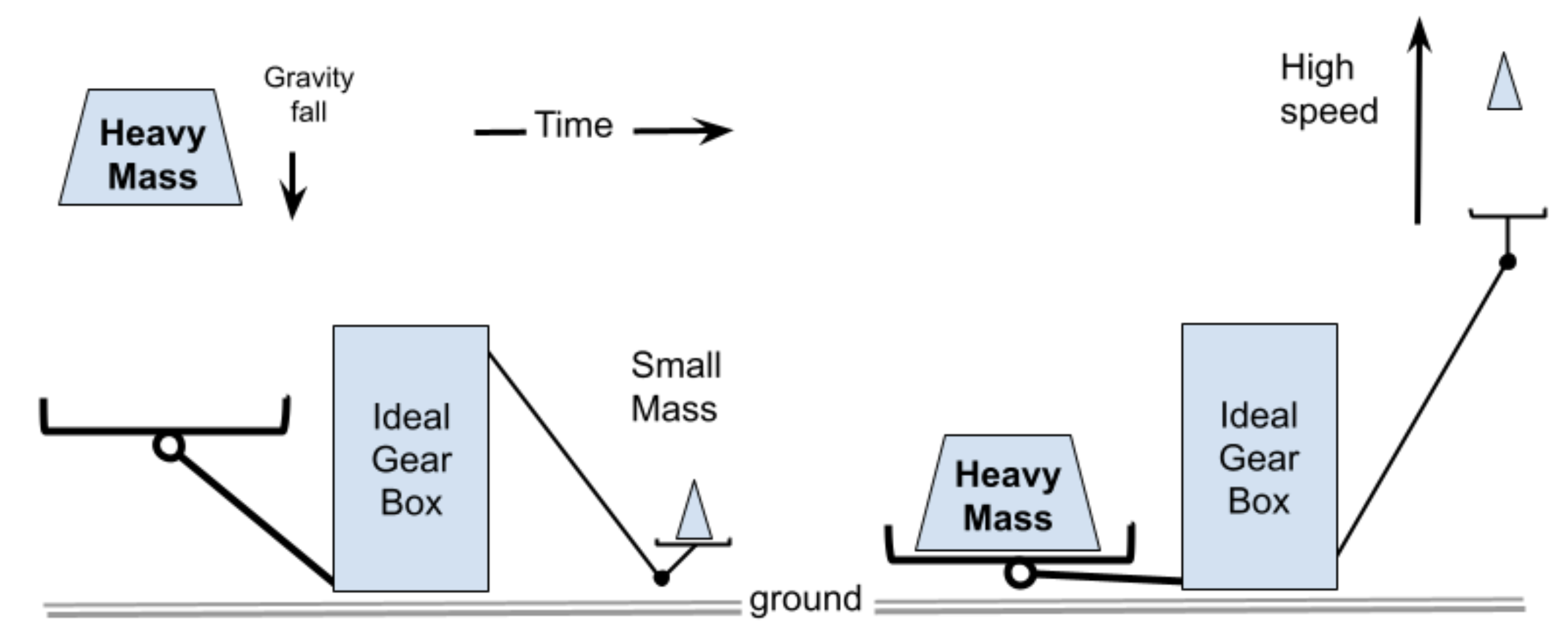}
\caption{The transfer in the abstract: a decelerating heavy mass, an accelerating light mass, and a transmission between them whose ratio must rise throughout.}
\label{fig:concept}
\end{figure}

Consider a source mass $M$ entering the stroke at velocity $v_0$ and decelerating while a payload $m$ accelerates (Figure~\ref{fig:concept}). Uniform payload force $F$ is the natural target: for a given transferred energy it minimises the peak force, and so the strength and mass of the payload structure, the tension member and the frame; for a given permissible peak force it minimises the stroke length and duration. With uniform $F$ imposed, the payload velocity $v_p$ follows from the work $Fy$ done on it and the source velocity $v_h$ from its own energy balance, $\tfrac12 M v_h^2 = \tfrac12 M v_0^2 + Mgd - Fy$, and the ratio the transmission must present is their quotient:
\begin{equation}
\Gstar(d) = \frac{dy}{dd} = \frac{dy/dt}{dd/dt} = \frac{v_p}{v_h} = \frac{\sqrt{2Fy/m}}{\sqrt{v_0^2 + 2gd - 2Fy/M}},
\label{eq:ode}
\end{equation}
a first-order ordinary differential equation for $y(d)$ in the \emph{displacement} domain. It is closed in $y$ and $d$ alone, contains no discrete events and no contact conditions, and its solution is smooth. The $2gd$ term is the work done by gravity on a descending source; for a ram or a spring it is replaced by the corresponding work term, or dropped, and nothing else changes. Because $y = 0$ is a fixed point the integration is seeded from the small-$d$ expansion $y \simeq (2F/m)\,d^2/(4v_0^2)$. That the target is a function of displacement rather than time is what makes the synthesis of Section~\ref{sec:synth} a curve fit: the array ratio of Section~\ref{sec:mech} is likewise a closed-form function of displacement, and the two are matched directly.

\subsection{Specifying the deceleration, not the force}

\begin{figure}[!htbp]
\centering
\includegraphics[width=\linewidth]{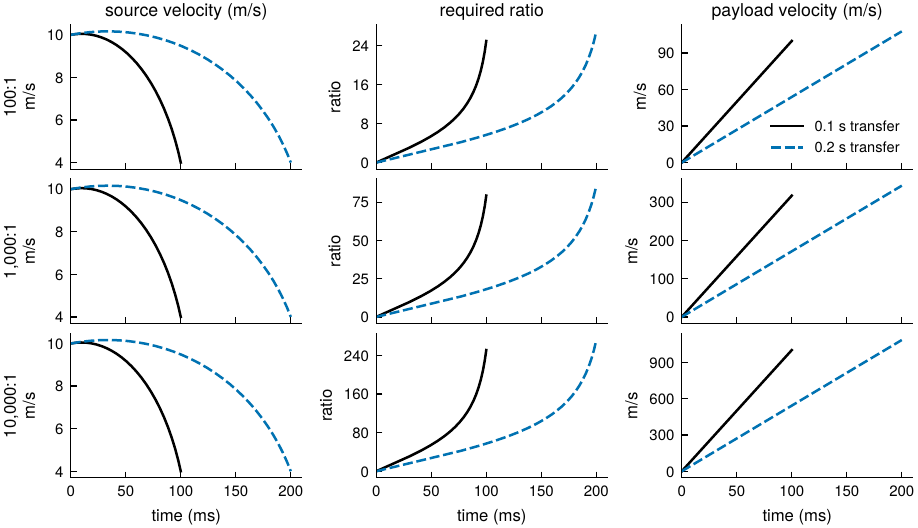}
\caption{Ratio profiles producing uniform payload force, for mass ratios of 100:1, 1,000:1 and 10,000:1 and two transfer durations. Left: source velocity; centre: required ratio; right: payload velocity, each against time. The required ratio rises at an accelerating rate throughout, and its terminal value scales as $\sqrt{M/m}$.}
\label{fig:targets}
\end{figure}

Equation~\eqref{eq:ode} is not defined until one fixes how much of its energy the source gives up, and that choice is best made by stating the deceleration and solving for the force that produces it, rather than choosing a force and accepting the deceleration that results: a target that asks more than the transmission can deliver degrades the machine (Section~\ref{sec:synth}). Throughout, the specification is that \textbf{the source decelerates from 10~m/s to 4~m/s over a 0.1~s stroke}, and the uniform force $F$ that achieves it is found by a one-dimensional root find, to within 0.002~m/s of the specified final speed. Table~\ref{tab:spec} gives the result at the three mass ratios that bracket the design space, and Figure~\ref{fig:targets} plots the profiles. The braking distance is common to all three because the deceleration is; force, exit velocity and terminal ratio scale with mass ratio as Section~\ref{sec:scaling} makes explicit. The final speed of 4~m/s is a design input chosen at the knee of a trade that Supplementary Section~\ref{si:knee} sweeps. Stopping the source harder than this lets the tension member unload during the stroke and the peak force rises steeply; stopping it softer buys little further force uniformity while giving up exit velocity and widening the array, at 1,000:1 from 2.44~m at 4~m/s to 5.51~m at 5~m/s for a peak force lowered from 1.11 to 1.03 times design, in the sweep's refits at a reduced restart budget; the reference design of Table~\ref{tab:single}, fitted with the full budget, is 2.42~m wide, the difference being the multi-start spread of Section~\ref{sec:fit}.

\begin{table}[!htbp]
\centering
\caption{The design force solved from the specified deceleration, and the profile it implies. The first two rows are the specification, identical in all three cases; everything below follows from it. Payload mass is 1~kg throughout; force scales linearly with payload mass.}
\label{tab:spec}
\begin{adjustbox}{max width=\linewidth}
\begin{tabular}{lccc}
\toprule
 & 100:1 & 1,000:1 & 10,000:1 \\
\midrule
\emph{specified} deceleration & 10.0 $\to$ 4.0 m/s & 10.0 $\to$ 4.0 m/s & 10.0 $\to$ 4.0 m/s \\
\emph{specified} stroke duration & 0.1 s & 0.1 s & 0.1 s \\
solved payload force $F$ & 999.0 N & 3,169.6 N & 10,033.8 N \\
braking distance $\dmax$ & 0.854 m & 0.854 m & 0.854 m \\
ideal exit velocity & 99.9 m/s & 317.0 m/s & 1,003.5 m/s \\
terminal ratio $\Gstar(\dmax)$ & 25.0 & 79.3 & 251.0 \\
\bottomrule
\end{tabular}
\end{adjustbox}
\end{table}

The energy accounting is taken against the energy put into the machine, which for a gravity-driven source is not $\tfrac12 M v_0^2$: before the next launch the source must be raised again, through the braking distance as well as the height that gave it its entry speed, so the input per cycle is $E_{\mathrm{in}} = Mg(h_0 + \dmax)$ with $h_0 = v_0^2/2g$, a total fall of 5.96~m in all three cases. With $gh_0 = \tfrac12 v_0^2$, the ratio of payload energy to input is
\begin{equation}
\eta = \frac{\tfrac12\,(v_0^2 - v_{\mathrm{end}}^2) + g\,\dmax}{\tfrac12\,v_0^2 + g\,\dmax} = \frac{1 + \gamma - \nu^2}{1 + \gamma}, \qquad \nu = \frac{v_{\mathrm{end}}}{v_0}, \quad \gamma = \frac{2g\,\dmax}{v_0^2},
\label{eq:eta}
\end{equation}
in which the source mass has cancelled: transfer efficiency is fixed by entry speed, final speed and braking distance alone, and mass ratio sets the force, the exit velocity and the ratio but cannot appear here. For the specification of Table~\ref{tab:spec}, $\eta = 86.3\%$ in every case; at 1,000:1 that is 50,371~J to the payload from 58,371~J of input, with 8,000~J retained by the source at 4~m/s. A horizontal ram or spring, $\gamma = 0$, returns $1 - \nu^2 = 84\%$ at the same deceleration. Charged over seconds to minutes and discharged in a tenth of a second, the machine is also a power amplifier: at 1,000:1, 50.6~kJ is delivered in 100~ms, a peak of 506~kW, while raising the source over 60~s requires under 1~kW.

\subsection{Dimensionless scaling and the universal transfer profile}
\label{sec:scaling}

Equation~\eqref{eq:ode} reduces to a scale-free form in two dimensionless groups. With the payload energy as a fraction of the source's entry kinetic energy and the displacement as a fraction of the braking distance,
\begin{equation}
\hat y = \frac{2Fy}{M v_0^2}, \qquad \delta = \frac{d}{\dmax},
\label{eq:hatdefs}
\end{equation}
energy conservation gives the source velocity as $v_h = v_0\sqrt{1 + \gamma\delta - \hat y}$ and the payload velocity as $v_0\sqrt{M/m}\,\sqrt{\hat y}$, and \eqref{eq:ode} becomes
\begin{equation}
\frac{d\hat y}{d\delta} = \Phi\,\sqrt{\frac{\hat y}{1 + \gamma\delta - \hat y}}, \qquad
\Gstar(\delta) = \sqrt{\frac{M}{m}}\,\sqrt{\frac{\hat y}{1 + \gamma\delta - \hat y}},
\label{eq:dimless}
\end{equation}
with the force number and the gravity number
\begin{equation}
\Phi = \frac{2F\,\dmax}{\sqrt{Mm}\,v_0^2}, \qquad \gamma = \frac{2g\,\dmax}{v_0^2}.
\label{eq:groups}
\end{equation}
The endpoint condition $\hat y = 1 + \gamma - \nu^2$ at $\delta = 1$, where the source has slowed to $\nu v_0$, fixes $\Phi$, and the stroke duration is $T = (\dmax/v_0)\,\tau$ with $\tau = \int_0^1 d\delta/\sqrt{1 + \gamma\delta - \hat y}$; both $\Phi$ and $\tau$ depend on $(\nu, \gamma)$ alone. Three scale-free properties follow.

\begin{itemize}
\item \textbf{Universal target profile.} At fixed $(\nu, \gamma)$ the normalised profile $\Gstar(\delta)/\sqrt{M/m}$ is the same at every mass ratio. In particular the terminal ratio is $\sqrt{M/m}\,\sqrt{1 + \gamma - \nu^2}/\nu$, the exit velocity $v_0\sqrt{M/m}\,\sqrt{1 + \gamma - \nu^2}$, and the payload travel $\dmax\sqrt{M/m}\,(1 + \gamma - \nu^2)/\Phi$. At $\nu = 0.4$ and $\gamma = 0.167$ these reproduce the integrated terminal ratios and exit velocities of Table~\ref{tab:spec} to within 0.5\% at all three mass ratios (Supplementary Section~\ref{si:scaling}).
\item \textbf{Force scaling.} The design force is $F = \Phi\sqrt{Mm}\,v_0^2/(2\dmax)$, so payload acceleration $F/m$ rises as $\sqrt{M/m}$, as the exit velocity does, while the force as a fraction of source weight, $F/Mg$, falls as $1/\sqrt{M/m}$, from 1.02 to 0.10 across Table~\ref{tab:spec}: the heavier the source relative to the payload, the smaller the payload force against the source's weight and the steeper the transmission that must multiply it.
\item \textbf{Geometric similarity.} The array ratio of Section~\ref{sec:mech} depends on $R_i$ and $s_i$ only through $R_i/\dmax$ and $s_i/\dmax$, so an array synthesised for one stroke scales as a shape to any stroke with the same $(\nu, \gamma)$; and since each engaged member contributes at most 2, the member count needed for the terminal ratio grows as $\sqrt{M/m}/k$.
\end{itemize}

\subsection{Closed-form target for a ram, spring or horizontal source}
\label{sec:closed}

When gravity does no work on the source during the stroke, as for a ram, a spring, a flywheel or a horizontal source, $\gamma = 0$ and \eqref{eq:dimless} integrates exactly. Energy conservation is then $(v_h/v_0)^2 + \hat y = 1$, so the partition of energy between source and payload is an angle: writing $\hat y = \sin^2\theta$ gives $v_h = v_0\cos\theta$ and a payload velocity $v_0\sqrt{M/m}\,\sin\theta$, with $\theta$ running from 0 at entry to $\arccos\nu$ at release, and the required ratio is the tangent of that angle:
\begin{equation}
\theta + \tfrac12 \sin 2\theta = \Phi\,\delta, \qquad \Gstar(\delta) = \sqrt{\frac{M}{m}}\,\tan\theta,
\label{eq:closed}
\end{equation}
and the whole transmission follows from the velocity ratio $\nu$:
\begin{equation}
\begin{aligned}
\Phi &= \arccos\nu + \nu\sqrt{1 - \nu^2}, &\qquad \tau &= \frac{2\sqrt{1 - \nu^2}}{\Phi}, \\
\dmax &= \frac{\Phi\,v_0 T}{2\sqrt{1 - \nu^2}}, &\qquad F &= \frac{\sqrt{Mm}\,v_0\sqrt{1 - \nu^2}}{T}.
\end{aligned}
\label{eq:closed-design}
\end{equation}
At $\nu = 0.4$ this gives $\Phi = 1.526$ and $\tau = 1.201$; the gravity-driven specification of Table~\ref{tab:spec} has $\Phi = 1.712$ and $\tau = 1.171$, gravity raising the force number and shortening the normalised duration by a few percent, recovered by integrating \eqref{eq:dimless} numerically at the corresponding $\gamma$.

The design recipe is four lines: \eqref{eq:closed-design} gives $\Phi$, $\dmax$ and $F$ from the specification; the terminal ratio is $\sqrt{M/m}\,\sqrt{1-\nu^2}/\nu$ and the transfer efficiency $1 - \nu^2$; the member floor at stage ratio $k$ is that ratio over $2k$; and the geometry of Section~\ref{sec:synth} scales with $\dmax$. A gravity-driven source replaces the first line by the numerical solution of \eqref{eq:dimless} and changes nothing else.

% ============================================================================
\section{The RopeComb mechanism}
\label{sec:mech}

\subsection{One engagement member}

A tension member spans two fixed supports $2R$ apart. An engagement member on the carriage descends into the span to a depth $D$, so that the tension member follows two segments of length $\sqrt{R^2 + D^2}$ and the length drawn in is $Y(D) = 2(\sqrt{R^2 + D^2} - R)$. The instantaneous displacement ratio of the fold is
\begin{equation}
\frac{dY}{dD} = \frac{2D}{\sqrt{R^2 + D^2}} = \frac{2}{\sqrt{(R/D)^2 + 1}},
\label{eq:member}
\end{equation}
which rises continuously from zero at $D = 0$ toward an asymptote of 2 (Supplementary Section~\ref{si:figs}): a member contributes at most 2, at a \emph{rate} governed by $R$, a narrow span rising rapidly and a wide span slowly. Writing $f(D)$ for the right-hand side of \eqref{eq:member},
\begin{equation}
f''(D) = \frac{-6R^2 D}{(R^2 + D^2)^{5/2}} < 0 \quad \text{for all } D > 0,
\label{eq:concave}
\end{equation}
so every engaged member is concave: its rate of contribution falls from the moment it engages, and an array of such members approximates the convex targets of Figure~\ref{fig:targets} only by staggering onsets, each engagement injecting an upward step in slope with the curve concave between. This is the origin of the force ripple of Section~\ref{sec:sim}: force quality is governed not by how closely the ratio curve is reproduced but by the arc length between successive engagements.

\subsection{The array: two freedoms per member}

\begin{figure}[!htbp]
\centering
\includegraphics[width=\linewidth]{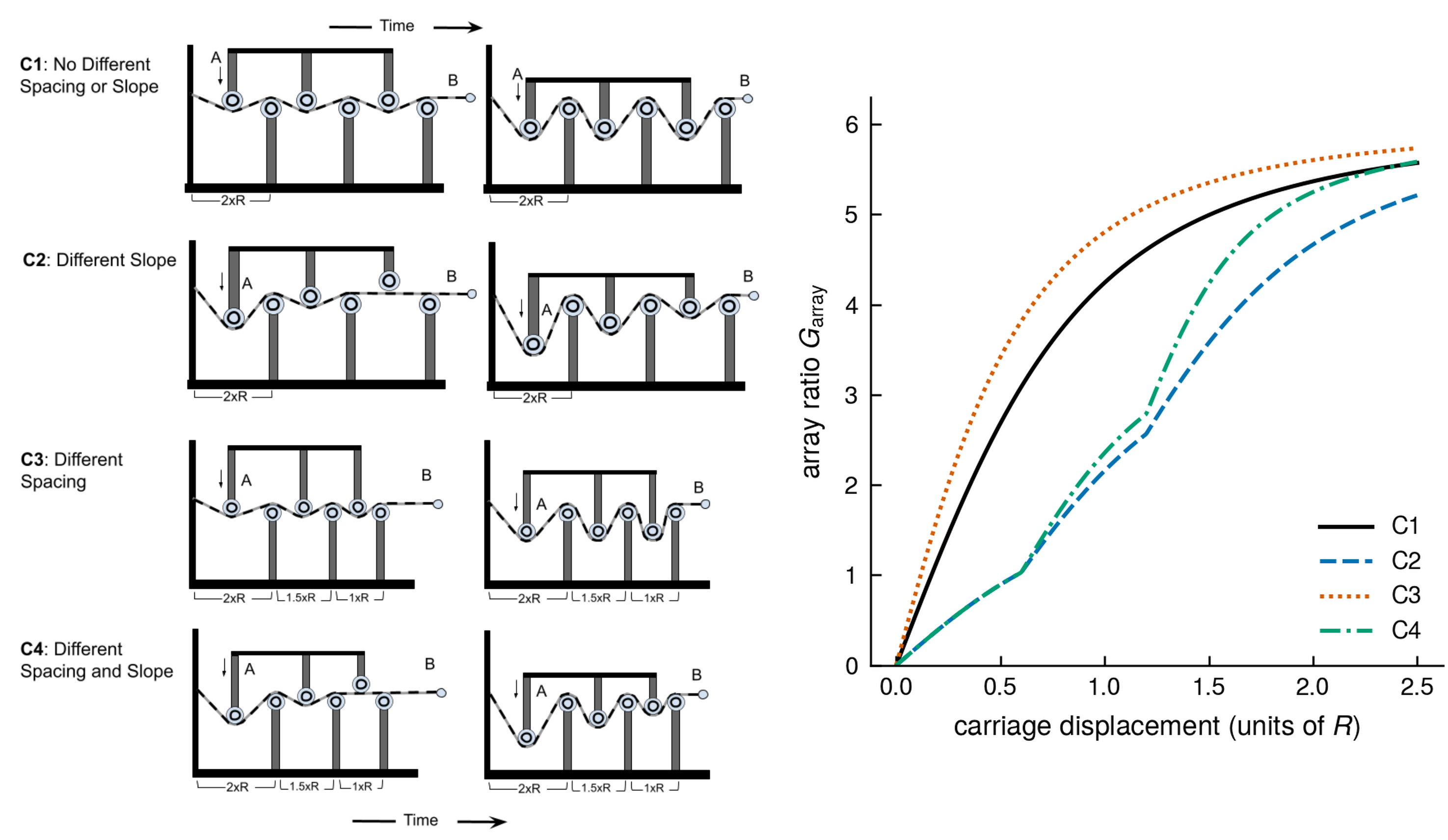}
\caption{How a comb shapes a profile. Left: four three-member configurations, C1 to C4, each at two instants of the descent; A is the comb, B the run to the stage. C1 has equal spans and heights; C2 staggers the heights; C3 narrows the successive spans; C4 does both. Right: the array ratio each produces, differing in shape and not merely in magnitude; only the staggered C2 and C4 rise at an accelerating rate over the early stroke, as the target demands.}
\label{fig:configs}
\end{figure}

$N$ members on the source-driven carriage, each descending into its own span of half-width $R_i$ and first touching the tension member at carriage displacement $s_i$, give the array displacement ratio
\begin{equation}
\Garr(d) = \sum_{i=1}^{N} \frac{2D_i}{\sqrt{R_i^2 + D_i^2}}, \qquad D_i = \max(0,\; d - s_i).
\label{eq:array}
\end{equation}
Equation~\eqref{eq:array} is exact: a polyline through the support and member coordinates, finite-differenced against carriage displacement, reproduces it to $1.5\times10^{-5}\%$ in ratio and $7\times10^{-4}\%$ in take-up for 4 to 20 members (Supplementary Section~\ref{si:polyline}). Each member offers \textbf{two independent geometric parameters}, $(R_i, s_i)$: the span half-width, which sets how fast it contributes, and the engagement offset, which sets when. Figure~\ref{fig:configs} shows what the pair does with four three-member combs: uniform spacing and height gives the concave sum of three identical folds; staggering the heights delays the later members and makes the early rise convex; narrowing the later spans makes each rise faster once engaged; and the two together give independent control of when each member contributes and how fast, as the accelerating target demands. This pair of parameters, not the chaining alone, lets an arbitrary rising profile be synthesised from individually concave primitives: the steeply convex targets of Figure~\ref{fig:targets}, rising to 25, 79 and 251, are reproduced by 8 to 13 such folds to within 0.6 to 1.5\% of full scale (Section~\ref{sec:singles}), with no continuous contour anywhere in the machine. An array of $N$ members has a ceiling of $2N$.

\subsection{The fixed-ratio stage: two lines and \texorpdfstring{$k = 2p+1$}{k = 2p+1}}
\label{sec:stage}

Reaching a high ratio by chaining alone would need $N \gtrsim \Gnet/2$ members, with their inertia, friction and cost. A fixed-ratio stage of ratio $k$ between array and payload multiplies the ratio with few added components and has a second, more important function: its blocks translate at $1/k$ of payload velocity, so the massive elements of the transmission stay at low speed, storing $1/k^2$ of the kinetic energy they would at payload speed, and only the output member moves at payload velocity. With the stage, the member floor of a single array is $N \ge \Gnet/2k$.

\begin{figure}[!htbp]
\centering
\includegraphics[width=0.9\linewidth]{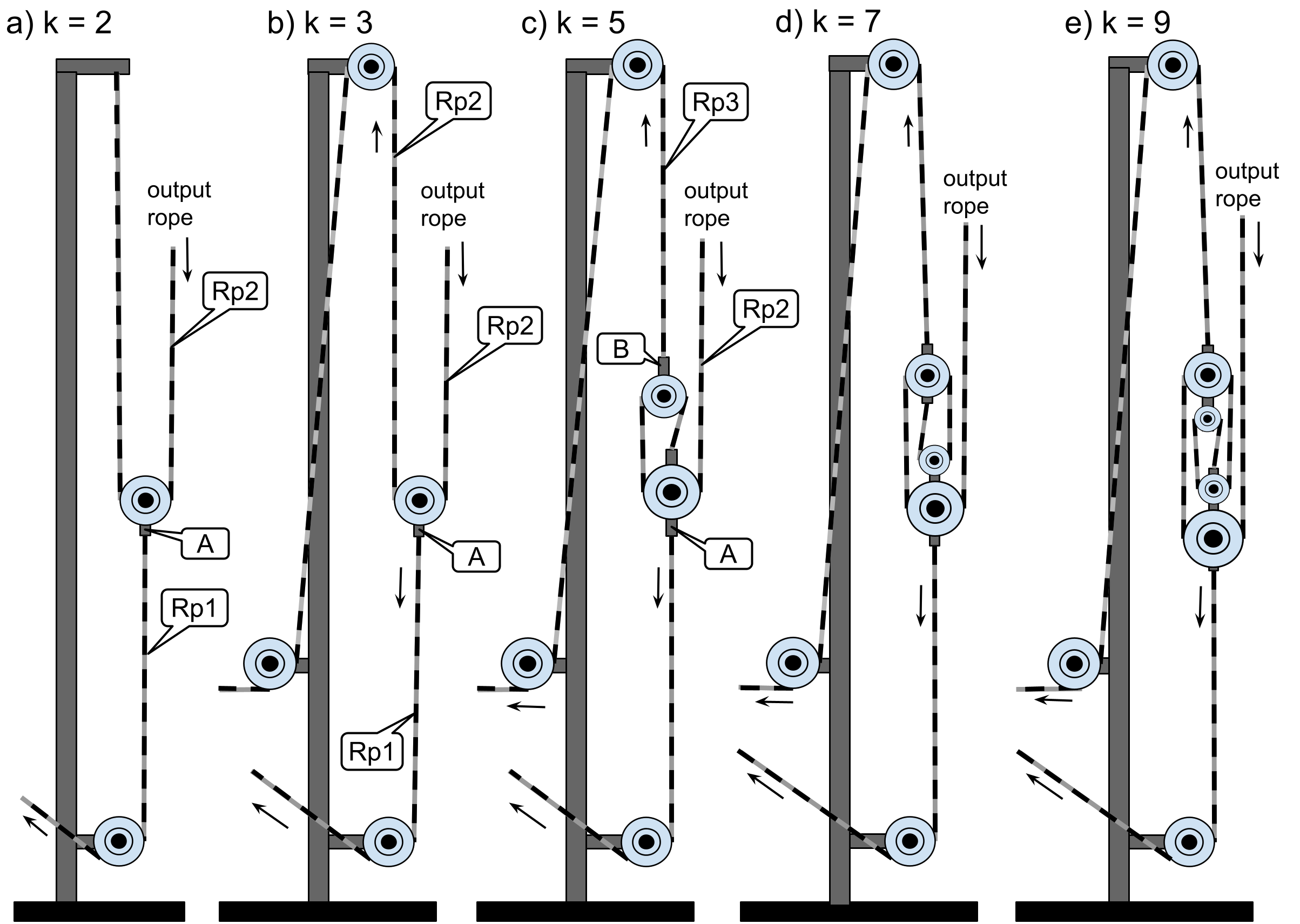}
\caption{Fixed-ratio stages at $k = 2$, 3, 5, 7 and 9. Rp1 and Rp3 are the tension members, drawn in by the arrays (arrows at the foot of each panel); Rp2 is the output member to the payload. Blocks A and B travel in opposite directions, A drawn down by Rp1 and B drawn up by Rp3, so the pair separates as the arrays take in rope. The travelling sheaves number one, two, three and four in panels (b) to (e), giving $k = 2p + 1 = 3$, 5, 7 and 9. At $k = 3$ there is no second block: the output member Rp2 itself continues over the top sheave to the second array and serves as its tension member. Panel (a) is the one simple even case, $k = 2$: a single block A with one sheave and the output member dead-ended to the frame. Arrows show the direction of travel.}
\label{fig:stage}
\end{figure}

The stage used throughout carries two travelling blocks that move in \emph{opposite} directions, each drawn by its own tension member (Figure~\ref{fig:stage}), and the output member is reeved with $n_1$ falls about one block and $n_2$ about the other. With $p$ sheaves distributed between the blocks the output member has $p+1$ segments, every one but the last touches both blocks, and
\begin{equation}
n_1 + n_2 = k = 2p + 1 .
\label{eq:parity}
\end{equation}
The blocks carry one to four sheaves in panels (b) to (e) of Figure~\ref{fig:stage}, giving stage ratios of 3, 5, 7 and 9; the rule is read off the drawing. A stage reeved otherwise, with a single travelling block or with the free end of the output member anchored to the frame, does not in general obey \eqref{eq:parity}; the one simple even case is $k = 2$, a single block with one sheave and the output member dead-ended to a stationary point (panel a), and beyond that forcing even parity costs sheaves without buying ratio. The counter-moving arrangement is chosen on kinematic grounds. Two blocks separating at speed $u$ open their gap at $2u$, so the output member is drawn in at twice the rate a single block at $u$ would give; a single block must move twice as fast to consume rope at the same rate, and since kinetic energy goes as velocity squared each of the two blocks stores a quarter of what the single fast block would, so the stage holds about half the parasitic energy despite carrying two blocks. Adding sheaves is likewise the right direction: stage energy goes as $p/k^2 \approx 1/(4p)$, so going from $k = 3$ to $k = 5$ doubles the sheave count and still cuts stage energy by 28\%, $k = 7$ by 45\% and $k = 9$ by 56\%, and at equal sheave count the odd reeving beats the even one by 21 to 36\%. Odd stage ratios are therefore not a corner case but what the stage naturally produces, and Section~\ref{sec:dual} shows what that implies for balance.

The two tension members may run through one array or two. Through one array, side by side over the same members and supports, each dead-ended on the guide, they share a geometry, $G_1 = G_2 = \Garr$; through two arrays they may differ. In either case the tension in line $j$ is $n_j F$ rather than $kF$, and each line's path costs two contacts per member it passes, one per fall it serves and the entry, about $2N + n_j + 1$ elements against $2N + k + 1$ for a single line serving all $k$ falls. Because friction along a rope path compounds, each contact costing a fraction of the tension that passes it \citep{jung2008,feyrer2015}, what governs efficiency is the number of elements each unit of transmitted power passes, and the two-line stage removes $\lfloor k/2 \rfloor$ of them from the busiest path whatever $N$ is (Table~\ref{tab:contacts}). A larger $k$ is wanted on three counts, fewer members, slower blocks and a shorter first stage; with one line these gains are opposed because the one rope path that matters lengthens with $k$ contact for contact, with two lines it does not, and the useful range of $k$ moves upward, the benefit compounding since a higher $k$ needs fewer members, which shortens the path again. Table~\ref{tab:contacts} counts elements at a nominal loss each; Section~\ref{sec:friction} carries a per-contact model through the integrators and finds the saving real but about half the nominal figure, since the elements a second line removes are the lightly loaded ones.

\begin{table}[!htbp]
\centering
\caption{Elements removed from the busiest rope path by dividing the $k$ falls between two separately anchored tension members. The saving is $\lfloor k/2 \rfloor$ and is independent of member count.}
\label{tab:contacts}
\begin{adjustbox}{max width=\linewidth}
\begin{tabular}{lccccc}
\toprule
$k$ & 5 & 7 & 9 & 11 & 13 \\
\midrule
elements saved & 2 & 3 & 4 & 5 & 6 \\
efficiency gain at 2\% per element & 4.1\% & 6.2\% & 8.4\% & 10.6\% & 12.9\% \\
\bottomrule
\end{tabular}
\end{adjustbox}
\end{table}

Writing $T_j$ for the tension delivered to block $j$ and $\theta_i$ for the angle of the segments at member $i$ from the direction of travel, the reaction of one array on the carriage is $T_j\sum_i 2\cos\theta_i = T_j G_j$, since $2\cos\theta_i = 2D_i/\sqrt{R_i^2 + D_i^2}$ is exactly the term member $i$ contributes to \eqref{eq:array}; the falls multiply it, and the whole carriage reaction is
\begin{equation}
F_{\mathrm{carriage}} = n_1 T_1 G_1(d) + n_2 T_2 G_2(d),
\label{eq:carriage}
\end{equation}
which at uniform output tension $T_1 = T_2 = F$ reduces to $F\,\Gnet$, conservation of power. The tension in a segment and the reaction it contributes differ by $\cos\theta$, which varies through the stroke; they are not interchangeable.

% ============================================================================
\section{Synthesis by fitting}
\label{sec:synth}

\subsection{The design problem is a curve fit}
\label{sec:fit}

Both the array ratio \eqref{eq:array} and the target \eqref{eq:ode} are functions of source displacement, so determining a single array's geometry is the minimisation
\begin{equation}
\min_{\{R_i,\,s_i\}} \int_0^{\dmax} \bigl|\, k\,\Garr(d) - \Gstar(d) \,\bigr|^2\, dd
\label{eq:single-obj}
\end{equation}
over $2N$ parameters. The objective is smooth except at the engagement points, and bounded least squares solves it in seconds; no dynamic simulation enters the fit, the dynamics being used afterwards to select among fitted candidates and to verify the chosen geometry. That the problem is a fit rather than a search matters, because the optimum is not one a designer would find by inspection. Supplementary Section~\ref{si:search} compares methods on a 1,000:1 target with eight members and sixteen free parameters: least squares reaches a residual of 0.089, 0.18\% of the full-scale ratio range, in about two thousand evaluations, 7.7 times better than 300,000 random samples. Fitted geometries have span widths that are non-monotonic, modest at the first member, maximal at the second and thereafter decreasing severalfold, with engagement intervals that contract steadily. Neither pattern is expressible in two or three parameters, so the low-dimensional families a designer would reach for, uniform spacing, uniform slope or a geometric progression, do not contain the optimum. The solution manifold is nonetheless broad: fits from different random starts converge to geometries differing by 6 to 13\% in their parameters at comparable residual, so the apparatus is correspondingly tolerant of departures from any nominal geometry.

\subsection{Reachable targets and practical bounds}

The fit behaves this well only when the target is reachable, which is why Section~\ref{sec:target} specifies the deceleration rather than the force. A target that asks more energy than the array can deliver does not merely fall short: the least-squares landscape becomes multi-modal, the residual stops predicting anything, and the fitter crowds members into narrow spans to climb faster, so engagement becomes abrupt and traction is lost early. Under the target of Table~\ref{tab:spec} the landscape is effectively unimodal, restarts agree on compliant \ptm{} force to within 2.4\%, 0.7\% and 0.1\% at the three mass ratios, and the plain squared residual needs no weighting, one-sided or smoothness terms. Two practical bounds enter. A span must clear the sheave diameter, the tension member and working clearance, and a floor of 10~cm ($R_i \ge 50$~mm) is imposed throughout; it \emph{improves} the designs, since unconstrained fitting produces spans as narrow as 2~cm with slope discontinuities $2/R$ of order 200, violent enough at engagement to lose traction early. And the total width of each array is bounded by a budget $\Lw$, since the array rides on the carriage and sets the size and mass of the moving structure.

\subsection{Selecting member count and stage ratio}
\label{sec:select}

Member count and stage ratio are not separable: fixing $k$ and sweeping $N$ returns a different and worse design at every mass ratio. The single-array reference designs of Section~\ref{sec:sim} are therefore produced by a joint grid over $N \in \{2,\dots,16\}$ and $k \in \{1,2,3,5,7,9\}$ with fifty restarts of \eqref{eq:single-obj} per cell, the five lowest-residual restarts per cell simulated on both models, admissibility requiring a compliant \ptm{} force of at most 1.3 and at least 99\% of the ideal exit velocity, and selection by a score that rewards exit velocity, penalises \ptm{} force and discounts member count, stage ratio and array width, the three quantities that set what the machine costs to build: 4,500 fits and 450 simulations per mass ratio, affordable because no simulation enters the fit (Supplementary Section~\ref{si:select}). For a single array the selection returns $k = 5$ at 1,000:1 because array width passes through a minimum there: a larger $k$ needs fewer members, but each must then span a wider gap to cover the same rise.

% ============================================================================
\section{The dual-array embodiment}
\label{sec:dual}

\subsection{Architecture and the fall-count weighting}
\label{sec:arch}

\begin{figure}[!htbp]
\centering
\includegraphics[width=\linewidth]{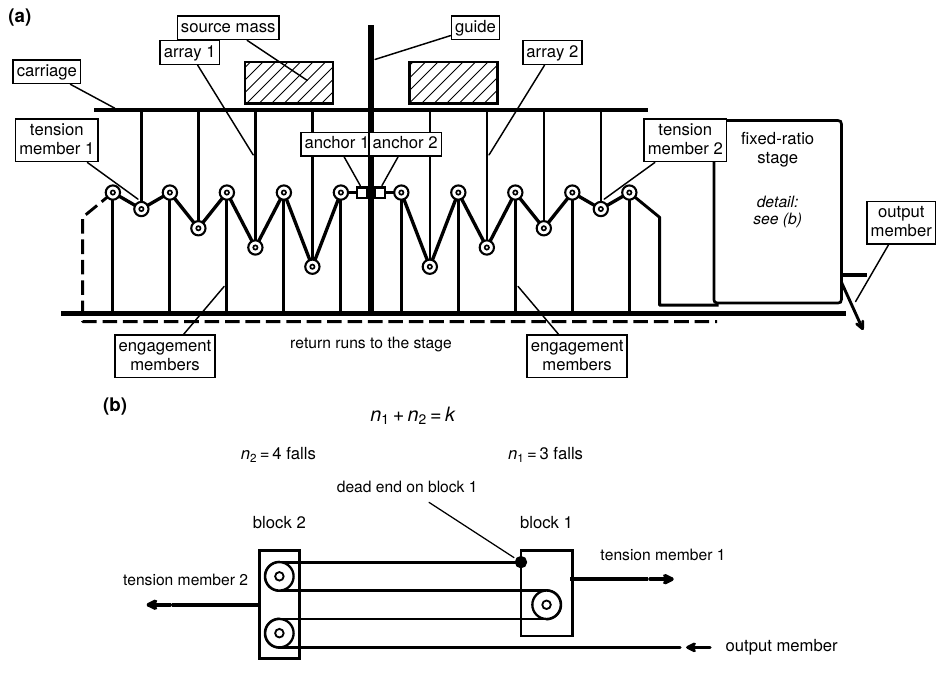}
\caption{The architecture. (a) The dual array with its stationary guide, carriage, source-mass halves and both tension members dead-ended on the guide. (b) The fixed-ratio stage at $k=7$: one continuous output member dead-ends on block~1, passes about the upper sheave of block~2, returns about the single sheave of block~1, passes about the lower sheave of block~2 and leaves for the payload. The segments touching each block number $n_1 = 3$ and $n_2 = 4$; with no stationary sheave, every sheave counts toward $p$ and $k = 2p+1 = 7$.}
\label{fig:arch}
\end{figure}

The transmission of Figure~\ref{fig:hero} carries two arrays, one on each side of the central guide (Figure~\ref{fig:arch}a). Each has its own tension member, dead-ended at a stationary anchor on the guide, running outboard through its own array and returning to the fixed-ratio stage of Section~\ref{sec:stage}, where it drives its own counter-moving block; the output member is reeved with $n_1$ falls about the first block and $n_2$ about the second, $n_1 + n_2 = k$. Separate anchors and separate blocks are what leave the two arrays free to draw in rope at different rates; were the two lines portions of one rope, or acting on one block, any difference in geometry would put one member slack. The order in which the spans are laid along each array is fixed by the rope flow, and Section~\ref{sec:spanorder} states it. Taking up a length $Y_1$ of the first tension member advances the output member by $n_1 Y_1$, and likewise for the second, so the net displacement ratio is
\begin{equation}
\Gnet(d) = n_1\,G_1(d) + n_2\,G_2(d),
\label{eq:gnet}
\end{equation}
where $G_1$ and $G_2$ are the array ratios \eqref{eq:array} of the two arrays. The sum is exact, by the segment counting that gives \eqref{eq:parity}: each segment of the output member changes in length by the travel of the block it touches, and $n_j$ segments touch block $j$.

\subsection{Span order along the array}
\label{sec:spanorder}

Rope enters each array at its outboard end and flows toward the anchor, so a fold is fed through every station outboard of it. With rolling sheaves nothing slides, but each metre of rope passing a station of radius $r_s$ turns it by $\ell_i / r_s$ against a bearing load $2T\sin(\theta_i/2)$, $\theta_i$ the wrap, and the rope takes a bending cycle at tension $T$. The rope passed over station $i$ during the stroke is
\begin{equation}
\ell_i = \tfrac{1}{2} Y_i + \sum_{j\ \text{anchor-side of}\ i} Y_j, \qquad Y_i = 2\left(\sqrt{R_i^2 + D_i^2} - R_i\right),
\label{eq:ropeflow}
\end{equation}
a support between two members passing the sum for the members beyond it. Tension is highest at the anchor (Section~\ref{sec:friction}) and the wrap largest at the deepest fold, so revolutions and bending cycles, and with them the bearing friction and the rope wear, which are what the revolutions and the cycles cost, are all minimised by laying the spans in engagement order from the anchor outward: the first, deepest member at the dead end, where only its own take-up passes over it, and the late, shallow members at the entry, where the whole array's rope passes through the smallest wraps. Over all orders of every configured array this is the minimum of wrap-weighted rope passed, and reversing it about doubles the rope over the deepest member (Supplementary Section~\ref{si:guide}). Every drawing in the paper is laid out so. Applied to both arrays the rule is also a balance rule: the arrays lie mirror-fashion about the guide, their centres of pressure sweep outward together, 1.40 and 1.23~m at the end of the stroke for $\cfg{8{:}6}{7}$ (Section~\ref{sec:moment}), and the balance of axial reactions that this section synthesises is, to that approximation, a balance of moments as well.

\subsection{The four configurations, and what the second array costs}
\label{sec:configs}

Four configurations of the same kinematics recur in this paper, and Table~\ref{tab:configs} sets them beside one another. One array and one line serving all $k$ falls gives $n_2 = 0$ and $\Gnet = k\,\Garr$: the \emph{one-line single array}. Two lines through one array give $G_1 = G_2 = \Garr$, and \eqref{eq:gnet} returns $k\,\Garr$ for any split summing to $k$: the \emph{two-line single array}, which halves the tension per line and shortens the paths but cannot change the force profile and reacts the whole carriage load on one side. Two arrays constrained to be mirror images give the \emph{mirror-symmetric pair}, the two-line single array unfolded onto both sides of the guide: identical in ratio, tension per line and contacts per line, differing only in that the reaction is shared across the guide. The weighting in \eqref{eq:gnet} becomes visible only in the fourth configuration, the \emph{unequal pair}, when the arrays differ.

\begin{table}[!htbp]
\centering
\caption{The four configurations of the same kinematics. $N$ is the member count per array; each engagement member and support carries one groove per line that passes it. The two reaction rows are at uniform tension, $T_1 = T_2$, under which the mirror pair is balanced at even $k$; Section~\ref{sec:odd} derives the mirror residual and treats the tension term that remains when the two line paths differ.}
\label{tab:configs}
\footnotesize
\begin{adjustbox}{max width=\linewidth}
\begin{tabular}{@{}l>{\centering\arraybackslash}p{2.2cm}>{\centering\arraybackslash}p{2.4cm}>{\centering\arraybackslash}p{2.4cm}>{\centering\arraybackslash}p{2.6cm}@{}}
\toprule
& one-line single & two-line single & mirror pair & unequal pair \\
\midrule
net ratio & $k\,\Garr$ & $k\,\Garr$ & $k\,G$ & $n_1 G_1 + n_2 G_2$ \\
independent geometric parameters & $2N$ & $2N$ & $2N$ & $4N$ \\
tension per line & $kF$ & $n_j F$ & $n_j F$ & $n_j F$ \\
elements on the busiest path & $2N + k + 1$ & $2N + \lceil \tfrac{k}{2}\rceil + 1$ & $2N + \lceil \tfrac{k}{2}\rceil + 1$ & $2N + \lceil \tfrac{k}{2}\rceil + 1$ \\
sheave grooves in the arrays & $2N+1$ & $2(2N+1)$ & $2(2N+1)$ & $2(2N+1)$ \\
load per engagement member & $kF$ & $kF$ & $n_j F$ & $n_j F$ \\
reaction on the guide & moment & moment & balanced at even $k$ & balanceable at any $k$ \\
residual reaction imbalance at odd $k$ & --- & --- & $|n_1 - n_2|/k$ & designer's choice \\
distinct span widths & $N$ & $N$ & $N$ & $2N$ \\
\bottomrule
\end{tabular}
\end{adjustbox}
\end{table}

The accounting of the second array is made against the two-line single array, which already has the stage benefits of Section~\ref{sec:stage}. That machine carries both lines on each of its $N$ members and $N+1$ supports, $2(2N+1)$ sheave grooves loaded at $n_1 F$ and $n_2 F$; the dual array has the same grooves on $2N$ members and $2N+2$ supports, each loaded by one line. Nothing changes in the fast path, since the output member, its sheaves and the two blocks are the same, and the arrays ride on the carriage as source-side mass rather than mass accelerated to launch speed. Nothing changes in the loss per unit power: each line passes $2N$ array contacts and the same stage contacts at the same tension, and bearing friction, being linear in load, is the same whether one groove carries $n_1 F$ or two grooves carry $n_1 F$ and $n_2 F$. What changes is that each line has its own $R_i$ and $s_i$, $4N$ geometric parameters in place of $2N$, or $2(N_1 + N_2)$ when the counts differ (Section~\ref{sec:cond}); that each member and support carries one line's load instead of two, roughly halving its bending and bearing loads and admitting a lighter line at the same safety factor, hence a lower bending stress over the same sheave, which sets the bend-over-sheave life of a fibre rope \citep{davies2015,feyrer2015}; and that the reaction is taken on both sides of the guide, which Section~\ref{sec:cond} shows is what allows it to be cancelled. The price is a second row of supports, a carriage loaded on both sides, and $2N$ distinct span widths and engagement heights where the mirror pair needs $N$. The second array doubles the design freedom at no cost in friction, fast-path inertia or sheave count; its cost is structure and distinct parts.

\subsection{The balance condition}
\label{sec:cond}

By \eqref{eq:carriage} the force each array applies to the carriage is the product of its fall count, the tension delivered to its block and its own displacement ratio, and it acts along the guide, since each fold is symmetric about its member and the transverse components of the rope tension cancel. The imbalance of the two axial reactions is the difference,
\begin{equation}
\Delta F(d) = \bigl|\, n_1 T_1 G_1(d) - n_2 T_2 G_2(d) \,\bigr|,
\label{eq:imbalance}
\end{equation}
and it vanishes at every point of the stroke where
\begin{equation}
n_1 T_1 G_1(d) = n_2 T_2 G_2(d), \qquad\text{that is}\qquad G_1 : G_2 = n_2 T_2 : n_1 T_1 .
\label{eq:balance}
\end{equation}
\textbf{The array serving more falls, or suffering less loss along its path, must be given the smaller displacement ratio}: the array doing more work at the stage is the one to be made weaker. With the output member at one uniform tension, $T_1 = T_2$, \eqref{eq:balance} reduces to $G_1 : G_2 = n_2 : n_1$, and with \eqref{eq:gnet} each array carries half the weighted target,
\begin{equation}
G_1(d) = \frac{\Gstar(d)}{2 n_1}, \qquad G_2(d) = \frac{\Gstar(d)}{2 n_2},
\label{eq:split}
\end{equation}
the form used for the designs of Section~\ref{sec:sim}, with equal member counts in Section~\ref{sec:duals} and counts in the same ratio in Section~\ref{sec:counts}. Mirror symmetry, $G_1 = G_2$, is the special case of \eqref{eq:balance} in which both the fall counts and the delivered tensions are equal, and only there. Since each engaged member contributes at most 2, \eqref{eq:split} fixes a floor $N_j \ge \Gstar(\dmax)/(4 n_j)$ on the members per array; Table~\ref{tab:floors} gives it for the 1,000:1 target, whose terminal ratio is 79.3, and raising the stage ratio lowers it on both arrays, the mechanism of Section~\ref{sec:stage} seen from the other side.

Saturation gives the same rule from the other end of the stroke. Deep in the stroke every engaged member approaches its asymptote of 2, so $G_1 \to 2N_1$ and $G_2 \to 2N_2$, and the reaction imbalance \eqref{eq:imbalance} tends to $|n_1 N_1 - n_2 N_2| / (n_1 N_1 + n_2 N_2)$ whatever the spans: with equal member counts it is the mirror residual $|n_1 - n_2|/k$ of Section~\ref{sec:odd}, because the counts, not the geometry, set the terminal ratios. Exact balance at the end of the stroke therefore requires
\begin{equation}
n_1 N_1 = n_2 N_2, \qquad\text{that is}\qquad N_1 : N_2 = n_2 : n_1,
\label{eq:counts}
\end{equation}
the discrete counterpart of \eqref{eq:balance}. At $k=9$ the floors of Table~\ref{tab:floors}, 5 and 4, already stand in that ratio; at $k=7$ the smallest counts above the floors that satisfy it are 8 and 6. The fit carries the balance through the middle of the stroke and the counts carry it at the end; Section~\ref{sec:counts} configures both cases.

Throughout, a \emph{reaction imbalance} stated as a percentage is the greatest value of \eqref{eq:imbalance} over the design stroke $0 \le d \le \dmax$ divided by the greatest value of $n_1 G_1 + n_2 G_2$ over the same stroke, at uniform tension: a peak against a peak on the fitted geometry, not an instantaneous ratio, which is meaningless near the start of the stroke where both contributions vanish. What the guide bearings react is not this force but its moment. Member $i$ of array $j$ pushes on the carriage with $f_i = n_j T_j\, g_i(d)$, $g_i = 2D_i/\sqrt{R_i^2 + D_i^2}$ its term in \eqref{eq:array}, at its position $x_i$ along the array, so the pitch moment about the guide axis is
\begin{equation}
\mathcal{M}(d) = \sum_{i \in 2} x_i f_i - \sum_{i \in 1} x_i f_i,
\label{eq:moment}
\end{equation}
reacted by two bearings at spacing $h$ as a couple $\mathcal{M}/h$. It equals $\bar{e}\,\Delta F$ only where the two arrays' centres of pressure $\bar{x}_j = \sum x_i f_i / \sum f_i$ coincide, and the positions $x_i$ are free: the ratio \eqref{eq:array} is a sum over members and does not depend on the order in which the spans are laid along the array. The synthesis of Section~\ref{sec:dualsynth} balances \eqref{eq:imbalance}; Section~\ref{sec:moment} evaluates \eqref{eq:moment} for the configured designs and shows what the placement buys.

\begin{table}[!htbp]
\centering
\caption{Minimum members per array set by \eqref{eq:split}, for the 1,000:1 target whose terminal net ratio is 79.3. Array~1 leads and serves the fewer falls (Section~\ref{sec:order}), so it carries the larger displacement ratio and the higher floor. At $k=9$ the two floors satisfy \eqref{eq:counts} exactly.}
\label{tab:floors}
\begin{adjustbox}{max width=\linewidth}
\begin{tabular}{cccccc}
\toprule
$k$ & falls $n_1/n_2$ & $G_1$ required & $G_2$ required & min.\ members, array 1 & min.\ members, array 2 \\
\midrule
5  & 2 / 3 & 19.82 & 13.22 & 10 & 7 \\
7  & 3 / 4 & 13.22 & 9.91  & 7  & 5 \\
9  & 4 / 5 & 9.91  & 7.93  & 5  & 4 \\
11 & 5 / 6 & 7.93  & 6.61  & 4  & 4 \\
\bottomrule
\end{tabular}
\end{adjustbox}
\end{table}

\subsection{Why symmetry cannot balance an odd stage}
\label{sec:odd}

\begin{figure}[!htbp]
\centering
\includegraphics[width=\linewidth]{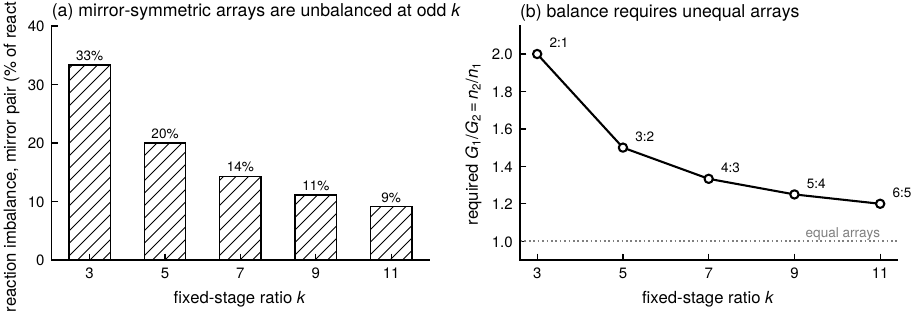}
\caption{Reaction imbalance and the balance condition. (a) The imbalance a mirror-symmetric machine carries as a function of stage ratio, \eqref{eq:residual}, zero only at even $k$. (b) The ratio $G_1/G_2$ the two arrays must hold to cancel it.}
\label{fig:imbalance}
\end{figure}

Where $k$ is odd the fall counts differ by one and $G_1 : G_2 = n_2 : n_1$ cannot be satisfied by equal arrays. A mirror-symmetric machine is then left with a residual reaction imbalance of
\begin{equation}
\frac{|\,n_1 - n_2\,|}{k}
\label{eq:residual}
\end{equation}
of the total: \textbf{33\% at $k=3$, 20\% at $k=5$, 14\% at $k=7$, 11\% at $k=9$}, zero only at even $k$ (Figure~\ref{fig:imbalance}a). This is a property of the reeving, not a tolerance: no adjustment, matching of components or care in construction reduces it, and it is far larger than any mismatch a built machine should otherwise show; a 20\% imbalance at 1,000:1 is 45.7~kN of unbalanced axial reaction, applied at the arrays' offset as a pitch moment on the guide. Since the counter-moving stage yields $k = 2p+1$, the odd case is the general one.

The assumption $T_1 = T_2$ is itself an idealisation: the output member loses tension at every element it passes, so the two blocks are not pulled equally, and the two tension members traverse different numbers of elements between anchor and stage and may differ in free length, elastic extension and accelerated rope mass. Mirror symmetry therefore does not fully balance a machine even at even $k$; equal fall counts cancel the fall-count term of \eqref{eq:balance} and leave the tension term, which mirror-image arrays, constrained to share one geometry, have no freedom to remove. Unequal arrays have it: the objective of Section~\ref{sec:dualsynth} extends by one term in the tension ratio, estimated from element counts and a per-element loss or measured on a built machine. The designs of Section~\ref{sec:sim} are configured on \eqref{eq:split} without that term, and Section~\ref{sec:friction} shows what it omits for the routing of Figure~\ref{fig:arch} to be under half a percentage point of reaction imbalance at 2\% loss per contact with equal member counts, and about one point once the member counts of Section~\ref{sec:counts} have removed the terminal residual that otherwise hides it, the two paths differing by a single stage segment; a dissimilar routing would see proportionately more.

\subsection{Engagement ordering: the low-fall array leads}
\label{sec:order}

The array ratio \eqref{eq:array} is continuous at the moment a member engages; it is the slope $dG/dd$ that steps. The reaction imbalance \eqref{eq:imbalance} therefore drifts toward whichever array engaged most recently, at a rate proportional to that array's fall count. \textbf{The array that engages first must be the one driving the block with the fewer falls}, $n_1 = \lfloor k/2 \rfloor$: leading with the low-fall array makes the first excursion grow at the lesser rate and lets the other array close it, while leading with the high-fall array does the opposite. Throughout this paper subscript~1 denotes the leading array, so $n_1 = \lfloor k/2 \rfloor$ and $n_2 = \lceil k/2 \rceil$: 2 and 3 at $k=5$, 3 and 4 at $k=7$, 4 and 5 at $k=9$.

Table~\ref{tab:leadorder} compares the two assignments. Each sits on its own accuracy-against-balance curve, so the comparison is between curves rather than at a single value of the balance weight $\lambda$ of Section~\ref{sec:dualsynth}; each entry is the best of fourteen restarts. At $k=7$ every high-fall-leading point is beaten on both axes by the low-fall point at the same $\lambda$. At $k=9$ the high-fall-leading column shows the lower reaction imbalance at five of six values of $\lambda$, but those points are bought with a worse residual, and a low-fall-leading fit at larger $\lambda$ reaches the same balance at better accuracy, 3.3\% at rms 1.057 against 3.7\% at rms 1.101; only at $\lambda = 10$ does the high-fall curve hold a point the low-fall curve does not dominate. The effect is strongest at small $k$, where the fall counts differ most in proportion: at $k=5$ they are 2 and 3, and leading with the wrong array roughly doubles the reaction imbalance, from 8.9\% to 14.5\% at $\lambda = 0.3$ and from 6.7\% to 15.4\% at $\lambda = 1$.

\begin{table}[!htbp]
\centering
\caption{The two lead assignments as trade curves, at the 1,000:1 target. Each row is one value of the balance weight $\lambda$ and each entry the best of fourteen restarts; rms is the residual against the target. Bold marks the better value of each pair. Comparison is between curves, not between rows.}
\label{tab:leadorder}
\footnotesize
\begin{adjustbox}{max width=\linewidth}
\begin{tabular}{lcccc}
\toprule
& \multicolumn{2}{c}{low-fall array leads} & \multicolumn{2}{c}{high-fall array leads} \\
\cmidrule(lr){2-3}\cmidrule(lr){4-5}
$\lambda$ & rms & reaction imbalance & rms & reaction imbalance \\
\midrule
\multicolumn{5}{l}{\emph{$k=7$, seven members per array}}\\
0.1  & \textbf{0.400} & \textbf{7.2\%} & 0.776 & 9.7\% \\
0.3  & \textbf{0.436} & \textbf{6.0\%} & 0.886 & 7.5\% \\
1    & \textbf{0.558} & \textbf{4.7\%} & 0.998 & 5.9\% \\
3    & \textbf{0.807} & \textbf{3.5\%} & 1.106 & 5.0\% \\
10   & \textbf{0.995} & \textbf{1.4\%} & 1.523 & 5.4\% \\
30   & \textbf{1.110} & \textbf{0.9\%} & 2.193 & 4.4\% \\
\midrule
\multicolumn{5}{l}{\emph{$k=9$, five members per array}}\\
0.1  & \textbf{0.662} & 6.6\% & 0.933 & \textbf{6.5\%} \\
0.3  & \textbf{0.704} & 6.1\% & 1.057 & \textbf{5.0\%} \\
1    & \textbf{0.916} & 4.4\% & 1.101 & \textbf{3.7\%} \\
3    & \textbf{1.057} & 3.3\% & 1.167 & \textbf{2.3\%} \\
10   & \textbf{1.159} & 1.8\% & 1.212 & \textbf{1.4\%} \\
30   & \textbf{1.235} & \textbf{1.1\%} & 1.256 & 1.1\% \\
\bottomrule
\end{tabular}
\end{adjustbox}
\end{table}

Alternation of the two arrays' engagements is a consequence of the correct lead, not a cause of balance: once the lead is assigned the fitted offsets alternate whether or not alternation is imposed. An unconstrained fit at $k=5$ produced the order LSLSLLSLSLLSLLLSS, with a longest run of three from either array, and imposing strict alternation on the same problem raised the reaction imbalance from 1.8\% to 3.1\%. What imposing it buys is in the fitting: parametrising the merged offset sequence by positive increments dealt alternately makes every candidate feasible by construction, so the seeded method below converges in a single solve where unconstrained restarts needed dozens. Coincident and near-coincident engagements remain permissible and occur in the configured designs. With the member counts of \eqref{eq:counts} the leading array has members in excess, and where they go matters: dealt first they raise the peak reaction imbalance of the $k=7$ design of Section~\ref{sec:counts} from 2.3\% to 14.4\%, dealt in the middle to 4.1\%, so the excess engages last, and the alternation itself survives a free solve: released from the order constraint and started from the alternating solution, the fit does not leave it, and every basin it otherwise reaches with two consecutive engagements of one array is worse in objective, by 1 to 7\%, the worst of them carrying 11 to 12\% reaction imbalance (Supplementary Section~\ref{si:counts}).

\subsection{Synthesis of the unequal pair}
\label{sec:dualsynth}

For the dual array the unknowns are the $2(N_1 + N_2)$ parameters of both arrays, and the requirement is that $n_1 G_1 + n_2 G_2$ approximate the target while $n_1 G_1$ and $n_2 G_2$ stay close to one another:
\begin{equation}
\min\; \bigl\| n_1 G_1 + n_2 G_2 - \Gstar \bigr\|^2 \;+\; \lambda\, \bigl\| n_1 G_1 - n_2 G_2 \bigr\|^2,
\label{eq:dual-obj}
\end{equation}
subject to the span floor, the width budget $\Lw$ per array, and the interleaving of Section~\ref{sec:order} imposed by construction. The balance weight $\lambda$ traverses a continuous family from best fit to exact cancellation; it is a dial the designer sets, not a value to be maximised. Table~\ref{tab:lambda} shows both ends of the family at an even stage ratio, $k=8$, where exact cancellation is reachable by mirror symmetry: at $\lambda = 0$ the two arrays, free to differ, track the target about three times more closely than mirror-image arrays of the same member count and width, at a reaction imbalance of 6.9\%; imposing balance drives them back onto the mirror solution and returns the accuracy with it. Where the doubled freedom is not needed for load cancellation it is available for profile accuracy.

\begin{table}[!htbp]
\centering
\caption{The accuracy--balance trade at $k=8$, seven members per array, on the 1,000:1 target. Imposing balance drives the two arrays back onto the mirror solution and returns the profile accuracy with it.}
\label{tab:lambda}
\begin{adjustbox}{max width=\linewidth}
\begin{tabular}{lccc}
\toprule
balance weight $\lambda$ & rms against target & reaction imbalance & mean difference in array spans \\
\midrule
0   & \textbf{0.197} & 6.9\% & 50 mm \\
0.3 & 0.603 & 0.2\% & 0.9 mm \\
1.0 & 0.603 & 0.0\% & 0.0 mm \\
\bottomrule
\end{tabular}
\end{adjustbox}
\end{table}

The objective \eqref{eq:dual-obj} is not convex, and two devices make it tractable. \emph{Two-stage seeding}: fit a single array to the same stage by \eqref{eq:single-obj}, initialise both arrays with that solution, and displace the second array's offsets to positions intermediate between those of the first; because $n_1 + n_2 = k$, the pair reproduces the single array's net ratio exactly and the seed is already feasible. For unequal member counts the same idea is applied to one array of $N_1 + N_2$ members fitted at the ratio $k/2$ and dealt alternately to the two arrays, the leading array taking the first member and the excess; with $G_1 = G_2$ the pair again reproduces the net ratio. This seed fit, at half the stage ratio and with a generous width cap, converges to one design from any start, so the restarts that the joint objective needs are supplied by perturbing the dealt seed; the plain seed itself ranks between third and nineteenth of the 31 starts and up to 35\% above the best in objective (Supplementary Section~\ref{si:counts}). For the $\cfg{7}{7}$ design of Section~\ref{sec:sim} one seeded fit reaches rms 0.380, better than the best of forty random restarts of the same objective, which plateau at 0.383, at about the cost of one restart. \emph{Staged solution}: solve for span widths with offsets held fixed, then for both jointly; the first phase is far better conditioned and stops two offsets migrating onto each other before the widths adapt. The whole procedure is Algorithm~1, and nothing in it is tuned: the designer supplies the two masses, the entry and terminal velocities, the stroke duration, the stage ratio, the members per array, a width budget and $\lambda$.

\begin{algobox}
\vspace{-4pt}
\begin{Verbatim}[fontsize=\scriptsize]
INPUT   M, m, v0, v_end, T    masses; entry velocity; final velocity; duration
        k, N1, N2, L_w, lam   stage ratio; members per array; width budget;
                              balance weight
OUTPUT  n1, n2;  R1[], s1[];  R2[], s2[]

1  TARGET      bisect on F until the ideal run, ended at v_h = v_end,
               takes time T  ->  F, d_max, Gstar(d)                   (Sec. 2)
2  FALLS       n1 = floor(k/2) for the LEADING array, n2 = ceil(k/2)  (Sec. 5.5)
               member floors  Nj >= Gstar(d_max) / (4 nj);
               terminal balance  n1 N1 = n2 N2                        (Sec. 5.3)
3  SEED        one array of N1 + N2 members fitted at ratio k/2:
               Rs[], ss[] = argmin SUM_d [ (k/2) Garray(d; R, s) - Gstar(d) ]^2,
               bounded least squares, width <= 3 L_w                  (Sec. 4.1)
               (for N1 = N2 equivalently N members at ratio k, width <= L_w)
4  TWO ARRAYS  deal the members in engagement order alternately to arrays
               1, 2, 1, 2, ..., array 1 first and last;  net ratio reproduced
               (N-member seed: R1 = R2 = Rs, s1 = ss, s2 at the midpoints of ss)
5  PARAMETRISE merged offsets from positive increments g = softplus(p_gap) + eps,
               s_all = d_max sigmoid(p_span) cumsum(g) / sum(g), dealt alternately
               to arrays 1, 2, 1, 2, ...;  half-widths R = R_floor + softplus(p_R)
6  RESTARTS    the dealt seed and 30 perturbations of it (log-normal widths
               and gaps, sigma 0.3); keep the best of 7-8 by J
7  PHASE 1     minimise J over p_R only, offsets frozen
8  PHASE 2     minimise jointly
               J = SUM_d [ n1 G1 + n2 G2 - Gstar ]^2
                 + lam SUM_d [ n1 G1 - n2 G2 ]^2
                 + w_L [ max(0, 2 sum R1 - L_w)^2 + max(0, 2 sum R2 - L_w)^2 ]
9  EVALUATE    simulate rigid and compliant; p2m over the first 99.5% of the
               trace; reaction imbalance = max |n1 G1 - n2 G2| / max (n1 G1 + n2 G2)
\end{Verbatim}
\end{algobox}

% ============================================================================
\section{Dynamic simulation and comparative verification}
\label{sec:sim}

\subsection{Simulation setup}
\label{sec:setup}

Two integrators are used, both given in full in Supplementary Sections~\ref{si:rigid} and~\ref{si:compliant}. The \emph{rigid} model treats the tension member as inextensible and enforces the unilateral contact condition, that the member may pull but not push, by advancing the payload by the greater of its free-flight displacement and the displacement required to remain in contact; it has no friction, rotational inertia, member mass or drag, and no free parameters. The \emph{compliant} model adds one element, the output member as a linear spring of stiffness $\Krope = EA/\Lrope$ pre-tensioned to the working load. For a 2~mm Dyneema output member, $E \approx 100$~GPa and $A \approx 2.8\times10^{-6}$~m$^2$ \citep{marlow,mckenna2004}, spanning the 15.8~m of payload travel of the 1,000:1 case plus about 2~m at the blocks and in reserve, $\Lrope \approx 17.8$~m and $\Krope \approx 1.6\times10^4$~N/m; the value used throughout is 16,471~N/m. The choice is not material-specific: with $A = F_{\mathrm{break}}/\sigma$, $\Krope = (E/\sigma)F_{\mathrm{break}}/\Lrope$, and $E/\sigma$ lies between about 34 and 66 for UHMWPE, steel wire rope, carbon fibre and PBO alike, so working strain is of order 1\% whatever the member \citep{davies2011}; Supplementary Section~\ref{si:stiffness} shows the results insensitive across that band. Pre-tension matters: a member starting slack rings the payload mode at about 20~Hz against a stroke under 100~ms, giving a \ptm{} force near 2.3 with the member unloaded for some 11\% of the stroke, whereas pre-tensioning to the working tension removes the step input. The ideal run against which every design is judged is computed by the same integration loop with the ratio solved for at each step, so that ideal and array share identical contact handling, reaction force and one-step lag. Time steps are $10^{-5}$~s rigid and $2\times10^{-5}$~s compliant; refining by a factor of twenty changes exit velocity by within 1.8\% and 0.07\%, and \ptm{} force by within 4.3\% and 0.23\% (Supplementary Section~\ref{si:numerics}).

\paragraph{Conventions} Every stroke is terminated by the traction-loss instability of Section~\ref{sec:instab}, and the payload is released immediately before the terminal transient it produces, so all \ptm{} figures are taken over the first 99.5\% of the simulated time. The convention matters for the rigid figures only: on the $\cfg{7}{7}$ design the rigid \ptm{} is 1.93 measured this way, 6.14 if the final sample, the traction-loss spike itself, is included, and 1.93 alike at the first 99.9, 99.5, 99, 98 and 95\%. Reaction imbalances are geometric quantities from the fitted ratios over the whole design stroke, as defined in Section~\ref{sec:cond}; restricting them to the 99.8\% of $\dmax$ that the compliant release window reaches would lower the reported figures of the equal-count designs, whose peaks sit at the end of the stroke, by 0.4 and 0.2 percentage points, to 7.7\% at $\cfg{7}{7}$ and 6.8\% at $\cfg{5}{9}$, and leave the count-balanced designs unchanged; the full-stroke figures are quoted as the conservative choice and the ones the fit itself minimises.

\subsection{Single-array reference designs at three mass ratios}
\label{sec:singles}

\begin{figure}[!htbp]
\centering
\includegraphics[width=\linewidth]{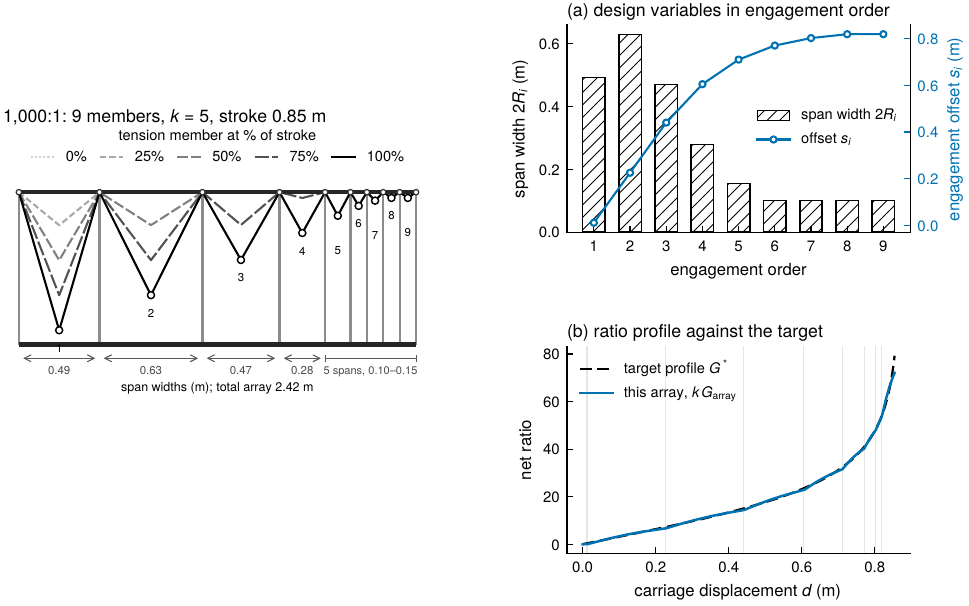}
\caption{The single-array reference at 1,000:1: nine members at $k=5$ in a 2.42~m array. The tension member is drawn at five points through the stroke, span widths are dimensioned beneath, and the design variables are plotted in engagement order against the target profile $\Gstar(d)$. Span width peaks at the second member and narrows to the 10~cm floor by the last.}
\label{fig:single-comb}
\end{figure}

The one-line single array, $n_2 = 0$, provides the reference designs against which the dual array is measured, worked at all three mass ratios of Table~\ref{tab:spec} by the joint selection of Section~\ref{sec:select}. Table~\ref{tab:single} gives the three designs, each fitted by plain least squares to the target of Section~\ref{sec:target} with a 1~kg payload and the 10~cm span floor; every fitted member engages in every case. Their residuals of 0.57 to 1.48\% are set by the width budget of the selection, not by the fit: with the same span floor and no practical width limit, Supplementary Section~\ref{si:search} matches a 1,000:1 target with eight members to 0.18\%. Figure~\ref{fig:single-comb} shows the 1,000:1 design, the reference for the dual designs of Sections~\ref{sec:duals} and~\ref{sec:counts}, and Figure~\ref{fig:single-compliant} its dynamics; the 100:1 and 10,000:1 designs are drawn in Supplementary Section~\ref{si:cases}, which also renders the single-array machine itself.

\begin{table}[!htbp]
\centering
\caption{Single-array reference designs at the three mass ratios, each selected by the joint $N$--$k$ procedure of Section~\ref{sec:select}. Peak forces are multiples of the design force over the release window; compliant figures are for an output member pre-tensioned to the working tension at 16,471~N/m. The last row applies the fast-path sheave correction of Section~\ref{sec:sheave} at $\neff = 3$ to $2$.}
\label{tab:single}
\small
\begin{adjustbox}{max width=\linewidth}
\begin{tabular}{l>{\centering\arraybackslash}p{2.7cm}>{\centering\arraybackslash}p{2.7cm}>{\centering\arraybackslash}p{2.7cm}}
\toprule
 & 100:1 & 1,000:1 & 10,000:1 \\
\midrule
design payload force $F$ & 999 N & 3,170 N & 10,034 N \\
engagement members $N$ & \textbf{8} & \textbf{9} & \textbf{13} \\
fixed-stage ratio $k$ & \textbf{2} & \textbf{5} & \textbf{9} \\
array width & 2.82 m & 2.42 m & 1.92 m \\
spans, in engagement order (cm) & 65, 83, 57, 31, 16, 10, 10, 10 & 49, 63, 47, 28, 15, 10, 10, 10, 10 & 24, 29, 25, 19, 14, 10 ($\times 8$) \\
fit residual, \% of full scale & 0.57\% & 0.78\% & 1.48\% \\
ideal exit velocity & 99.9 m/s & 317.0 m/s & 1,003.5 m/s \\
exit velocity, rigid & 89.8 m/s & 284.8 m/s & 885.9 m/s \\
exit velocity, compliant & 100.2 m/s & 318.2 m/s & 1,004.8 m/s \\
\ptm{} force, rigid & 2.05 & 2.07 & 2.10 \\
\ptm{} force, compliant & \textbf{1.12} & \textbf{1.09} & \textbf{1.19} \\
compliant exit, sheave-corrected & 87.9--91.5 m/s & 279.1--290.5 m/s & 881.3--917.2 m/s \\
\bottomrule
\end{tabular}
\end{adjustbox}
\end{table}

Three features of Table~\ref{tab:single} carry into the dual-array evaluation. Every design reaches essentially all of its ideal exit velocity on the compliant model and meets the \ptm{} bound of 1.3 there, while no rigid configuration does: across the 1,350 candidates simulated in the three grids the lowest rigid \ptm{} force is 1.31, and the three selected designs sit at 2.05, 2.07 and 2.10. The force criterion is therefore a requirement on compliance, not on geometry, and a pre-tensioned compliant output member is a necessary feature of the apparatus, not a refinement. Member count and stage ratio rise with mass ratio but far more slowly, $N$ from 8 to 13 and $k$ from 2 to 9 across a hundredfold change against terminal ratios of 25, 79 and 251, the $\sqrt{M/m}/k$ scaling of Section~\ref{sec:scaling} at work. And at 1,000:1 the selection returns $k = 5$ because array width passes through a minimum there: with member floors $\Gnet/2k$ of 39.6, 19.8, 13.2, 7.9, 5.7 and 4.4 at $k = 1$, 2, 3, 5, 7 and 9, the narrowest admissible array is 1.56~m at $k=3$, 1.44~m at $k=5$, 2.63~m at $k=7$ and 3.12~m at $k=9$. The stage ratios of 7 and 9 at which the dual designs are worked are thus ones a single array handles poorly, which is the point of the comparison in Section~\ref{sec:compare}.

\begin{figure}[!htbp]
\centering
\includegraphics[width=\linewidth]{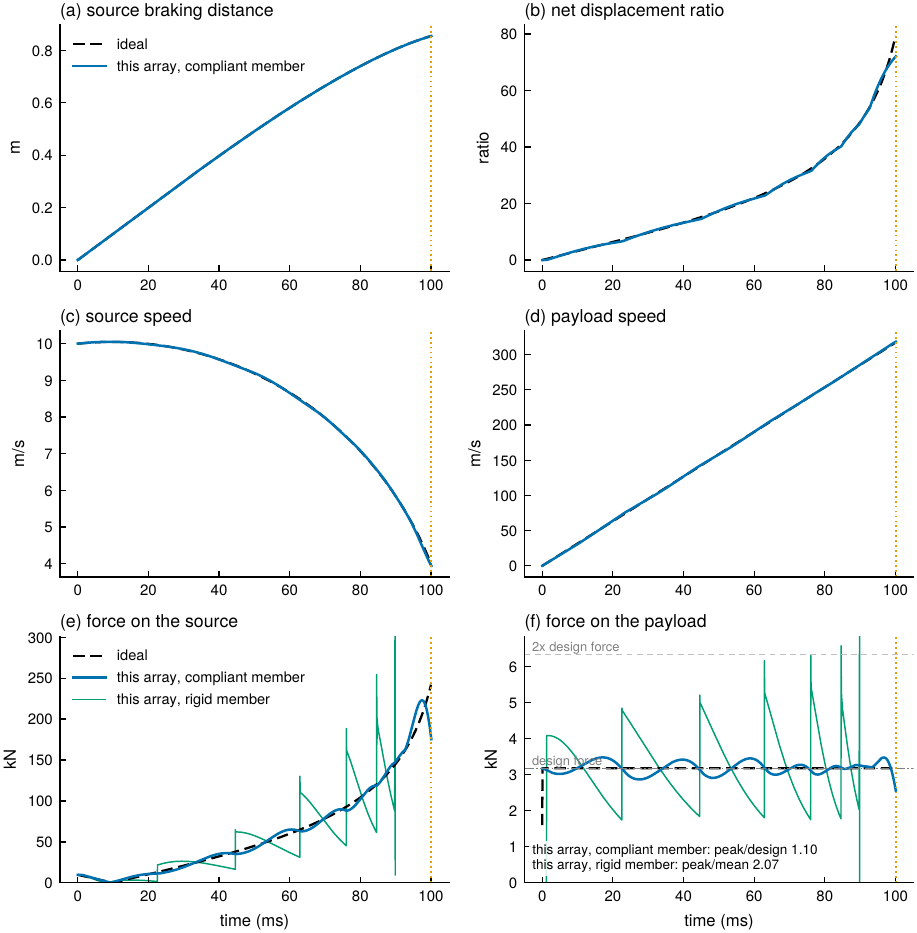}
\caption{The 1,000:1 reference array with a compliant, pre-tensioned output member at 16,471~N/m. Dashed curves are the ideal run, a uniform 3,170~N on the payload, computed by the same integration loop with the ratio solved for at each step. The green trace in (e) and (f) is the same array with an inextensible member (Supplementary Section~\ref{si:figs}): one bounded sawtooth per engagement, ending in the traction-loss spike at 90~ms. Compliance replaces it with a smooth ripple between 2.8 and 3.5~kN about the design force; (f) is drawn from zero so the ripple can be read against the absolute scale.}
\label{fig:single-compliant}
\end{figure}

\subsection{Equal-count dual-array designs at 1,000:1}
\label{sec:duals}

Two equal-count pairs are configured by Algorithm~1 for the 1,000:1 case, and Section~\ref{sec:counts} adds two whose member counts satisfy \eqref{eq:counts}: source mass 1,000~kg, payload 1~kg, entering at 10~m/s and leaving at 4~m/s, uniform payload force 3,169.6~N, braking distance 0.854~m. Both equal-count pairs are fitted at $\lambda = 0.03$ and simulated at 16,471~N/m. They are denoted by $\cfg{N}{k}$: $\cfg{7}{7}$, seven members per array at $k=7$, and $\cfg{5}{9}$, five members per array at $k=9$, the second with its leading array at the member floor of Table~\ref{tab:floors}.

\begin{figure}[!htbp]
\centering
\includegraphics[width=\linewidth]{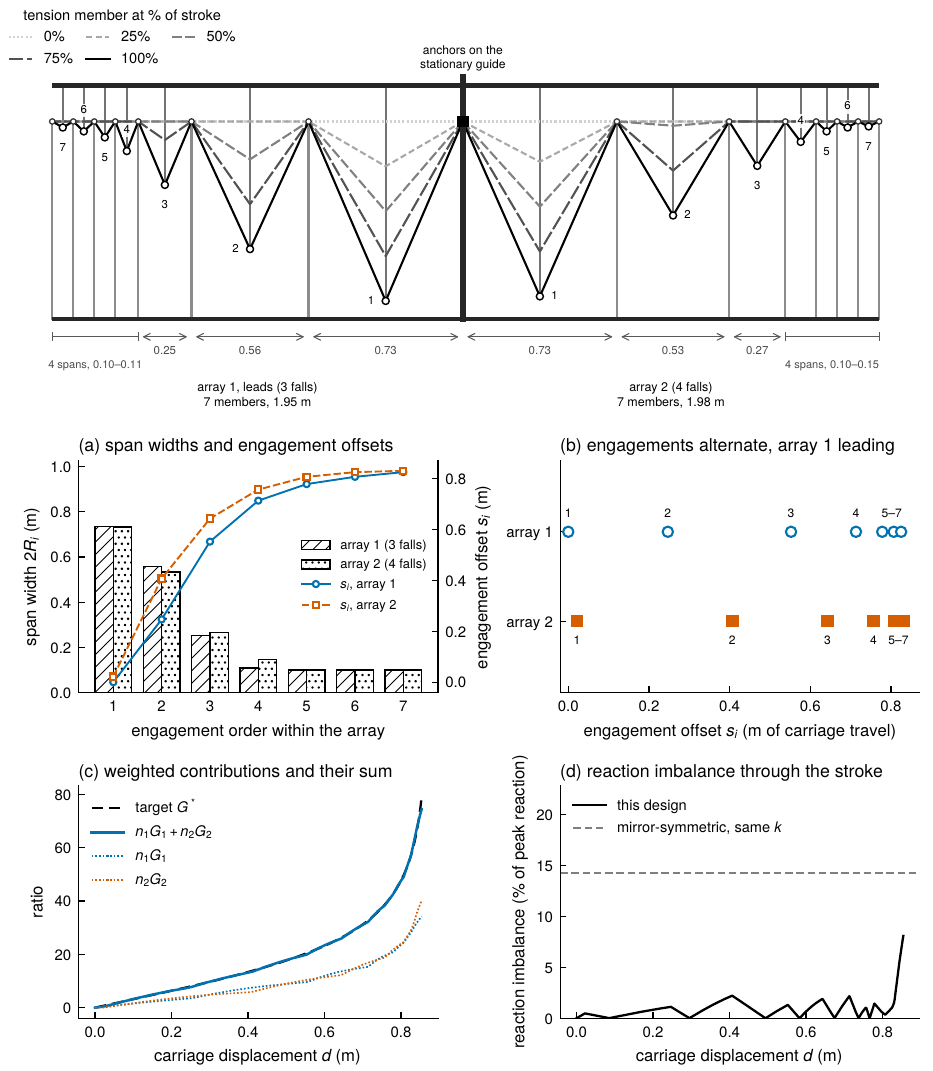}
\caption{The $\cfg{7}{7}$ design. Top: to scale, both arrays either side of the central guide, anchors on the guide, tension members at 0, 25, 50, 75 and 100\% of the stroke, members numbered in engagement order, spans dimensioned beneath. (a) Span widths and engagement offsets per array; (b) the engagement sequence along the stroke, alternating with array~1 leading; (c) the weighted contribution of each array and their sum against the target; (d) reaction imbalance through the stroke against the mirror-symmetric value.}
\label{fig:k7}
\end{figure}

\paragraph{Seven members per array, $k=7$} Three falls about the block driven by the leading array, four about the other; Table~\ref{tab:k7} and Figure~\ref{fig:k7}. The engagement order alternates throughout, the three-fall array leading. Peak reaction imbalance is 8.1\% of the peak carriage reaction of 236~kN, 19.1~kN, against the whole 236~kN that a single array places on one side of the guide and the 33.7~kN (14.3\%) of the mirror-symmetric benchmark at the same stage ratio.

\begin{table}[!htbp]
\centering
\caption{Configured design $\cfg{7}{7}$: seven members per array at $k=7$, 1,000:1 mass ratio, fitted at $\lambda = 0.03$. The greater array width is 1.98~m; the terminal ratio reaches 94\% of target. Full-precision coordinates are given in Supplementary Section~\ref{si:coords}.}
\label{tab:k7}
\begin{adjustbox}{max width=\linewidth}
\begin{tabular}{lll}
\toprule
& array 1, serving 3 falls & array 2, serving 4 falls \\
\midrule
span half-widths $R_i$ (mm)  & 367, 279, 127, 55, 50, 50, 50 & 367, 267, 134, 73, 50, 50, 50 \\
engagement offsets $s_i$ (mm) & 0, 247, 552, 713, 778, 807, 825 & 22, 407, 643, 757, 807, 825, 831 \\
\bottomrule
\end{tabular}
\end{adjustbox}
\end{table}

\begin{figure}[!htbp]
\centering
\includegraphics[width=\linewidth]{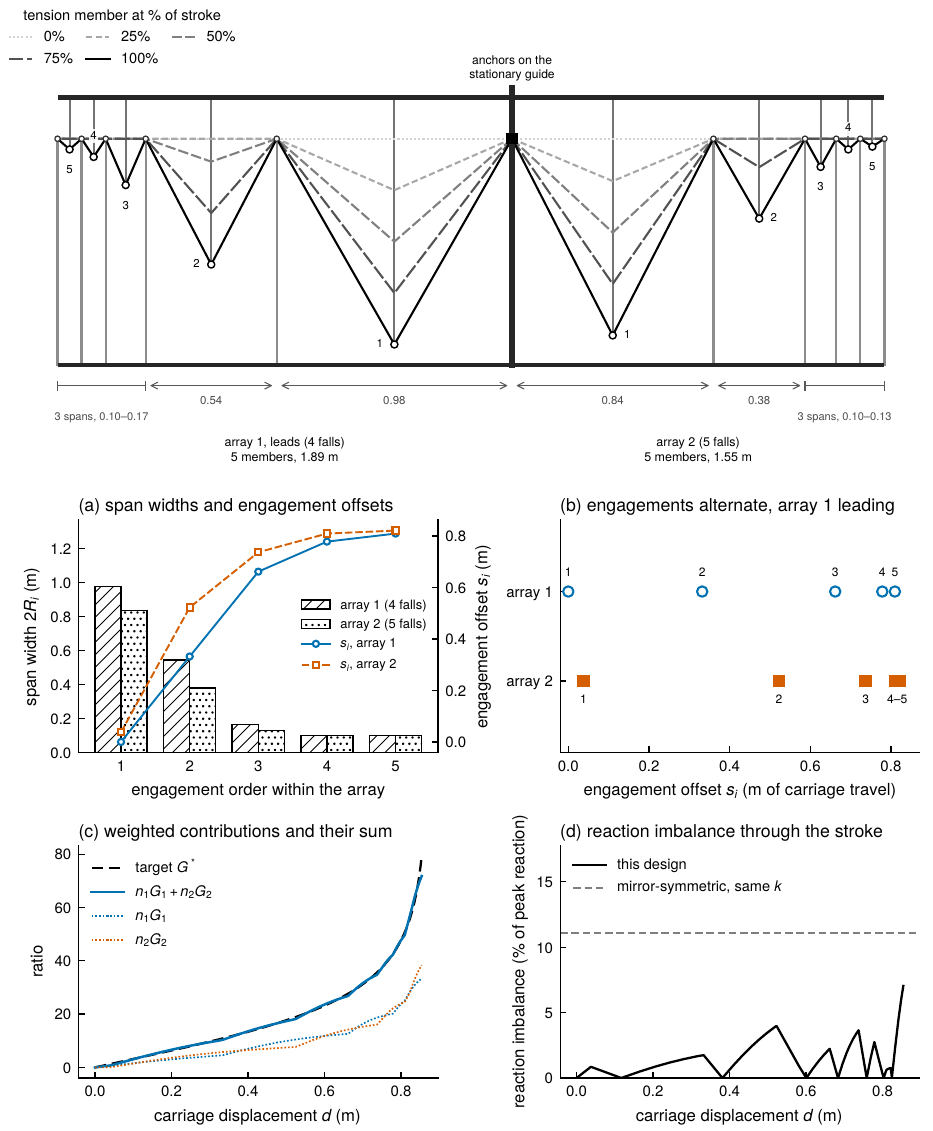}
\caption{The $\cfg{5}{9}$ design, panels as Figure~\ref{fig:k7}.}
\label{fig:k9}
\end{figure}

\paragraph{Five members per array, $k=9$} Four falls about the block driven by the leading array, five about the other; Table~\ref{tab:k9} and Figure~\ref{fig:k9}. Peak reaction imbalance is 7.0\% of the peak carriage reaction of 228~kN, 16.0~kN, against 25.3~kN (11.1\%) for the mirror-symmetric benchmark.

\begin{table}[!htbp]
\centering
\caption{Configured design $\cfg{5}{9}$: five members per array at $k=9$, 1,000:1 mass ratio, fitted at $\lambda = 0.03$. The greater array width is 1.89~m; the terminal ratio reaches 91\% of target.}
\label{tab:k9}
\begin{adjustbox}{max width=\linewidth}
\begin{tabular}{lll}
\toprule
& array 1, serving 4 falls & array 2, serving 5 falls \\
\midrule
span half-widths $R_i$ (mm)  & 488, 272, 83, 50, 50 & 418, 190, 65, 50, 50 \\
engagement offsets $s_i$ (mm) & 0, 332, 662, 779, 810 & 38, 523, 737, 810, 822 \\
\bottomrule
\end{tabular}
\end{adjustbox}
\end{table}

The departure from symmetry that achieves this balance is modest: corresponding span half-widths of the two arrays differ by at most 19~mm at $k=7$ and 82~mm at $k=9$, on arrays about 2~m wide. Both designs sit at a chosen point on the balance trade, not at exact cancellation: fitted at a $\lambda$ that favours profile accuracy, they reach $G_1/G_2$ of 1.13 at $k=7$ where \eqref{eq:split} asks 1.33, and 1.09 at $k=9$ against 1.25, leaving the 8.1\% and 7.0\% residuals. Both residuals sit at the end of the stroke (Figures~\ref{fig:k7}d and~\ref{fig:k9}d), where by \eqref{eq:counts} equal counts cannot balance: as the members saturate $G_1/G_2 \to 1$ whatever the spans, and the mirror residual re-emerges. Closer cancellation along the sweep of Table~\ref{tab:lamsweep}, 6.4\% at $k=7$ at a residual 14\% worse, works against that limit; Section~\ref{sec:counts} removes it by the member counts, and Section~\ref{sec:friction} traces the $\lambda$ trade to its end, where at this width budget the force profile becomes inadmissible below about 6\% reaction imbalance and further balance is bought with array width rather than with $\lambda$ alone.

\subsection{Balance by member count}
\label{sec:counts}

\begin{figure}[!htbp]
\centering
\includegraphics[width=\linewidth]{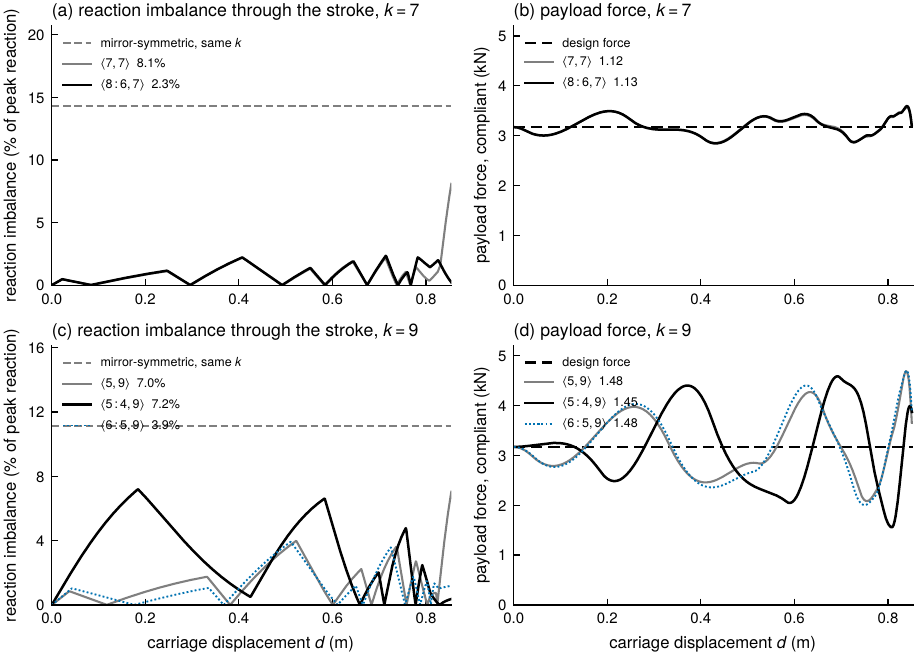}
\caption{Balance by member count at 1,000:1 and 16,471~N/m. (a), (c) Reaction imbalance through the stroke, the equal-count designs against the count-balanced ones and, at $k=9$, the comparison count $\cfg{6{:}5}{9}$, the mirror residual dashed: with equal counts the load returns to the mirror value as the members saturate at the end of the stroke, and with counts in the ratio $n_2 : n_1$ it does not. (b), (d) Compliant payload force over the release window, design force dashed. Legend figures are the peak reaction imbalance and the compliant \ptm{} force.}
\label{fig:counts}
\end{figure}

Two further pairs apply \eqref{eq:counts}, and a third count is tabulated beside them for comparison. At $k=7$ the fourteen members of $\cfg{7}{7}$ are redistributed as eight on the leading array and six on the other, at the same 2.11~m budget and the same $\lambda = 0.03$: $\cfg{8{:}6}{7}$, the notation giving the leading array's count first. At $k=9$ the counts are the floors of Table~\ref{tab:floors} themselves, five and four, nine members against the ten of $\cfg{5}{9}$: $\cfg{5{:}4}{9}$, fitted at $\lambda = 0.02$ for the reason given below. Both are seeded by the merged fit of Algorithm~1 and solved with the engagement order enforced; released from it, neither moves. Table~\ref{tab:counts} and Figure~\ref{fig:counts} compare them with the equal-count designs, and Supplementary Section~\ref{si:counts} draws them, gives their coordinates and their balance-weight ladders.

\begin{table}[!htbp]
\centering
\caption{Equal-count and count-balanced designs at 1,000:1 and 16,471~N/m, fitted at $\lambda = 0.03$ except $\cfg{5{:}4}{9}$ at 0.02. The last row is the comparison count of Section~\ref{sec:counts}, not a count-balanced design. Reaction imbalance as the peak, the mean over the stroke and the value at its end, in percent of the peak carriage reaction, and the peak in kN; \ptm{} forces over the release window; minimum tension in the compliant trace as a fraction of the design force. Widths are the leading array and the other.}
\label{tab:counts}
\footnotesize
\begin{adjustbox}{max width=\linewidth}
\begin{tabular}{@{}lcccccccccc@{}}
\toprule
& & & & & \multicolumn{2}{c}{reaction imbalance} & \multicolumn{2}{c}{\ptm{}} & min. & exit \\
\cmidrule(lr){6-7}\cmidrule(lr){8-9}
& $N_1 / N_2$ & widths (m) & rms & terminal & peak / mean / end & kN & compl. & rigid & tension & (m/s) \\
\midrule
$\cfg{7}{7}$ & 7 / 7 & 1.95 / 1.98 & 0.380 & 94\% & 8.1 / 0.9 / 8.1\% & 19.1 & 1.12 & 1.93 & 0.90 & 318.7 \\
$\cfg{8{:}6}{7}$ & 8 / 6 & 2.05 / 1.83 & 0.413 & 93\% & \textbf{2.3} / 0.9 / \textbf{0.2}\% & \textbf{5.4} & 1.13 & 1.90 & 0.90 & 318.8 \\
\midrule
$\cfg{5}{9}$ & 5 / 5 & 1.89 / 1.55 & 0.640 & 91\% & 7.0 / 1.4 / 7.0\% & 16.0 & 1.48 & 2.11 & 0.66 & 321.2 \\
$\cfg{5{:}4}{9}$ & 5 / 4 & \textbf{1.13 / 0.98} & 0.965 & 86\% & 7.2 / 3.2 / \textbf{0.4}\% & 15.6 & \textbf{1.45} & 2.56 & 0.49 & 314.2 \\
$\cfg{6{:}5}{9}$ & 6 / 5 & 1.96 / 1.68 & 0.446 & 94\% & \textbf{3.9} / 1.2 / 1.2\% & \textbf{9.2} & 1.48 & 1.89 & 0.63 & 321.6 \\
\bottomrule
\end{tabular}
\end{adjustbox}
\end{table}

At $k=7$ the count-balanced design is the equal-count design with one member transferred. Every span and offset the two share differs by under 15~mm; the seventh member of the trailing array, a 10~cm span engaging 23~mm before the end of the stroke, is moved to the leading array, where it engages 28~mm before the end. That member contributes about 1 to its array's ratio at the end of the stroke, weighted by four falls on one side and three on the other, and its transfer removes the terminal imbalance: the peak reaction imbalance falls from 8.1\% to 2.3\%, 19.1 to 5.4~kN on a peak reaction of 235~kN, with 0.2\% at the end of the stroke, and the peak moves to $d = 0.72$~m, where it is the drift between engagements of Section~\ref{sec:order}. Measured against the single array, which places its whole reaction on one side of the guide, 97.7\% of the one-sided load is cancelled; the moment the bearings react is taken up in Section~\ref{sec:moment}. Nothing else changes: \ptm{} 1.13 against 1.12 compliant and 1.90 against 1.93 rigid, minimum tension 0.90 of design in both, exit velocity 318.8 against 318.7~m/s, rms 0.413 against 0.380 and terminal ratio 93 against 94\%. The mean reaction imbalance over the stroke is 0.9\% in both: the count fixes the end of the stroke and the fit had already fixed the middle.

At $k=9$ the floor count buys minimalism rather than balance. $\cfg{5{:}4}{9}$ has nine members in arrays of 1.13 and 0.98~m against ten in 1.89 and 1.55~m, matches $\cfg{5}{9}$ on compliant \ptm{} force, 1.45 against 1.48, and on peak reaction imbalance, 7.2 against 7.0\%, and moves that peak from the end of the stroke, where $\cfg{5}{9}$ carries 7.0\%, to $d = 0.18$~m, the interval before the trailing array's first member engages, leaving 0.4\% at the end; 92.8\% of the single array's one-sided load is cancelled. Its costs are in the table: with no member to spare it reaches 86\% of the terminal ratio against 91\%, gives up 2\% of exit velocity, 314 against 321~m/s, carries a mean reaction imbalance of 3.2\% against 1.4\% and a rigid \ptm{} of 2.56 against 2.11, and its minimum tension of 0.49 of design is the lowest in Table~\ref{tab:counts}. The balance weight is a real choice here: at $\lambda = 0.03$ the fit crosses to a wider basin, 1.60 and 1.22~m, with a peak reaction imbalance of 5.5\% and a mean of 1.6\% but a compliant \ptm{} of 1.78, the payload mode rung by a leading span of 0.88~m, so $\lambda = 0.02$, which holds the force at 1.45, is the design reported. The comparison row $\cfg{6{:}5}{9}$, the nearest count above the floor, misses \eqref{eq:counts} by one part in 49 and gives 3.9\% at a \ptm{} of 1.48, one member more than $\cfg{5}{9}$ for a peak 44\% lower (Supplementary Section~\ref{si:counts}).

Under per-contact friction (Section~\ref{sec:friction}, Table~\ref{tab:friction}) the count-balanced designs retain what their equal-count counterparts retain, 76.5\% against 76.6\% at $k=7$ and $\eta_c = 0.98$, and more at $k=9$, 86.7\% against 80.6\%, the nine-member machine having the fewest contacts of any design in the paper, while the eleven-member $\cfg{6{:}5}{9}$ retains 78.4\% (Supplementary Section~\ref{si:friction}); the path-loss asymmetry moves the count-balanced designs' reaction imbalances to 3.4\% and 7.9\%.

\subsection{The guide moment}
\label{sec:moment}

The reaction imbalance of Table~\ref{tab:counts} is what the synthesis balances; the guide bearings react the moment \eqref{eq:moment}. Table~\ref{tab:moment} evaluates it for the five designs of Table~\ref{tab:counts} from their frozen coordinates, each member's force at its position along the array, first with the spans laid in engagement order outward from the guide, the convention of Figures~\ref{fig:k7} and~\ref{fig:k9}, and then at the span order that minimises the peak moment, found by exhaustive search over the orders of both arrays where the counts permit ($5!\,5!$, $5!\,4!$ and $6!\,5!$ orders at $k=9$) and by pairwise-swap descent from 31 starts at $k=7$; Supplementary Section~\ref{si:guide} gives the orders. The single-array reference of Table~\ref{tab:single} sets the scale: its centre of pressure sweeps from 0.25~m at first engagement to 1.58~m at the end of the stroke, its peak moment is 360~kN\,m, and no fixed offset of the driving mass removes a sweeping moment (Supplementary Section~\ref{si:guide}).

\begin{table}[!htbp]
\centering
\caption{The guide moment from the frozen coordinates, at 1,000:1, for the designs of Table~\ref{tab:counts} and the single-array reference. "Drawing" lays the spans in engagement order outward from the guide (Figures~\ref{fig:k7} and~\ref{fig:k9}); "searched" is the span order minimising the peak moment, a bound that departs from the rope-flow order of Section~\ref{sec:spanorder}. The offset is the peak moment divided by the peak reaction; the reduction is the single-array reference's peak moment, 360~kN\,m, divided by the design's (stroke integrals in Supplementary Section~\ref{si:guide}).}
\label{tab:moment}
\footnotesize
\begin{adjustbox}{max width=\linewidth}
\begin{tabular}{@{}lccccccccc@{}}
\toprule
& \multicolumn{2}{c}{reaction imbalance} & \multicolumn{3}{c}{moment, drawing placement} & \multicolumn{3}{c}{moment, searched placement} \\
\cmidrule(lr){2-3}\cmidrule(lr){4-6}\cmidrule(lr){7-9}
& kN & \% & kN\,m & offset (mm) & reduction & kN\,m & offset (mm) & reduction \\
\midrule
$\cfg{9}{5}$ single & --- & --- & 360 & 1,578 & 1 & --- & --- & --- \\
$\cfg{7}{7}$ & 19.1 & 8.1 & 21.3 & 90 & 17 & 7.5 & 32 & 48 \\
$\cfg{5}{9}$ & 16.0 & 7.0 & 32.0 & 141 & 11 & 5.0 & 22 & 72 \\
$\cfg{8{:}6}{7}$ & \textbf{5.4} & \textbf{2.3} & 18.7 & 80 & 19 & \textbf{3.8} & \textbf{16} & \textbf{95} \\
$\cfg{5{:}4}{9}$ & 15.6 & 7.2 & 11.2 & 52 & 32 & 8.6 & 40 & 42 \\
$\cfg{6{:}5}{9}$ & 9.2 & 3.9 & 21.5 & 91 & 17 & 6.8 & 29 & 53 \\
\bottomrule
\end{tabular}
\end{adjustbox}
\end{table}

Two things follow. First, force balance and moment balance are different quantities, and the drawing placement spends the difference. $\cfg{8{:}6}{7}$, whose reaction imbalance is 2.3\%, carries a peak moment of 19~kN\,m in that placement, the same as the 21~kN\,m of $\cfg{7}{7}$, because the member transferred to the leading array sits outermost, 2.0~m from the guide, where its 9.5~kN has the longest arm. The placement is free, and the search removes most of it: 3.8~kN\,m for $\cfg{8{:}6}{7}$, 7.5 for $\cfg{7}{7}$ and 5.0 for $\cfg{5}{9}$, equivalent to the peak reaction acting 16, 32 and 22~mm off the guide axis against 1.58~m for the single array. The searched orders share one feature: the late, floor-width members, which carry the largest forces at the end of the stroke, sit nearest the guide, and the two arrays' centres of pressure track one another, 0.83 and 0.82~m for $\cfg{8{:}6}{7}$ at the end of the stroke against 1.40 and 1.23~m in the drawing. Second, on the quantity the bearings react the dual array's advantage over the single array is a factor of 11 to 32 in peak moment at the drawing placement and 42 to 95 at the searched one; the 97.7\% of Section~\ref{sec:counts} is the force figure, and $\cfg{8{:}6}{7}$ removes 95\% of the single array's peak moment in the drawing placement and 99\% at the searched one. The searched placements are bounds, not recommendations. They put the deep, early members at the outboard entry, where the rope drawn for every fold behind them passes through the largest wraps, about doubling the rope passed over the deepest member against the order of Section~\ref{sec:spanorder}, a cost the order-blind per-contact model of Section~\ref{sec:friction} does not see. The drawing placement is the one the designs are laid out in, and its factor of 11 to 32 is the figure to build on. The stroke integral of the moment, which sets the guide friction, the rope passed through the contacts in each order, and the effect of placement on the friction model are in Supplementary Section~\ref{si:guide}.

\subsection{Compliant versus rigid dynamics}
\label{sec:dyn}

\begin{figure}[!htbp]
\centering
\includegraphics[width=\linewidth]{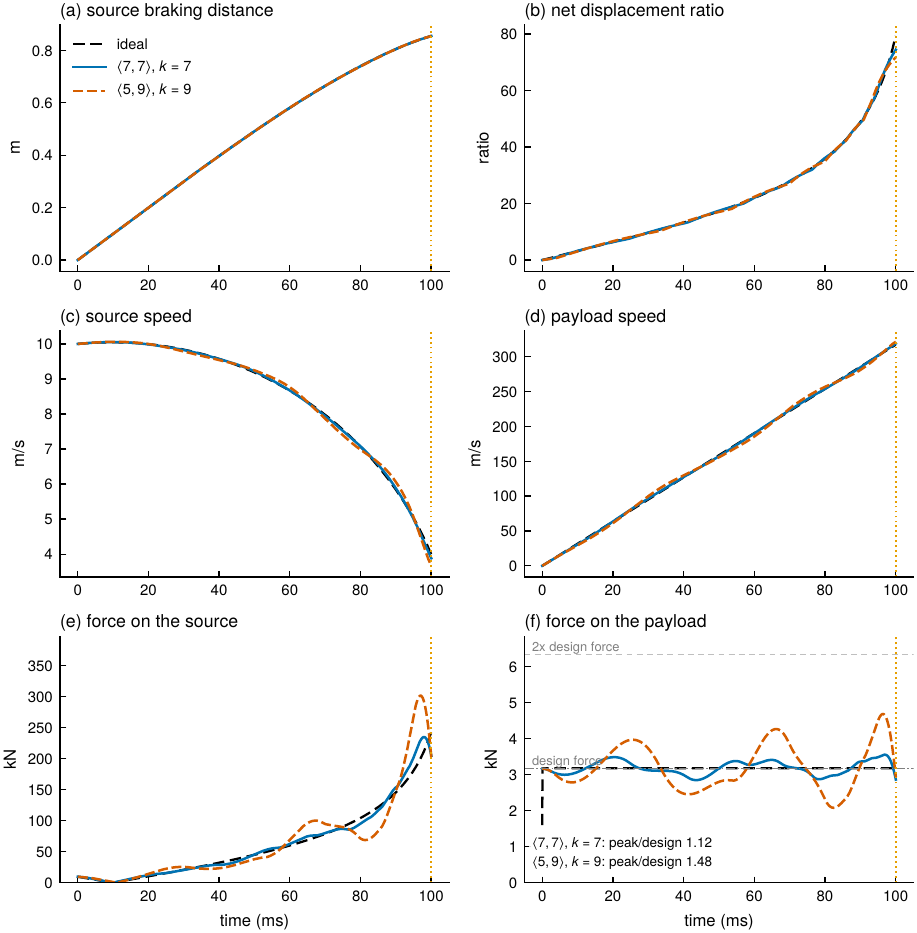}
\caption{The two equal-count designs with a compliant, pre-tensioned output member at 16,471~N/m. The rigid sawtooth (Supplementary Section~\ref{si:figs}) is replaced by a smooth ripple about the design force. Panel (f) is drawn from zero so the ripple can be read against the absolute scale.}
\label{fig:runscompliant}
\end{figure}

Figure~\ref{fig:runscompliant} and Supplementary Section~\ref{si:figs} show both equal-count designs run against the ideal profile, compliant and rigid, in the layout of Figure~\ref{fig:single-compliant}. Panels (a) to (d) are nearly indistinguishable between the two equal-count designs and the ideal, which is the point: the geometry is fitted to reproduce the ideal ratio profile and does so. The designs separate in panel (f). Under the rigid integrator both show the bounded sawtooth of discrete engagement, one tooth per member, larger at $\cfg{5}{9}$ because five members must cover the profile that seven cover at $\cfg{7}{7}$. With the compliant member the sawtooth becomes a smooth ripple about the design force, and the $\cfg{5}{9}$ ripple is nearly four times the amplitude of the $\cfg{7}{7}$ one, which is the whole of the difference between their compliant \ptm{} forces of 1.48 and 1.12. Table~\ref{tab:lamsweep} places the two equal-count designs on the balance trade. Larger $\lambda$ buys reaction balance and costs force uniformity, and the two \ptm{} measures do not move together: at $\cfg{5}{9}$ the rigid figure is best at $\lambda = 0.1$ and the compliant at $\lambda = 0.03$. The compliant figure is the one an acceleration-sensitive payload sees and the one the pre-tensioned member delivers, so $\lambda = 0.03$ is taken for both equal-count designs, and for $\cfg{8{:}6}{7}$; the count-balanced designs are placed on the same trade in Supplementary Section~\ref{si:counts}.

\begin{table}[!htbp]
\centering
\caption{Both equal-count designs against the balance weight, at 16,471~N/m. Larger $\lambda$ buys reaction balance and costs force uniformity; the count-balanced designs' ladders are in Supplementary Section~\ref{si:counts}.}
\label{tab:lamsweep}
\begin{adjustbox}{max width=\linewidth}
\begin{tabular}{lccccc}
\toprule
$\lambda$ & rms & reaction imbalance & \ptm{} rigid & \ptm{} compliant & exit (m/s) \\
\midrule
\multicolumn{6}{l}{\emph{$\cfg{7}{7}$, seven members per array, $k=7$}}\\
0.03 & 0.380 & 8.1\% & 1.93 & \textbf{1.12} & 318.7 \\
0.1  & 0.399 & 7.2\% & 1.92 & 1.14 & 319.0 \\
0.3  & 0.432 & 6.4\% & 2.23 & 1.23 & 319.9 \\
\midrule
\multicolumn{6}{l}{\emph{$\cfg{5}{9}$, five members per array, $k=9$}}\\
0.03 & 0.640 & 7.0\% & 2.11 & \textbf{1.48} & 321.2 \\
0.1  & 0.659 & 6.6\% & 1.88 & 1.62 & 322.2 \\
0.3  & 0.699 & 6.0\% & 2.00 & 1.82 & 324.6 \\
\bottomrule
\end{tabular}
\end{adjustbox}
\end{table}

\subsection{Comparison at matched member count and stage ratio}
\label{sec:compare}

\begin{figure}[!htbp]
\centering
\includegraphics[width=\linewidth]{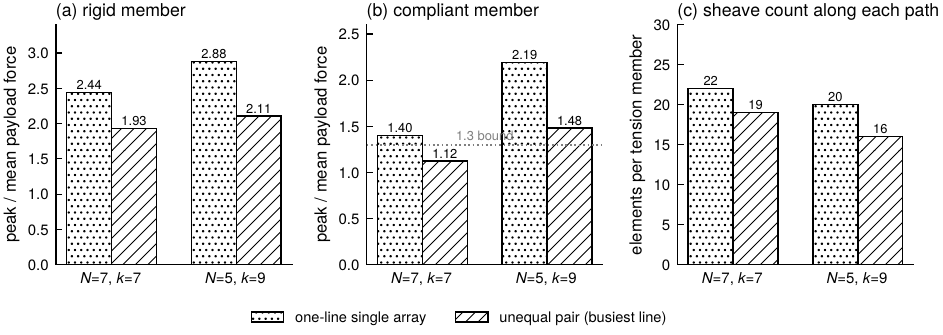}
\caption{Comparison against a single array at the same member count and stage ratio. (a) Peak-to-mean payload force, rigid; (b) the same, compliant, against the 1.3 bound; (c) elements traversed per tension member.}
\label{fig:compare}
\end{figure}

The relevant comparison is against a single array of the \emph{same member count and stage ratio}, since the second array is added hardware, and by Table~\ref{tab:configs} it covers the two-line single array and the mirror-symmetric pair at once: all three share the ratio $k\,\Garr$ and so deliver the same force profile, which only the unequal pair can change. Table~\ref{tab:compare} and Figure~\ref{fig:compare} give the comparison, the single arrays fitted by \eqref{eq:single-obj} at the member count, stage ratio and width budget of the dual design beside them.

\begin{table}[!htbp]
\centering
\caption{Unequal pairs against a single array at matched member count and stage ratio. The force and velocity rows of the single-array columns hold for the one-line single array, the two-line single array and the mirror pair alike, which share the ratio $k\,\Garr$; the contact and reaction-imbalance rows are for the one-line single array, the two-line single and the mirror pair matching the dual array's contacts. Peak-to-mean forces over the release window; all cases at 16,471~N/m. The last column is the single-array reference of Table~\ref{tab:single}, free to choose $N$ and $k$.}
\label{tab:compare}
\small
\begin{adjustbox}{max width=\linewidth}
\begin{tabular}{lcccc|c}
\toprule
& \multicolumn{2}{c}{$\cfg{7}{7}$} & \multicolumn{2}{c|}{$\cfg{5}{9}$} & $\cfg{9}{5}$ \\
\cmidrule(lr){2-3}\cmidrule(lr){4-5}\cmidrule(lr){6-6}
& single & unequal pair & single & unequal pair & single \\
\midrule
\ptm{}, rigid                & 2.44  & \textbf{1.93}    & 2.88  & \textbf{2.11} & 2.07 \\
\ptm{}, compliant            & 1.40  & \textbf{1.12}    & 2.19  & \textbf{1.48} & 1.09 \\
exit velocity, compliant (m/s) & 321.0 & 318.7          & 328.1 & 321.2 & 318.2 \\
peak reaction imbalance (mirror pair / unequal) & 14.3\% & \textbf{8.1\%}  & 11.1\% & \textbf{7.0\%} & --- \\
elements per line (one-line single / unequal) & 22 & \textbf{18 / 19} & 20 & \textbf{15 / 16} & 24 \\
contact efficiency, 2\% per element & 0.641 & \textbf{0.688} & 0.668 & \textbf{0.731} & 0.616 \\
\bottomrule
\end{tabular}
\end{adjustbox}
\end{table}

Three things follow. First, at matched configuration the unequal pair improves the force profile on both models, from 2.44 to 1.93 and 2.88 to 2.11 rigid and from 1.40 to 1.12 and 2.19 to 1.48 compliant, and since the single-array column is equally the mirror column these are improvements over mirror symmetry as much as over the single array: the doubled geometric freedom makes a member count admissible at a stage ratio where a symmetric machine of the same member count is not. Second, the improvement is the difference between an admissible machine and an inadmissible one. Held at $\cfg{7}{7}$, a configuration the single-array selection would not choose, the unequal pair returns the machine to the class of Table~\ref{tab:single}, 1.12 compliant and 1.93 rigid against the reference family's 1.09 to 1.19 and 2.05 to 2.10; at $\cfg{5}{9}$ neither arrangement reaches the compliant bound of 1.3, but the unequal pair cuts the excess over it by four fifths, from 0.89 to 0.18. Third, what the second array provides that no symmetric arrangement can is in the first four rows, the force profile and the balance; the three and four elements saved on the busiest path belong to the two-line stage, which the two-line single array and the mirror pair share. The single-array column is not a restatement of the reference designs, which select $N$ and $k$ jointly and return nine members at $k=5$ at this mass ratio with a compliant \ptm{} of 1.09; the single arrays of the first four columns are \emph{constrained} to the member count and stage ratio of the dual designs beside them, so the comparison is like for like. What the unequal pair offers relative to $\cfg{9}{5}$ is not a better force profile at 1,000:1 but the same class of profile at a higher stage ratio, with fewer members per array, half the load on each, slower stage blocks and a balanced guide. The count-balanced designs of Section~\ref{sec:counts} leave the force rows of Table~\ref{tab:compare} essentially where they are, 1.13 and 1.45 compliant, and differ in the reaction-imbalance row alone.

\subsection{Distance from the ideal}
\label{sec:ideal}

The designs are motivated by ideals at every level, and Table~\ref{tab:ideal} lists each ideal beside what the configured machines reach. Uniform force is the optimum for peak force at given energy, and the lossless transfer efficiency is fixed by the specification alone; the fits reproduce the target to within a few percent of full scale and the compliant machines reach the ideal exit velocity; compliance carries the force to within 12\% of uniform at $\cfg{7}{7}$; and the reaction imbalance, ideally zero, is held to 7 to 8\% of the peak reaction at the accuracy-favouring $\lambda$ chosen with equal member counts, to 2.3\% with the counts of \eqref{eq:counts}, or lower at the designer's discretion. Every entry is a bound and a measured distance from it, a more useful statement of a design's quality than any single figure of merit.

\begin{table}[!htbp]
\centering
\caption{Ideals and the distance the configured designs stand from them, at 1,000:1 and 16,471~N/m. The $\cfg{8{:}6}{7}$ column is the count-balanced design of Section~\ref{sec:counts} and the last column the near-balanced design of Section~\ref{sec:friction}.}
\label{tab:ideal}
\footnotesize
\begin{adjustbox}{max width=\linewidth}
\begin{tabular}{@{}p{3.4cm}p{2.3cm}>{\centering\arraybackslash}p{1.8cm}>{\centering\arraybackslash}p{1.8cm}>{\centering\arraybackslash}p{1.8cm}>{\centering\arraybackslash}p{1.8cm}>{\centering\arraybackslash}p{2.1cm}@{}}
\toprule
quantity & ideal & $\cfg{9}{5}$ single & $\cfg{7}{7}$ unequal pair & $\cfg{5}{9}$ unequal pair & $\cfg{8{:}6}{7}$ count-balanced & $\cfg{9}{7}$ unequal pair, 3.3 m, $\lambda = 10$ \\
\midrule
transfer efficiency of the target profile & 86.3\% (fixed by $\nu$, $\gamma$) & 86.7\% realised & --- & --- & --- & --- \\
exit velocity, compliant & 317.0 m/s & 318.2 m/s & 318.7 m/s & 321.2 m/s & 318.8 m/s & 319.1 m/s \\
terminal ratio reached, \% of target & 100\% & --- & 94\% & 91\% & 93\% & 97\% \\
\ptm{} payload force, compliant & 1.00 & 1.09 & 1.12 & 1.48 & 1.13 & 1.29 \\
\ptm{} payload force, rigid & 1.00 & 2.07 & 1.93 & 2.11 & 1.90 & 1.77 \\
peak reaction imbalance & 0 & moment, not balanced & 8.1\% & 7.0\% & 2.3\% & 0.5\% \\
\bottomrule
\end{tabular}
\end{adjustbox}
\end{table}

\subsection{Friction, and the price of balance}
\label{sec:friction}

Two of the paper's claims rest on nominal accounting rather than simulation: that dividing the falls between two lines saves friction, counted at 2\% per element in Table~\ref{tab:contacts}, and that the residual reaction imbalances of Section~\ref{sec:duals} can be driven toward zero along the $\lambda$ trade. This section puts both to the integrators. For the first, a per-contact efficiency $\eta_c$ is carried through the reeving: along the output member the tension steps by $1/\eta_c$ at every sheave from the payload end, so the two blocks are pulled unequally, and along each tension member it steps again at the entry sheave and at every support and member from the block back to the anchor, so members nearer the anchor carry higher tension. The reaction each member applies to the carriage is its two segment tensions resolved along the guide, and the integrators use that reaction in place of $n_j F G_j$ with the kinematics unchanged. The model reduces to the lossless one at $\eta_c = 1$ and resolves the tension term of \eqref{eq:balance} that Section~\ref{sec:odd} left unquantified; Supplementary Section~\ref{si:friction} gives the construction. Table~\ref{tab:friction} evaluates the configured designs, the count-balanced designs of Section~\ref{sec:counts}, the matched single arrays and the reference design at $\eta_c$ of 0.995, 0.99 and 0.98.

\begin{table}[!htbp]
\centering
\caption{Per-contact friction applied to the frozen designs at 1,000:1, the count-balanced designs of Section~\ref{sec:counts} included. Energy delivered to the payload as a fraction of the lossless value at each $\eta_c$; the remaining columns at $\eta_c = 0.98$, except the first reaction-imbalance column, which is the lossless figure. The two-line single array and the mirror pair share every entry. Reaction imbalance is the geometric figure of Section~\ref{sec:cond} with the friction-modified reactions. Compliant model at 16,471~N/m; geometry not refitted for the losses.}
\label{tab:friction}
\footnotesize
\begin{adjustbox}{max width=\linewidth}
\begin{tabular}{@{}lccccccc c@{}}
\toprule
& \multicolumn{3}{c}{energy retained at $\eta_c$} & exit & \ptm{} & \multicolumn{2}{c}{reaction imbalance} & peak reaction \\
\cmidrule(lr){2-4}\cmidrule(lr){7-8}
& 0.995 & 0.99 & 0.98 & (m/s) & compliant & $\eta_c = 1$ & 0.98 & (kN) \\
\midrule
$\cfg{9}{5}$ one-line single & 92.6\% & 85.5\% & 72.2\% & 270 & 1.09 & --- & --- & 300 \\
$\cfg{9}{5}$ two-line single / mirror & 92.9\% & 86.1\% & 73.4\% & 273 & 1.09 & 20.0\% & 19.5\% & 295 \\
$\cfg{7}{7}$ one-line single & 92.7\% & 85.3\% & 72.0\% & 272 & 1.31 & --- & --- & 302 \\
$\cfg{7}{7}$ two-line single / mirror & 93.3\% & 86.5\% & 73.9\% & 276 & 1.31 & 14.3\% & 13.8\% & 294 \\
$\cfg{7}{7}$ unequal pair & 93.9\% & 88.0\% & 76.6\% & 279 & 1.09 & 8.1\% & 7.9\% & 296 \\
$\cfg{8{:}6}{7}$ count-balanced & 93.9\% & 87.9\% & 76.5\% & 279 & 1.09 & 2.3\% & 3.4\% & 294 \\
$\cfg{5}{9}$ one-line single & 93.6\% & 88.0\% & 77.9\% & 290 & 2.11 & --- & --- & 279 \\
$\cfg{5}{9}$ two-line single / mirror & 94.4\% & 89.5\% & 80.6\% & 294 & 2.10 & 11.1\% & 10.6\% & 269 \\
$\cfg{5}{9}$ unequal pair & 95.6\% & 90.8\% & 80.6\% & 288 & 1.23 & 7.0\% & 6.8\% & 274 \\
$\cfg{5{:}4}{9}$ count-balanced & 96.4\% & 93.0\% & 86.7\% & 292 & 1.40 & 7.2\% & 7.9\% & 258 \\
\bottomrule
\end{tabular}
\end{adjustbox}
\end{table}

Four things follow. First, per-contact friction at the level the nominal accounting assumes is a large term: at 2\% per contact the machines deliver 72 to 87\% of their lossless payload energy and exit 22 to 49~m/s slower, a loss of the same order as the fast-path sheave inertia of Section~\ref{sec:sheave}, and the peak carriage reaction rises by 19 to 32\% because the source must supply the losses, which the guide of Supplementary Section~\ref{si:guide} and the loads of Table~\ref{tab:loads} must be sized for; at 1\% per contact the energy loss is 7 to 15\%, at 0.5\% it is 4 to 7\%. Second, the two-line stage does save energy, but about a third of what Table~\ref{tab:contacts} counts: 1.2, 1.9 and 2.7 percentage points at $\eta_c = 0.98$ for $k = 5$, 7 and 9, against nominal 4.1, 6.2 and 8.4\%. The nominal count charges every element the full 2\%, whereas in the reeving the losses on the output member are borne by segments at increasing tension from the payload end and those in the array by members at increasing tension from the stage end, so the elements a second line removes are on average the cheaper ones. The unequal pairs retain within three points of their two-line singles, above at $k=7$ and level at $k=9$, and the nine-member $\cfg{5{:}4}{9}$ six points above its ten-member counterpart, having the fewest contacts of any design here. Third, friction does not degrade the force profile: the compliant \ptm{} figures fall by 0.01 to 0.09 as the losses damp the ripple, and by 0.25 at the $\cfg{5}{9}$ unequal pair. Fourth, the tension term of the balance condition is small: with the block pulls unequal by the path losses, the reaction imbalance of the equal-count designs moves from 8.1 to 7.9\% and from 7.0 to 6.8\% at $\eta_c = 0.98$, and that of the mirror pairs from 14.3 to 13.8\% and 11.1 to 10.6\%, under half a percentage point, because the two lines' paths differ by one stage segment and are otherwise alike; the count-balanced designs, whose terminal residual no longer hides it, move by about one point, from 2.3 to 3.4\% and from 7.2 to 7.9\%. A machine with dissimilar routings would see more, and the correction enters the objective through the same reaction functions.

The second claim is the reach of the balance trade. Table~\ref{tab:frontier} traces it for the $\cfg{7}{7}$ family and two wider families, each point a seeded fit of Algorithm~1 at the stated $\lambda$ with its dynamics simulated; Supplementary Section~\ref{si:frontier} gives the full sweeps. At the configured budget of 2.11~m per array the trade is steep and its far end inadmissible: along the sweep the reaction imbalance falls from 8.1\% to 1.2\%, but the compliant \ptm{} force rises from 1.12 to 1.97 and crosses the bound of 1.3 between 6 and 5\% reaction imbalance, so about 6\% is as far as $\lambda$ can usefully be taken at this budget. The peaks are the counterpart of tension dips: the minimum tension in the compliant trace falls from 0.90 of the design force at $\lambda = 0.03$ to 0.80, 0.48 and 0.05 at $\lambda = 0.3$, 1 and 10, the last at the edge of slack, and the full sweeps continue into designs whose member goes slack for up to a fifth of the stroke and whose next engagement snatches the payload; across the 56 designs of the sweeps and their variants the compliant \ptm{} force correlates with the minimum tension at $-0.88$. Smooth tracking error is not the cause, since the ideal profile perturbed by smooth 1\% errors of several shapes returns \ptm{} figures of 1.02 to 1.07. The dips come from local flats in the fitted profile, stretches of a few centimetres in which its slope falls to about 40\% of the target slope (0.37 at $\lambda = 10$ against 0.68 at $\lambda = 0.03$, and above 0.5 in every admissible design): there the rope end decelerates with the source while the payload coasts, and because the target slope late in the stroke mostly compensates the source's own deceleration, 41 per metre at mid-stroke and 184 at 90\% of it, a flat there empties the 0.19~m design stretch in milliseconds. The rigid integrator, whose \ptm{} figure is set by the engagement steps and stays near 2 along the sweep, cannot register this, which is why the trade must be read on the compliant figure. The flats follow from Table~\ref{tab:floors}: in the accuracy-favouring fit the trailing array, with the larger fall count, supplies the larger share of the terminal rise; balance requires the two arrays to rise in proportion, and with seven members, the floor for the leading array at $k=7$, the leading array cannot rise further, so the fitter reaches balance by pulling the trailing array down, the terminal ratio falls from 94 to 90\% of target, and the crowded final engagements leave flats between them. Adding members at the same width does not help, since the spans narrow and the members cease to overlap, the width argument of Section~\ref{sec:width}. What buys balance at equal member counts is width, because wider members overlap and the flats close; the count condition \eqref{eq:counts} removes the terminal residual before either is spent, and the $\cfg{8{:}6}{7}$ design of Section~\ref{sec:counts} reaches 2.3\% at the configured budget and a \ptm{} of 1.13 by redistributing the same fourteen members. With nine members and 3.3~m per array, $\lambda = 10$ reaches a reaction imbalance of 0.5\%, 1.2~kN against the mirror pair's 33.7~kN, at a compliant \ptm{} of 1.29 and a minimum tension of 0.68 of design, inside the bound, and 319~m/s, the whole of the ideal exit velocity; $\lambda = 3$ gives 1.2\% at 1.23. Reweighting the objective does not move this frontier: a relative residual, which removes the early-stroke under-tracking of the large-$\lambda$ fits, and a slope-regularised residual both leave the dips in place (Supplementary Section~\ref{si:frontier}), so the frontier is a property of the geometry at the given width, not of the weighting in Algorithm~1. The price of near-exact balance at this mass ratio and stage ratio, with equal member counts, is therefore about 1.3~m of additional array width per side and two members per array; with the counts set by \eqref{eq:counts} the first six points come free. Under path-loss asymmetry at 2\% per contact the 0.5\% design rises to 0.8\% (2.6~kN) and the 1.2\% design to 2.0\%, the size of the tension term at exact geometric balance. Its coordinates are given in Supplementary Section~\ref{si:frontier} as a third reference design.

\begin{table}[!htbp]
\centering
\caption{The balance trade: seeded fits of Algorithm~1 at the stated $\lambda$ with their simulated dynamics, at 1,000:1 and 16,471~N/m. The $\lambda = 0.03$ row of the first family is the configured design of Table~\ref{tab:k7}; "width" is the wider of the two arrays. Full sweeps, including $\cfg{5}{9}$ and the 2.7~m families, are in Supplementary Section~\ref{si:frontier}.}
\label{tab:frontier}
\footnotesize
\begin{adjustbox}{max width=\linewidth}
\begin{tabular}{@{}llcccccc@{}}
\toprule
family, budget & $\lambda$ & imbalance & (kN) & \ptm{} compl. & \ptm{} rigid & exit (m/s) & width (m) \\
\midrule
$\cfg{7}{7}$, 2.11 m & 0.03 & 8.1\% & 19.1 & 1.12 & 1.93 & 318.7 & 1.98 \\
$\cfg{7}{7}$, 2.11 m & 0.3 & 6.4\% & 14.9 & 1.23 & 2.08 & 319.9 & 2.11 \\
$\cfg{7}{7}$, 2.11 m & 1 & 5.0\% & 11.5 & 1.59 & 2.13 & 323.3 & 2.11 \\
$\cfg{7}{7}$, 2.11 m & 10 & 1.2\% & 2.6 & 1.97 & 2.14 & 324.8 & 2.12 \\
$\cfg{7}{7}$, 3.3 m & 3 & 2.9\% & 6.6 & 1.22 & 1.98 & 318.4 & 3.30 \\
$\cfg{7}{7}$, 3.3 m & 10 & 1.5\% & 3.4 & 1.23 & 1.94 & 317.9 & 3.30 \\
$\cfg{9}{7}$, 3.3 m & 0.03 & 3.5\% & 8.7 & 1.03 & 1.47 & 317.5 & 3.30 \\
$\cfg{9}{7}$, 3.3 m & 1 & 2.0\% & 4.9 & 1.17 & 1.55 & 318.9 & 3.30 \\
$\cfg{9}{7}$, 3.3 m & 3 & 1.2\% & 3.0 & 1.23 & 1.96 & 318.4 & 3.30 \\
$\cfg{9}{7}$, 3.3 m & 10 & 0.5\% & 1.2 & 1.29 & 1.77 & 319.1 & 3.30 \\
\bottomrule
\end{tabular}
\end{adjustbox}
\end{table}

\subsection{Loads a builder must provide for}
\label{sec:loads}

Table~\ref{tab:loads} collects, for the equal-count designs, the $\cfg{8{:}6}{7}$ design and the single-array reference, the loads and speeds that size the components, all following from the design force $F$, the fall counts and the fitted geometry by the relations of Sections~\ref{sec:mech} and~\ref{sec:dual}. The tension in line $j$ is $n_j F$, also the load at its anchor and the pull on its block; the per-array reaction is $n_j F G_j(d)$, peaking late in the stroke; the block speed is the exit velocity over $k$; and the output member, at payload speed, carries $F$ and turns any fixed-axle sheave at that speed. Rope breaking loads are given at a safety factor of 3, the factor at which Section~\ref{sec:critical} evaluates the critical-speed margin; the pre-tension of the output member is the design force.

\begin{table}[!htbp]
\centering
\caption{Component loads and speeds at 1,000:1, from the design force 3,169.6~N and the fitted geometries. Reactions are peaks over the stroke; rope loads are minimum breaking loads at a safety factor of 3. Release tolerance from Section~\ref{sec:instab}.}
\label{tab:loads}
\footnotesize
\begin{adjustbox}{max width=\linewidth}
\begin{tabular}{@{}p{5.6cm}>{\centering\arraybackslash}p{2.2cm}>{\centering\arraybackslash}p{2.4cm}>{\centering\arraybackslash}p{2.4cm}>{\centering\arraybackslash}p{2.4cm}@{}}
\toprule
& $\cfg{9}{5}$ single & $\cfg{7}{7}$ unequal pair & $\cfg{5}{9}$ unequal pair & $\cfg{8{:}6}{7}$ count-balanced \\
\midrule
tension in line 1 / line 2 ($n_j F$) & 15.8 kN & 9.5 / 12.7 kN & 12.7 / 15.8 kN & 9.5 / 12.7 kN \\
anchor load on the guide, line 1 / line 2 & 15.8 kN & 9.5 / 12.7 kN & 12.7 / 15.8 kN & 9.5 / 12.7 kN \\
minimum breaking load, lines, at SF 3 & 47.5 kN & 28.5 / 38.0 kN & 38.0 / 47.5 kN & 28.5 / 38.0 kN \\
output member tension; pre-tension & 3.17 kN & 3.17 kN & 3.17 kN & 3.17 kN \\
minimum breaking load, output member, at SF 3 & 9.5 kN & 9.5 kN & 9.5 kN & 9.5 kN \\
peak carriage reaction & 228 kN & 236 kN & 228 kN & 235 kN \\
peak reaction per array & 228 kN & 108 / 128 kN & 106 / 122 kN & 117 / 118 kN \\
peak reaction imbalance $\Delta F$ & one-sided & 19.1 kN & 16.0 kN & 5.4 kN \\
peak guide moment, drawing / searched placement & 360 kN m & 21 / 7.5 kN m & 32 / 5.0 kN m & 19 / 3.8 kN m \\
block speed at exit ($v/k$) & 63.6 m/s & 45.5 m/s & 35.7 m/s & 45.5 m/s \\
100 mm fixed-axle sheave on the output member & 60,800 rev/min & 60,900 rev/min & 61,400 rev/min & 60,900 rev/min \\
release tolerance for 2\% of exit velocity & 11 mm, 2.5 ms & 11 mm, 2.5 ms & 11 mm, 2.5 ms & 11 mm, 2.5 ms \\
\bottomrule
\end{tabular}
\end{adjustbox}
\end{table}

\section{Physical operating limits}
\label{sec:limits}

\subsection{Traction-loss divergence}
\label{sec:instab}

\begin{figure}[!htbp]
\centering
\includegraphics[width=\linewidth]{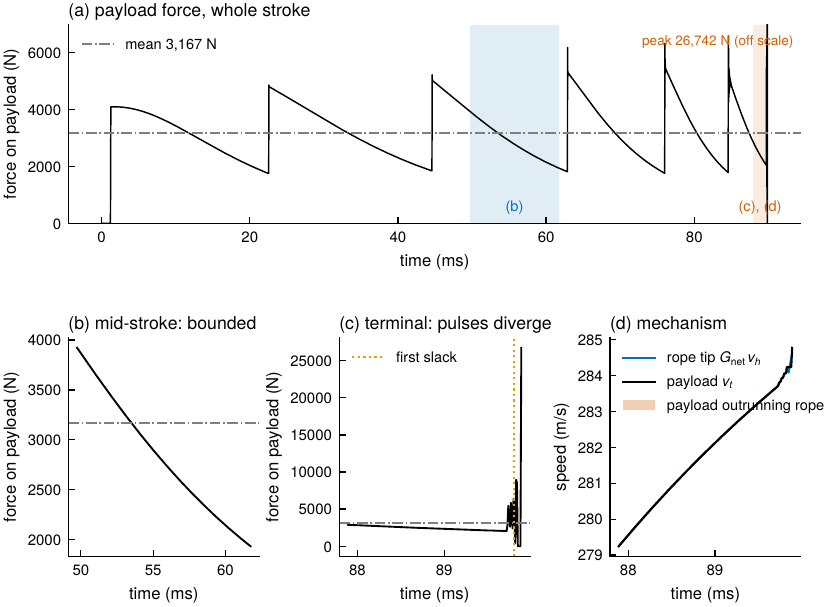}
\caption{The traction-loss instability, 1,000:1 single-array reference. (a) Payload force over the whole stroke. (b) A mid-stroke engagement: bounded. (c) The terminal window: successive pulses diverge. (d) The mechanism: payload velocity rising above rope-tip velocity $\Gnet v_h$, shaded where the member is slack.}
\label{fig:instab}
\end{figure}

Every configuration examined terminates in the same way. Through the stroke, each engagement produces a bounded sawtooth in payload force: the force steps up as a member engages, then decays as the array ratio momentarily lags. Toward the end of the stroke these excursions grow until the payload velocity exceeds the rope-tip velocity $\Gnet v_h$. The member goes slack, the source, no longer resisted, re-accelerates, the rope tip catches up, and contact is regained as an impulse; each cycle delivers a larger impulse than the last and the sequence diverges within a few milliseconds (Figure~\ref{fig:instab}). For the 1,000:1 single-array reference, first slack occurs at 89.81~ms of an 89.89~ms rigid stroke, after which successive re-contacts deliver roughly three and then eight times the mean force; those multiples are indicative only, moving by about 20\% between time steps of $10^{-5}$ and $10^{-6}$~s, and the whole divergence occupies the final 0.1\% of the stroke. Full-stroke \ptm{} figures that include the pulse exceed the release-window figures by a factor of 1.9 at 10,000:1, 4.1 at 1,000:1 and 21.7 at 100:1, which is why the release convention of Section~\ref{sec:setup} excludes it: the terminal pulse is the one part of the trace that is neither experienced by a payload released before it nor time-step converged.

Severity follows the discreteness of engagement, the arc length between successive members relative to the stroke, not how closely the ratio curve is reproduced or the mass ratio. Across the three reference designs the fit residual varies by a factor of 2.6, from 0.57\% to 1.48\%, while rigid \ptm{} force varies by 2.5\%, from 2.05 to 2.10; nor does residual order the compliant column, where the best-fitting case is not the mildest. A controlled test agrees: at 1,000:1, improving the residual sevenfold changed the time to divergence by under 4\%. This is the mechanism of Section~\ref{sec:mech}: every engaged member is concave, so between engagements the slope of $\Gnet$ decays and the payload gains on the rope, and it is the spacing of the fresh slope steps, not the fidelity of the curve between them, that sets how far the excursions grow. The dual array's finer interleaving of two arrays' engagements acts on exactly this quantity, one reading of the rigid improvement in Table~\ref{tab:compare}.

The consequence for operation is a useful stroke terminating at the onset of divergence and a release operating at or before that point. Releasing early costs less exit velocity than the geometry suggests, because the payload force is uniform by design, so $v = \sqrt{2Fy/m}$ and exit velocity follows the square root of payload travel: releasing 5\% early on the payload's run costs about 2\% of exit velocity, 2.5\%, 2.4\% and 1.7\% at the three mass ratios against 2.53\% from the square-root law, and 10\% early costs about 5\%. Five percent of the payload's run is, in every case, 11~mm of the source's 0.854~m braking distance or 2.5~ms of timing. The tolerance is therefore centimetre-scale on the source stroke, decimetre-scale or more on the launch track and of order a millisecond, undemanding for a mechanical or pyrotechnic release at these tensions; conversely, the release position sets exit velocity at the same rate.

\subsection{Fast-path sheave rotational inertia}
\label{sec:sheave}

For a rotating element of mass $m_s$ with its mass concentrated near the rim, at a rim velocity equal to the member velocity $v$,
\begin{equation}
E_{\mathrm{rot}} = \tfrac12 I \omega^2 = \tfrac12 m_s v^2,
\label{eq:erot}
\end{equation}
the translational kinetic energy of the same mass at member velocity. Rim speed is the speed of the member \emph{relative to the sheave's own axle}, so only fixed-axle sheaves on the output member turn at full payload speed; the stage blocks translate with the member and turn at $1/k$ of it, storing $1/k^2$ as much. Because sheave energy and payload energy both go as $v^2$, fast-path inertia acts as added payload mass, and the exit velocity of the lossless models is corrected by
\begin{equation}
v = v_{\mathrm{model}} \sqrt{\frac{m}{m + \neff\, m_s}}, \qquad \neff = \sum_i \left(\frac{v_i}{v}\right)^2,
\label{eq:sheave}
\end{equation}
where $v_i$ is the rim speed of sheave $i$. In the arrangement of Figure~\ref{fig:hero} two sheaves run at full speed, an orientation sheave and the first sheave taking the output member, and counting the slower sheaves gives $\neff \approx 2$ to $3$. Against a 1~kg payload that is a 20 to 30\% mass increase and a \textbf{9 to 12\% reduction in exit velocity}, the last row of Table~\ref{tab:single} and Supplementary Section~\ref{si:cases}: 100~g sheaves at 318~m/s store 10 to 15~kJ against a 50.6~kJ output. The correction does not wash out at larger payload, because a sheave must react the rope tension, so its mass scales with the force it carries and hence with payload mass, leaving $\neff m_s/m$ roughly invariant with scale; and it exceeds any difference between candidate geometries, all of which reach their ideal to within 1\%. The \ptm{} figures are ratios and survive a uniform added mass largely unchanged; only the velocity columns need reading as upper bounds. Each full-speed sheave removed recovers about 4\% of exit velocity, and the orientation sheave is a convenience of layout, not a requirement (Supplementary Section~\ref{si:cases} gives an arrangement without it). The rule is to minimise the number of elements the \emph{output member} touches, not the number in the machine: a sheave on the array side, at $1/k$ of payload velocity, stores $1/k^2$ of the energy, so one sheave moved out of the fast path is worth $k^2$ of them. The dual array adds nothing to the fast path: its two stage blocks translate at $1/k$, and the counter-moving reeving of Section~\ref{sec:stage} halves their stored energy relative to a single fast block.

The term also imposes a rotational-speed constraint. A 100~mm sheave at 318~m/s turns at 61,000~rev/min, and at 1,005~m/s at 192,000~rev/min. A rotating rim carries hoop stress $\sigma = \rho v^2$, so its burst speed is $\sqrt{\sigma/\rho}$, the same expression as the tension member's critical speed below; steel gives 358~m/s and carbon composite about 1,100~m/s, so sheaves in the fast path sit at the same fraction of their strength as the member does of its own. Larger sheaves reduce rotational speed but increase inertia, directly opposing the design objective. This term is applied here only as the post-hoc correction \eqref{eq:sheave}. Together with per-contact friction, which Section~\ref{sec:friction} shows to be of the same order at 2\% per contact, it is the largest omission from the lossless models; incorporating it in the integrator, and establishing $\neff$ for a given reeving, is the most important extension of the present work.

\subsection{Critical speed of the tension member}
\label{sec:critical}

A member of linear density $\rho_\ell = \rho A$ under tension $T$ traversing a curved guide requires centripetal force to follow the curve. The contact force per unit length is proportional to $T - \rho_\ell v^2$, vanishing at
\begin{equation}
c = \sqrt{T/\rho_\ell} = \sqrt{\sigma/\rho},
\label{eq:critical}
\end{equation}
above which the guide cannot redirect the member; this is the critical transport speed of an axially moving string \citep{wickert1990}, and transmitted power $P = (T - \rho_\ell v^2)\,v$ is maximised at $v = c/\sqrt{3} \approx 0.577c$ and falls to zero at $c$. It depends on working \emph{stress}, not cross-section, since a larger diameter raises $T$ and $\rho_\ell$ together; at breaking stress $c \approx 1{,}900$~m/s for UHMWPE \citep{marlow}, with comparable values for PBO and carbon fibre, which cluster in specific strength, so the ceiling is a property of high-performance fibre in general, near 2,000~m/s at breaking stress and considerably lower at a working safety factor. At a safety factor of 3 the three reference cases sit at $v/c$ of 0.09, 0.29 and 0.92; the 10,000:1 case is rope-limited, past the $0.577c$ power optimum, and no braking distance rescues it, since lengthening the stroke raises exit velocity as fast as it lowers tension. The dual array halves the tension in each array member, permitting a smaller cross-section or a higher safety factor there, but does not move this limit: it binds on the output member, which carries the full payload force at payload velocity however many array members feed it, while the array members, at $1/k$ of payload velocity, are nowhere near their critical speed.

\subsection{Array width}
\label{sec:width}

Reproducing a slow initial rise requires wide spans, since a member's rate of contribution goes as $1/R_i$. A machine needing a modest net ratio over a long braking distance therefore needs a proportionally wider array than one needing a high ratio over a short distance: in the three reference cases the ratio of array width to braking distance is 3.3, 2.8 and 2.3, the \emph{lowest} mass ratio demanding the widest array relative to its stroke. At a given total width, fidelity improves with member count and then plateaus at a level set by the width, so array width rather than member count is the governing practical constraint. The dual designs of Sections~\ref{sec:duals} and~\ref{sec:counts}, at 1.98~m and 1.89~m per array with equal counts and 2.05~m, 1.13~m and 1.96~m in Table~\ref{tab:counts}, are narrower than the 2.42~m single-array reference at the same mass ratio because their higher stage ratios ask less of each array; the price is two arrays rather than one, on opposite sides of the same carriage.

% ============================================================================
\section{Conclusion and implementation considerations}
\label{sec:concl}

The RopeComb is a new class of variable mechanical advantage transmission for high-impulse transfer. The conventional mechanisms for continuous ratio variation over a wide range, variable-radius drums, spiral winches and cams, become structurally prohibitive at launch scale: their radii must grow in proportion to the ratio, their rotational inertia grows as the fourth power of radius, and a rim running at payload velocity meets the same $\sqrt{\sigma/\rho}$ stress ceiling as the rope of Section~\ref{sec:critical}. The RopeComb bypasses these limits by synthesising a wide-range ratio profile, from zero to 79 at 1,000:1 and to 251 at 10,000:1, from an array of discrete constant-radius sheaves acting through transverse deflections of a tension member. Because the comb travels with the driver, its mass is source-side driving weight rather than parasitic launch inertia. Because the fold ratio \eqref{eq:member} vanishes at first contact, each engagement is a step in slope rather than in ratio: no member imparts a velocity to the tension member at the instant it engages, so a carriage entering at 10~m/s engages its members without snatch and without clutches or dampers. And because the array ratio is an exact closed-form sum in carriage displacement \eqref{eq:array}, an array of individually concave primitives is synthesised by bounded least squares in seconds to track convex, accelerating targets to within 0.6 to 1.5\% of full scale at 8 to 13 members, a precision set by the member and width budget rather than by the fit, with a two-parameter scaling in $(\nu, \gamma)$ that makes every design a shape scaled by the braking distance, a force scaled by $\sqrt{Mm}$ and a member count scaled by $\sqrt{M/m}/k$.

At launch scale the carriage reaction reaches 23 times the driving weight, about 230~kN at 1,000:1, and how the guide carries it becomes the governing structural problem. Dividing the transmission into two arrays on opposite sides of a central guide, each driving its own counter-moving block of a fixed-ratio stage, resolves it. Partitioning the $k$ output falls between two independent tension members gives the exact weighted net ratio $\Gnet = n_1 G_1 + n_2 G_2$, removes $\lfloor k/2 \rfloor$ elements from the busiest rope path and halves the load on every engagement member. Because the two arrays are independent, the guide can be balanced: the condition is $G_1 : G_2 = n_2 T_2 : n_1 T_1$, with member counts in the same ratio, $n_1 N_1 = n_2 N_2$, since the counts alone fix the terminal ratios, and with the array serving the fewer falls engaging first so that the drift between engagements is bounded. A counter-moving stage produces odd stage ratios, $k = 2p+1$, so the balanced arrays differ; constraining them to mirror images would forfeit the freedom and leave a reaction imbalance of $|n_1 - n_2|/k$, 14\% at $k=7$.

Algorithm~1 configures both arrays in a single solve by seeding from a single-array fit, dealt alternately and perturbed for restarts, and staged optimisation. In simulated 1,000:1 launches with a pre-tensioned compliant output member, the dual array at $k=7$ cancels 97.7\% of the one-sided reaction a single array places on the guide, leaving a reaction imbalance of 2.3\% of the peak, 5.4~kN of 235~kN, and a guide moment equivalent to that peak acting 80~mm off the guide axis in the rope-flow placement of the drawings, and 16~mm at the placement bound, against 1.58~m for the single array, while cutting the compliant \ptm{} payload force from 1.40 to 1.13 against a single array at the same stage ratio and width budget and raising the payload energy retained at 2\% loss per contact from 72 to 77\%; at $k=9$ nine members in arrays under 1.2~m cancel 93\% at a \ptm{} of 1.45. The member counts carry the end of the stroke, where the fit cannot, and the fit carries the middle. Beyond the counts, the balance trade shows that driving the reaction imbalance lower while keeping the force profile admissible is governed by array width, not by reweighting the objective: wider spans overlap and close the local slope flats that cause tension dips. Widening the arrays from 2.0 to 3.3~m with nine members brings the reaction imbalance to 0.5\% (1.2~kN) at a compliant \ptm{} force of 1.29 and 97\% of the terminal ratio.

Four constraints govern the translation of this architecture into hardware.

\begin{itemize}
\item \textbf{Compliance and pre-tension.} A compliant output member pre-tensioned to the working load is mandatory. With an inextensible member the discrete engagements produce a sawtooth whose \ptm{} is 1.9 to 2.6 in the configured designs and no lower than 1.3 in any design examined; the working strain of about 1\% of high-modulus fibre such as UHMWPE or PBO acts as a mechanical low-pass filter that turns the sawtooth into a smooth ripple, at no cost in exit velocity.
\item \textbf{Release timing.} The stroke must end immediately before the terminal traction-loss instability of Section~\ref{sec:instab}. Because payload force is uniform by design, exit velocity follows the square root of payload travel, so releasing 5\% early costs about 2\% of exit velocity while allowing a release tolerance of 11~mm, or 2.5~ms, on the carriage stroke.
\item \textbf{Inertial and frictional penalties.} Fixed-axle sheaves in the fast path act as added payload mass and reduce exit velocity by 9 to 12\%. Per-contact friction at 2\% per contact removes 13 to 28\% of the delivered payload energy and raises the peak carriage reaction by 19 to 32\%, which the guide must be sized for; the two-line stage recovers 1.2 to 2.7 percentage points of the loss, and the path-loss asymmetry between the two lines moves the reaction balance by under half a percentage point with equal member counts and by about one point once the counts have removed the terminal residual.
\item \textbf{Critical speed.} The transverse wave speed $c = \sqrt{\sigma/\rho}$ sets a ceiling near 2,000~m/s at breaking stress. At a safety factor of 3 it binds on the output member at 10,000:1, where $v/c = 0.92$; the array-side members, at $1/k$ of launch speed, are nowhere near it.
\end{itemize}

The same geometry runs in reverse. With the carriage held low and the folds deep, a body that engages the output member draws the tension members out of the spans, and the ratio between its travel and the carriage lift falls through the stroke: a falling mechanical advantage that transfers the kinetic energy of a fast, light body to a slow resisting element and shapes the decelerating force by geometry alone, without programmed valves or active control. The braking member the body engages is itself a first fold, whose ratio \eqref{eq:member} rises from zero, so the resisting element is picked up from rest without a step in speed, the mirror of the snatch-free engagement of the launch. The target is \eqref{eq:ode} with the two sides exchanged, and the balance condition, the count condition and the span order carry over unchanged, since they depend only on the forces the arrays apply to the carriage. Supplementary Section~\ref{si:arrest} configures a worked example, a 50~kg body decelerated from 100 to 10~m/s in 30~m at a \ptm{} deceleration of 1.38, against 9.7 for the same hold-down mass at a fixed ratio.

Single arrays remain the practical choice for low-energy systems, where the overturning moment can be reacted by ordinary rolling-element guides. For heavy, high-impulse launch transmissions with multi-fall reevings, the independent dual array is the general kinematic class. Mirror symmetry is not a design choice within it but an outcome, a solution the same solve may return when the stage ratio is even and the two line routings are alike (Table~\ref{tab:lambda}); since two different arrays are no harder to build than two identical ones, there is no case for imposing it. All results presented here are numerical, and the immediate experimental step is the quasi-static measurement of an array's displacement ratio against the closed-form sum \eqref{eq:array}, which validates the geometric model before any dynamic test.

% ============================================================================

\section*{CRediT authorship contribution statement}

\textbf{Rami N. Mahdi:} Conceptualization, Methodology, Software, Validation, Formal analysis, Investigation, Data curation, Writing -- original draft, Writing -- review \& editing, Visualization, Project administration.

\section*{Funding}

This research did not receive any specific grant from funding agencies in the public, commercial, or not-for-profit sectors.

\section*{Code and data availability}

Everything needed to reproduce this work is at \url{https://github.com/ramimahdi/ropecomb} under the MIT licence. A documented Python package implements the rigid and compliant integrators, the target solver, the single-array fit and selection procedure, and the dual-array extension: the fall-count weighting, the balance condition and its discrete form on the member counts, the merged seed and its restarts, the staged solve, the order-preserving parametrisation, the stage-parity comparison, and the reaction imbalance and guide moment reported from a simulated run. The directory \texttt{paper3/} holds the frozen material for this paper: the configured designs of Section~\ref{sec:sim}, with their full spans, offsets, engagement orders and simulated performance, in \texttt{designs\_16471.json}, \texttt{designs\_counts\_16471.json} and \texttt{locked\_4ms.json}, together with the scripts that produced every number and figure and a script that runs them. An interactive browser tool for designing and simulating RopeComb arrays is available at \url{https://ropecomb.com}; it is a convenience interface to the same method, and the numbers reported here were produced by the package.

\section*{Licence}

Copyright \copyright{} 2026 Rami N. Mahdi, Interesting Machines LLC. All rights reserved. This preprint is distributed by arXiv under the arXiv.org perpetual, non-exclusive license to distribute (\url{https://arxiv.org/licenses/nonexclusive-distrib/1.0/}), which permits arXiv to distribute it and grants no other rights. It is not released under a Creative Commons licence. This version is dated 1~October 2026.

\section*{Competing interests and intellectual property}

The author is the sole member of Interesting Machines LLC and has a financial interest in the commercialisation of the transmission described here. Three United States provisional patent applications, all filed before this article was made public, are directed to aspects of it: No.~64/124,997, filed 3 August 2026, to the single-array architecture; No.~64/139,129, filed 23 August 2026, to the dual-array architecture, including the balance condition, the interleaving rule and the design method; and No.~64/158,432, filed 19 September 2026, to the member-count condition, the span order and placement along the arrays, the merged-seed method and operation in reverse as an arresting transmission. The licence under which this preprint is distributed and the MIT licence on the accompanying software are copyright licences; neither is intended to grant rights under any patent.

\section*{Declaration of generative AI and AI-assisted technologies in the writing process}

All novel mechanisms, simulation models and scripts, design rationales and interpretations in this work are solely the author's own. The author developed the initial codebase and manuscript draft. Large language models (Anthropic's Claude and Google's Gemini) were then used under the author's direct supervision for code refactoring and packaging, exploratory scripts, data logging, plotting, \LaTeX{} typesetting and text polishing. The author has independently verified all claims and data against the accompanying code and takes full responsibility for the integrity and contents of the manuscript.

\clearpage
\bibliographystyle{elsarticle-num-names}
\bibliography{references}

\clearpage
\appendix
\section*{Supplementary Material: Dual-Array RopeComb Derivations and Simulation Data}
\startsupplement

Section, equation, figure and table numbers without an S prefix refer to the main text.

\section{Rigid-body integrator}
\label{si:rigid}

\subsection*{Conventions and assumptions}
Displacements and velocities of the source mass are positive downward, in the direction of its travel, so gravity is a positive acceleration on the source. The net ratio \texttt{get\_gear\_fn(d)}, the array ratio multiplied by the fixed stage or, for a dual array, the weighted sum $n_1 G_1 + n_2 G_2$, is a pure function of source displacement, not of time or velocity. The payload is taken to travel against gravity, so the tension required is $m(g+a)$ and its free-flight deceleration is $g$; for a payload on a track inclined at $\phi$, substitute $g\sin\phi$ in both places. The tension member is inextensible and massless, sheaves are frictionless and of zero rotational inertia, and there is no aerodynamic drag. Integration is explicit and first-order in force, with trapezoidal displacement updates. The force acting on the source at the start of each step uses the ratio and reaction force carried over from the previous step; this one-step lag is intentional and is what makes the scheme explicit.

\subsection*{Pseudocode}
{\footnotesize
\begin{verbatim}
FUNCTION RopeCombRigidBodySimulation(
    heavy_weight,           // source mass M, kg
    target_weight,          // payload mass m, kg
    heavy_height,           // drop height setting entry speed, m
    max_time,               // stop if exceeded, s
    max_target_acc,         // stop if payload acceleration exceeds this, m/s^2
    get_gear_fn,            // net ratio as a function of source displacement
    time_unit = 1e-5        // fixed step, s
)
    g = 9.8

    // --- initial state at the start of the stroke ---------------------
    gear                  = 0.0      // no member engaged yet
    average_gear          = 0.0
    heavy_speed           = SQRT(2 * g * heavy_height)
    heavy_acc             = g
    target_speed          = 0.0
    target_acc            = 0.0
    target_reaction_force = 0.0

    heavy_dec_dist        = 0.0      // source displacement
    target_acc_dist       = 0.0      // payload displacement
    target_rope_end_dist  = 0.0      // displacement of the rope tip
    total_time            = 0.0

    // energy invariant: constant in a lossless model, used only as a check
    E0 = 0.5 * heavy_weight * heavy_speed^2

    WHILE (total_time < max_time AND target_acc < max_target_acc) DO

        total_time = total_time + time_unit

        // --- source mass -----------------------------------------------
        // Reaction is (ratio x payload tension), both carried over from the
        // previous step. Net force goes negative once reaction exceeds weight.
        net_force_on_heavy = (g * heavy_weight)
                             - (average_gear * target_reaction_force)
        heavy_acc          = net_force_on_heavy / heavy_weight

        new_heavy_speed      = heavy_speed + heavy_acc * time_unit
        heavy_dec_dist_delta = time_unit * (new_heavy_speed + heavy_speed) / 2
        heavy_dec_dist       = heavy_dec_dist + heavy_dec_dist_delta
        heavy_speed          = new_heavy_speed

        // --- ratio ------------------------------------------------------
        new_gear     = CALL get_gear_fn(heavy_dec_dist)
        average_gear = (gear + new_gear) / 2
        gear         = new_gear

        // rope tip advances by (source displacement x ratio)
        target_rope_end_dist = target_rope_end_dist
                               + heavy_dec_dist_delta * average_gear

        // --- payload: unilateral contact --------------------------------
        // The member may pull but not push, so the payload advances by
        // whichever is greater: its free flight, or the amount the rope
        // permits. When free flight wins, the member is slack.
        free_flight   = (target_speed - 0.5 * g * time_unit) * time_unit
        rope_limited  = target_rope_end_dist - target_acc_dist
        target_acc_dist_delta = MAX(free_flight, rope_limited)
        is_slack      = (free_flight > rope_limited)

        // invert the trapezoidal rule for the new velocity
        new_target_speed = 2 * (target_acc_dist_delta / time_unit) - target_speed
        new_target_acc   = (new_target_speed - target_speed) / time_unit

        target_speed    = new_target_speed
        target_acc_dist = target_acc_dist + target_acc_dist_delta

        // --- tension for the next step -----------------------------------
        // Averaged over the step; only positive accelerations correspond to
        // the member pulling.
        a_old = IF target_acc     > 0 THEN (g + target_acc)     ELSE 0
        a_new = IF new_target_acc > 0 THEN (g + new_target_acc) ELSE 0
        target_reaction_force = target_weight * (a_old + a_new) / 2

        target_acc = new_target_acc

        // --- diagnostics (no effect on the integration) -------------------
        E = 0.5 * heavy_weight * heavy_speed^2
            + 0.5 * target_weight * target_speed^2
            - heavy_weight * g * heavy_dec_dist
        RECORD total_time, heavy_dec_dist, target_acc_dist,
               heavy_speed, target_speed, gear,
               net_force_on_heavy, target_reaction_force,
               is_slack, (E - E0) / E0

    END WHILE

    RETURN recorded history
END FUNCTION
\end{verbatim}

}

\subsection*{Known limitations}
The contact switch \texttt{MAX(free\_flight, rope\_limited)} flips between two regimes without resolving the instant of transition; this is the principal source of the time-step sensitivity of Section~\ref{si:numerics}, and sub-stepping to the crossing would remove it. Termination lags by one step, since \texttt{target\_acc} is tested at the head of the loop using the previous value. The energy residual $(E - E_0)/E_0$ should remain at zero in exact arithmetic; it is recorded rather than enforced and provides a check on any reimplementation, drift indicating an integration error and non-monotonic drift under refinement indicating that the contact switch is being resolved inconsistently. Finally, \texttt{max\_target\_acc} conflates a physical release with a numerical blow-up; distinguishing them requires the release logic to be modelled explicitly.

\FloatBarrier
\section{Compliant integrator}
\label{si:compliant}

The compliant model differs from Section~\ref{si:rigid} in one element: the output member is a linear spring of stiffness $\Krope = EA/\Lrope$ between the rope tip and the payload, and the payload force is $\Krope$ times the extension of that spring rather than the contact tension of the rigid model. For the worked case $\Krope \approx 1.6\times10^4$~N/m, giving 0.24~m of extension at 4~kN, or 1.4\% strain. The member is pre-tensioned to the working load before the stroke. With a slack start the payload mode on that compliance, about 20~Hz against a stroke of under 100~ms, is excited by the step input at first engagement and gives a \ptm{} force near 2.3 with the member unloaded for some 11\% of the stroke; pre-tensioning removes the step and the same apparatus then gives a \ptm{} near 1.03 in this isolated test, with no unloaded interval and exit velocity within 0.1\% of ideal. The time step is $2\times10^{-5}$~s.

The model lumps all compliance into the output member. It neglects the array-side tension members, which are locally an order of magnitude stiffer and are where discrete engagement occurs, and it omits sheave rotational inertia, member mass and roller friction. Each of these is a plausible pulse mechanism, and the value of the compliant model is in isolating the effect of pre-tension rather than in predicting real force histories.

\FloatBarrier
\section{Single-array selection procedure}
\label{si:select}

The single-array reference designs of Table~\ref{tab:single} of the main text are produced by a joint search over member count and stage ratio, with the plain least-squares objective \eqref{eq:single-obj} and no other terms:

{\footnotesize
\begin{verbatim}
INPUT   M, m       source and payload mass     OUTPUT  k          fixed-stage ratio
        v0         entry velocity                      {R_i}      span half-widths
        v_end, T   deceleration to achieve             {s_i}      engagement offsets

1  SOLVE THE TARGET
   bisect on F until the ideal run, terminated at v_h = v_end, takes time T
   -> design force F, braking distance d_max, target profile G*(d)

2  GRID SEARCH
   for N = 2 .. 16:
     for k in {1, 2, 3, 5, 7, 9}:
       for restart = 1 .. 50:
         w      ~ U(0.2, 0.8)                       mean span for this restart
         R_i(0) = 0.5 * w * U(0.6, 1.4)             half-widths, i = 1..N
         s_i(0) = sort( U(0, 0.85 * d_max) )        engagement offsets
         {R,s}  = argmin  SUM_d ( k*G_array(d) - G*(d) )^2
                  bounded least squares in log R; no other terms in the objective
       keep the 5 restarts of lowest residual

3  EVALUATE each kept candidate
   simulate rigid      -> carriage travel d_r
   simulate compliant  -> carriage travel d_c
   prune members with s_i >= max(d_r, d_c)
   re-simulate the pruned array, rigid and compliant
   p2m = max(F_payload) / mean(F_payload) over the first 99.5% of the trace

4  SELECT
   admissible: p2m_compliant <= 1.3  AND  v_compliant >= 0.99 * v_ideal
   score = (55 - N_eff - k - W)/50 * v_rigid * v_compliant / sqrt(p2m_r + p2m_c)
           W = total array width in metres
   return the highest-scoring admissible candidate
\end{verbatim}
}

The score rewards exit velocity in both models, penalises \ptm{} force in both, and discounts cost through $N_{\mathrm{eff}} + k + W$, the three quantities that set what the machine costs to build; width enters in metres with unit weight, which at these scales makes a metre of array about as expensive as one engagement member. The admissibility filter is applied first because the score alone will trade away exit velocity: at 10,000:1 the highest-scoring candidate overall reaches only 97\% of its ideal and is excluded. The two-stage initialisation of step~2 matters: drawing every $R_i$ independently makes the total width concentrate near its mean, so wide arrays are never sampled. Pruning removed nothing in any of the 1,350 candidates simulated; under a reachable target the fitter does not park surplus members beyond the stroke. Of the 450 candidates simulated per case only about 320 are distinct, and within the selected cell the five kept restarts agree on compliant \ptm{} force to within 2.4\%, 0.7\% and 0.1\% at 100:1, 1,000:1 and 10,000:1.

\begin{table}[!htbp]
\centering
\caption{Member count against array width, peak force and exit velocity at 1,000:1 with $k=5$ held fixed. The trade is width against force smoothness, not member count against exit velocity: exit velocity saturates at six members, and further members buy a lower peak force at roughly 0.2~m of additional array each.}
\label{tab:si-nsweep}
\begin{tabular}{lccc}
\toprule
members & array width & peak force, compliant & exit velocity, compliant \\
\midrule
4 & 0.60 m & 1.82 & 269.7 m/s \\
5 & 0.84 m & 1.57 & 297.8 m/s \\
6 & 1.13 m & 1.39 & 310.1 m/s \\
7 & 1.48 m & 1.26 & 315.7 m/s \\
8 & 1.89 m & 1.16 & 318.1 m/s \\
\bottomrule
\end{tabular}
\end{table}

\begin{table}[!htbp]
\centering
\caption{Fixed-stage ratio against member floor, smallest qualifying member count, narrowest qualifying array and best compliant \ptm{} force, for the 1,000:1 case over the grid, counting only candidates that meet the 1.3 criterion and reach at least 99\% of the ideal exit velocity. Below $k=3$ the member floor cannot be satisfied within the sixteen-member limit; above it, array width passes through a minimum at $k=5$, which is the value the search selects for a single array.}
\label{tab:si-kfloor}
\begin{tabular}{lcccccc}
\toprule
$k$ & 1 & 2 & 3 & \textbf{5} & 7 & 9 \\
\midrule
member floor $\Gnet/2k$ & 39.6 & 19.8 & 13.2 & \textbf{7.9} & 5.7 & 4.4 \\
smallest $N$ meeting the criterion & --- & --- & 11 & \textbf{7} & 7 & 6 \\
narrowest array meeting it & --- & --- & 1.56 m & \textbf{1.44 m} & 2.63 m & 3.12 m \\
best compliant \ptm{} & --- & --- & 1.066 & \textbf{1.009} & 1.007 & 1.011 \\
\bottomrule
\end{tabular}
\end{table}

The smallest member count that meets the 1.3 bound while reaching at least 99\% of the ideal exit velocity is six at both 100:1 and 1,000:1 and twelve at 10,000:1; below those counts the bound is unreachable at any stage ratio in the grid, the best five-member design giving 1.43 at 100:1 and 1.61 at 1,000:1. The selected designs use eight, nine and thirteen members, one to three above the floor, which buys margin in peak force at a modest cost in array width.

\FloatBarrier
\section{Sensitivity to output-member stiffness}
\label{si:stiffness}

The stiffness $\Krope = EA/\Lrope$ spans a narrow band across the grades a designer would choose and the routing lengths the machine needs: 15,730~N/m for SK75 over 17.8~m, 16,471~N/m for the length adopted in the main text, and 19,663~N/m for a top-end SK99 or DM20 fibre at $E = 125$~GPa. Over that band the compliant \ptm{} of the configured dual designs moves by 0.02 at $\cfg{7}{7}$ and 0.04 at $\cfg{5}{9}$. The figure is not insensitive to stiffness in general: taken over an unphysical 8,000 to 30,000~N/m it ranges from 1.04 to 1.40 at $\cfg{7}{7}$ and from 1.11 to 1.61 at $\cfg{5}{9}$, and for the 1,000:1 single-array reference it is 1.04 at half the nominal stiffness, 1.18 at 1.5 times and 1.36 at twice. What makes it stable in practice is that material and length choices are themselves constrained. A candidate whose \ptm{} climbs steeply with stiffness is leaning on the assumption rather than on its geometry, and the sweep is a useful screen: at 100:1 a six-member $k=3$ candidate that looked competitive at nominal stiffness reached 2.49 at double stiffness and was rejected on that ground.

\FloatBarrier
\section{Numerical status}
\label{si:numerics}

All results use a time step of $10^{-5}$~s for the rigid model and $2\times10^{-5}$~s for the compliant one. Refining by a factor of twenty, to $10^{-6}$~s, changes them by the amounts in Table~\ref{tab:si-dt}; the compliant model, from which the reported peak forces come, is converged to a fraction of a percent, and the rigid model, which must resolve a hard contact switch, to a few percent. A step of $10^{-4}$~s is not adequate and was not used: at 100:1 it overstates exit velocity by 26\% with an energy error of 35\%. The terminal traction-loss pulse itself is not converged, but it is excluded from every reported figure by the release convention of Section~\ref{sec:setup}; establishing its magnitude and the precise time of divergence would require sub-stepping to the contact transition.

\begin{table}[!htbp]
\centering
\caption{Time-step convergence: change in each reported quantity on refining the step by a factor of twenty, to $10^{-6}$~s. Peak-to-mean is taken over the first 99.5\% of the trace.}
\label{tab:si-dt}
\begin{tabular}{lcc}
\toprule
& rigid & compliant \\
\midrule
exit velocity & within 1.8\% & within 0.07\% \\
\ptm{} payload force & within 4.3\% & within 0.23\% \\
energy drift & below 0.9\% & below 0.6\% \\
\bottomrule
\end{tabular}
\end{table}

The energy balance of each single-array reference run, payload energy plus the kinetic energy the source carries away, closes on the input $Mg(h_0 + \dmax)$ to 99.3\%, 100.0\% and 100.4\% at 100:1, 1,000:1 and 10,000:1, and the simulated designs deliver 86.0\%, 86.7\% and 86.4\% of that input against the analytic 86.3\%; the departures track the closure errors and the fact that each design brakes the source marginally away from exactly 4.00~m/s, not a transmission exceeding its lossless bound.

\FloatBarrier
\section{Complete design coordinates}
\label{si:coords}

Tables~\ref{tab:si-k7} to~\ref{tab:si-10000to1} give the fitted half-widths $R_i$ and engagement offsets $s_i$ of every design in the main text, in metres and in engagement order, at the precision of the frozen data files; the count-balanced designs of main-text Section~\ref{sec:counts} are tabulated in Section~\ref{si:counts}. The span floor of 10~cm is active in every design: each array ends in a run of members at $R_i = 0.050$~m.

\begin{table}[!htbp]\centering\caption{Full-precision coordinates (m) of the $\cfg{7}{7}$ dual design, $\lambda=0.03$, from \texttt{designs\_16471.json}. Array 1 serves 3 falls, array 2 serves 4. Fitted rms 0.3801; reaction imbalance 8.09\%; terminal ratio 74.44 (93.9\% of target); array widths 1.954 and 1.981 m.}\label{tab:si-k7}
\begin{tabular}{lcccc}\toprule
member & $R_{1,i}$ & $s_{1,i}$ & $R_{2,i}$ & $s_{2,i}$ \\ \midrule
1 & 0.3671 & 0.0000 & 0.3667 & 0.0218 \\
2 & 0.2785 & 0.2468 & 0.2671 & 0.4068 \\
3 & 0.1266 & 0.5524 & 0.1336 & 0.6429 \\
4 & 0.0547 & 0.7134 & 0.0732 & 0.7569 \\
5 & 0.0500 & 0.7785 & 0.0500 & 0.8070 \\
6 & 0.0500 & 0.8071 & 0.0500 & 0.8254 \\
7 & 0.0500 & 0.8254 & 0.0500 & 0.8306 \\
\bottomrule\end{tabular}\end{table}
\begin{table}[!htbp]\centering\caption{Full-precision coordinates (m) of the $\cfg{5}{9}$ dual design, $\lambda=0.03$, from \texttt{designs\_16471.json}. Array 1 serves 4 falls, array 2 serves 5. Fitted rms 0.6395; reaction imbalance 7.04\%; terminal ratio 71.77 (90.5\% of target); array widths 1.887 and 1.546 m.}\label{tab:si-k9}
\begin{tabular}{lcccc}\toprule
member & $R_{1,i}$ & $s_{1,i}$ & $R_{2,i}$ & $s_{2,i}$ \\ \midrule
1 & 0.4885 & 0.0000 & 0.4182 & 0.0376 \\
2 & 0.2721 & 0.3320 & 0.1901 & 0.5227 \\
3 & 0.0830 & 0.6619 & 0.0649 & 0.7374 \\
4 & 0.0500 & 0.7786 & 0.0500 & 0.8099 \\
5 & 0.0500 & 0.8099 & 0.0500 & 0.8216 \\
\bottomrule\end{tabular}\end{table}
\begin{table}[!htbp]\centering\caption{Full-precision coordinates (m) of the 100:1 single-array reference, $N=8$, $k=2$, from \texttt{locked\_4ms.json}. Array width 2.820 m; fit residual 0.57\% of full scale.}\label{tab:si-100to1}
\begin{tabular}{lcc}\toprule
member & $R_i$ & $s_i$ \\ \midrule
1 & 0.3238 & 0.0120 \\
2 & 0.4169 & 0.2641 \\
3 & 0.2846 & 0.4949 \\
4 & 0.1543 & 0.6551 \\
5 & 0.0803 & 0.7482 \\
6 & 0.0500 & 0.7980 \\
7 & 0.0500 & 0.8227 \\
8 & 0.0500 & 0.8278 \\
\bottomrule\end{tabular}\end{table}
\begin{table}[!htbp]\centering\caption{Full-precision coordinates (m) of the 1,000:1 single-array reference, $N=9$, $k=5$, from \texttt{locked\_4ms.json}. Array width 2.422 m; fit residual 0.78\% of full scale.}\label{tab:si-1000to1}
\begin{tabular}{lcc}\toprule
member & $R_i$ & $s_i$ \\ \midrule
1 & 0.2459 & 0.0119 \\
2 & 0.3139 & 0.2261 \\
3 & 0.2349 & 0.4407 \\
4 & 0.1392 & 0.6051 \\
5 & 0.0773 & 0.7102 \\
6 & 0.0500 & 0.7704 \\
7 & 0.0500 & 0.8025 \\
8 & 0.0500 & 0.8190 \\
9 & 0.0500 & 0.8190 \\
\bottomrule\end{tabular}\end{table}
\begin{table}[!htbp]\centering\caption{Full-precision coordinates (m) of the 10,000:1 single-array reference, $N=13$, $k=9$, from \texttt{locked\_4ms.json}. Array width 1.924 m; fit residual 1.48\% of full scale.}\label{tab:si-10000to1}
\begin{tabular}{lcc}\toprule
member & $R_i$ & $s_i$ \\ \midrule
1 & 0.1224 & 0.0121 \\
2 & 0.1459 & 0.1535 \\
3 & 0.1274 & 0.3099 \\
4 & 0.0971 & 0.4512 \\
5 & 0.0690 & 0.5646 \\
6 & 0.0502 & 0.6481 \\
7 & 0.0500 & 0.7028 \\
8 & 0.0500 & 0.7390 \\
9 & 0.0500 & 0.7649 \\
10 & 0.0500 & 0.7867 \\
11 & 0.0500 & 0.7999 \\
12 & 0.0500 & 0.8000 \\
13 & 0.0500 & 0.8013 \\
\bottomrule\end{tabular}\end{table}

\FloatBarrier
\section{Verification of the closed-form ratio}
\label{si:polyline}

The closed form \eqref{eq:array} was checked against rope length measured from an explicit polyline through the support and engagement-member coordinates, finite-differenced with respect to carriage displacement. Across three configurations spanning 4 to 20 members the worst relative disagreement in ratio was $1.5\times10^{-5}\%$ and cumulative take-up agreed to $7\times10^{-4}\%$. Agreement between a \emph{fitted} array profile and the target is not evidence of correctness, since the fit produces that agreement by construction; the meaningful checks are the geometric verification above and the agreement between simulated dynamics and the independently computed ideal velocity and force histories, which were not fitted to anything.

\FloatBarrier
\section{The single-array embodiment, the 100:1 and 10,000:1 references, and a minimal fast path}
\label{si:cases}

\begin{figure}[!htbp]
\centering
\includegraphics[width=\linewidth]{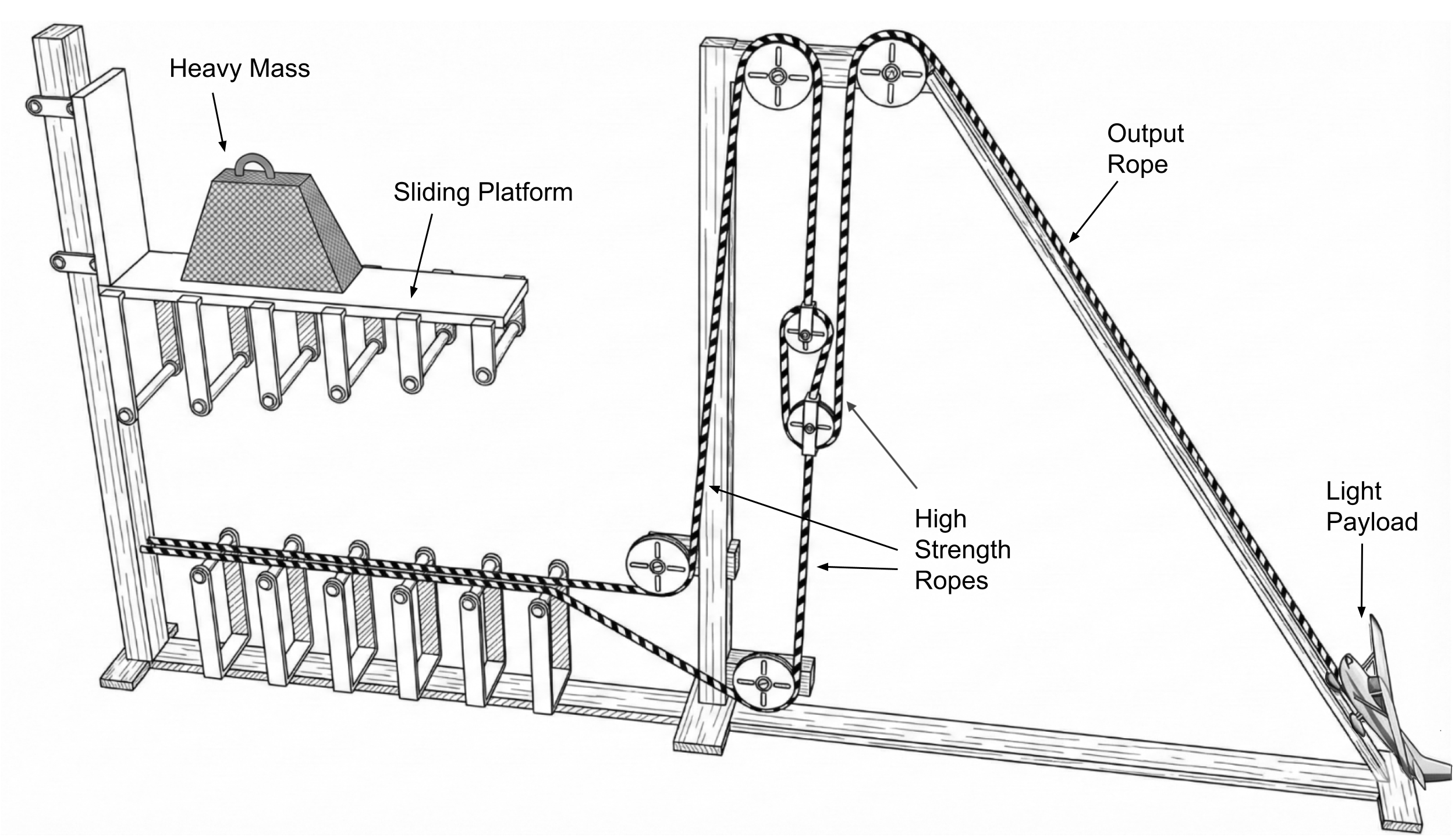}
\caption{The single-array RopeComb launcher. The heavy mass rides a sliding platform, the carriage, down a vertical guide; the engagement members beneath the platform deflect a single tension member into a row of fixed spans, each fold drawing in rope. The rope passes through a fixed-ratio stage at the mast, here 5:1, and the output rope leaves along the inclined rail to the light payload. Only the output rope and the sheaves that carry it move at payload speed; every other element travels at a fifth of that or slower. Member count and spacing are illustrative, not one of the designs of Table~\ref{tab:single} of the main text.}
\label{fig:si-single-hero}
\end{figure}

Figure~\ref{fig:si-single-hero} renders the single-array embodiment, the case $n_2 = 0$ of the kinematics of Section~\ref{sec:mech} of the main text and the machine that the reference designs of Table~\ref{tab:single} there configure. One array on one side of the guide drives one tension member into a fixed-ratio stage; the whole carriage reaction is carried at a lateral offset from the guide, which is the overturning moment of Section~\ref{si:guide} and the load that the dual array of Figure~\ref{fig:hero} of the main text places where it can be balanced.

Section~\ref{sec:singles} of the main text tabulates all three single-array references and illustrates the 1,000:1 configuration. The remaining two are drawn here. They bracket the design space at either end. The 100:1 array, eight members at $k=2$ in 2.82~m, is the widest of the three because a modest terminal ratio of 25 spread over a full stroke demands a slow initial rise, and a slow rise demands wide spans. The 10,000:1 array, thirteen members at $k=9$ in 1.92~m, requires a terminal ratio ten times larger and is half as wide, yet its fit residual of 1.48\% is the largest of the three: geometry becomes harder as mass ratio rises while the array becomes narrower, and at 10,000:1 the tension member is simultaneously approaching its critical speed, at $v/c = 0.92$ at a safety factor of 3. The two binding constraints thus run in opposite directions.

\begin{figure}[!htbp]
\centering
\includegraphics[width=\linewidth]{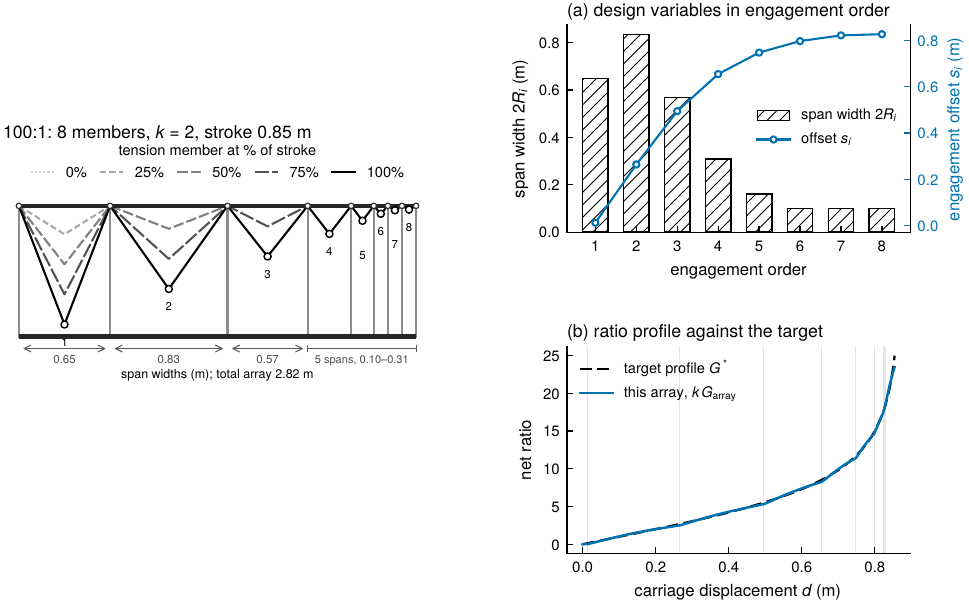}
\caption{The 100:1 single-array reference: eight members at $k=2$ in 2.82~m of array, spans of 65, 83, 57, 31, 16, 10, 10 and 10~cm.}
\end{figure}

\begin{figure}[!htbp]
\centering
\includegraphics[width=\linewidth]{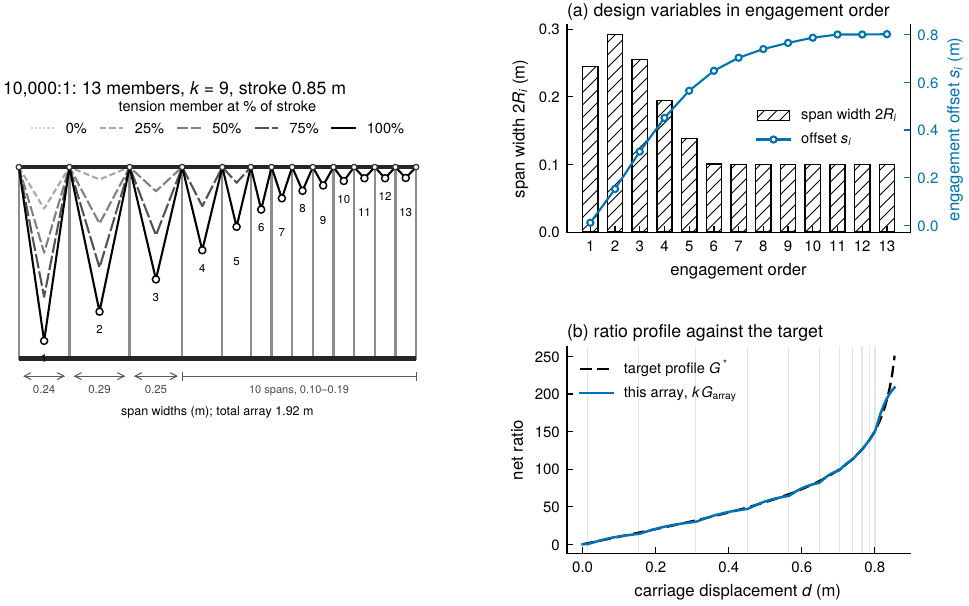}
\caption{The 10,000:1 single-array reference: thirteen members at $k=9$ in 1.92~m of array, spans of 24, 29, 25, 19, 14 and eight of 10~cm.}
\end{figure}

\begin{figure}[!htbp]
\centering
\includegraphics[width=\linewidth]{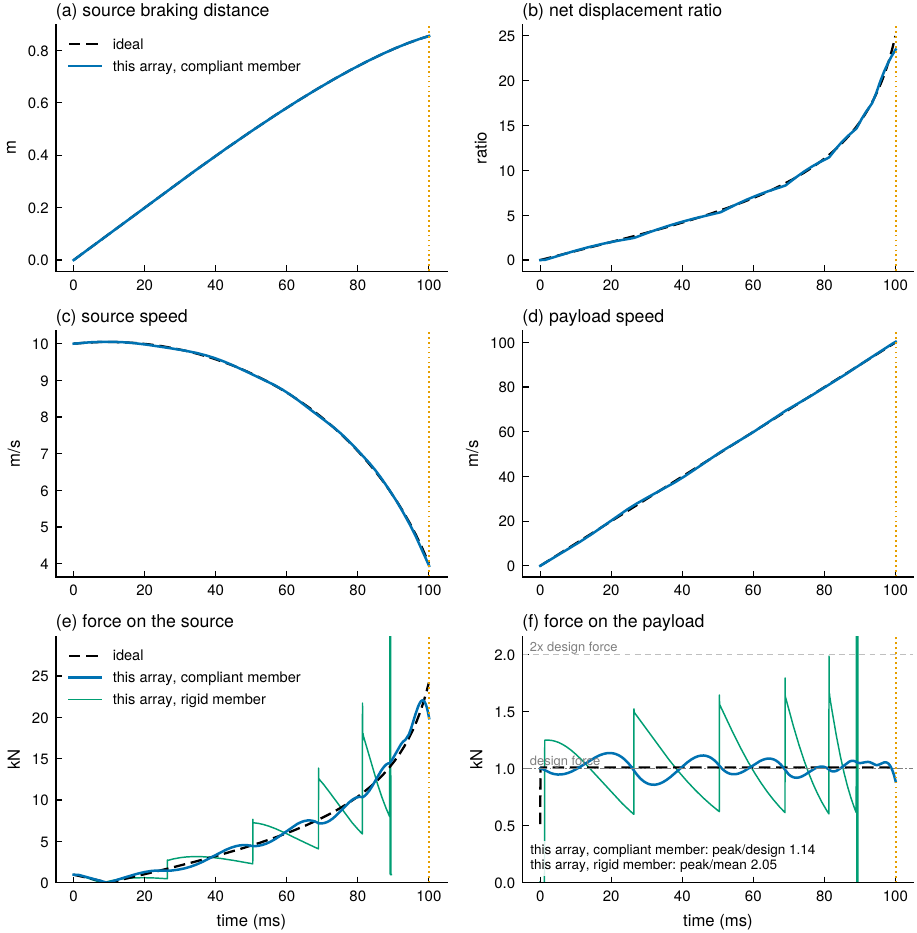}
\caption{The 100:1 reference with a compliant, pre-tensioned output member: 100.2~m/s at 1.12 times the design force; the green trace in (e) and (f) is the rigid member. Compliance closes the rigid figure of 2.05 by a factor of 1.82 here, 1.90 at 1,000:1 and 1.76 at 10,000:1.}
\end{figure}

\begin{figure}[!htbp]
\centering
\includegraphics[width=\linewidth]{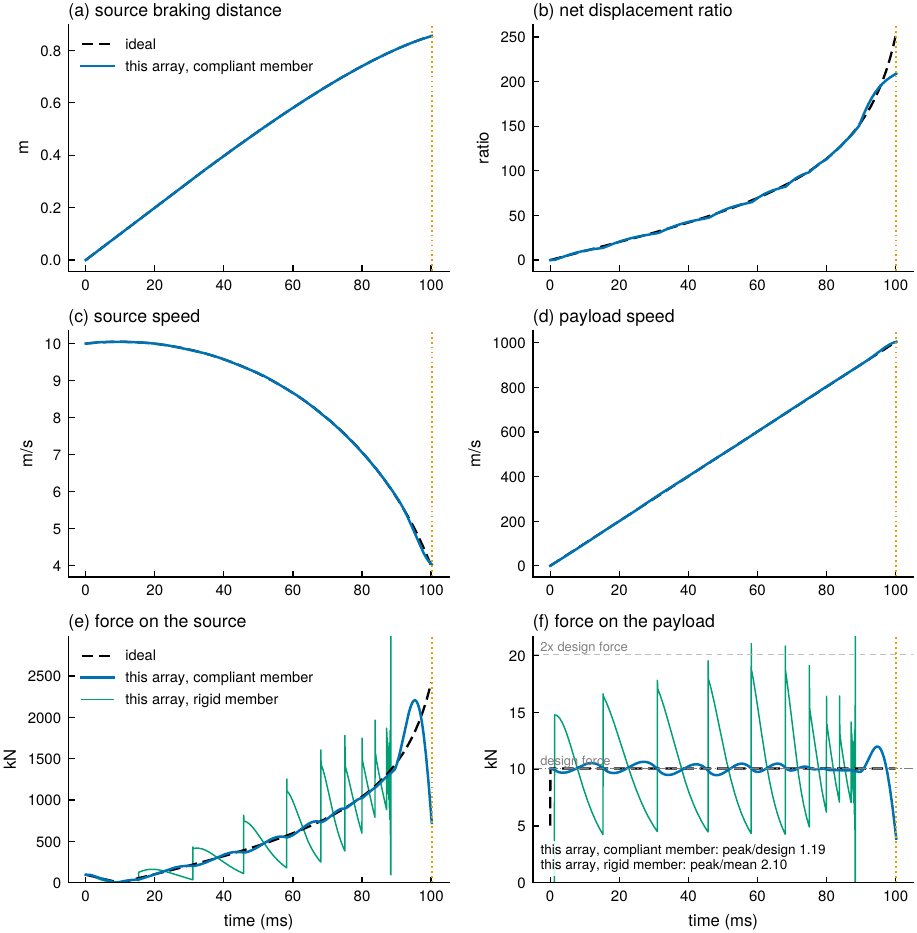}
\caption{The 10,000:1 reference, compliant: 1,004.8~m/s at 1.19 times the design force, 100\% of the ideal exit velocity; the green trace in (e) and (f) is the rigid member. Both force panels are drawn from zero so the traces can be read against the absolute scale.}
\end{figure}

\begin{figure}[!htbp]
\centering
\includegraphics[width=\linewidth]{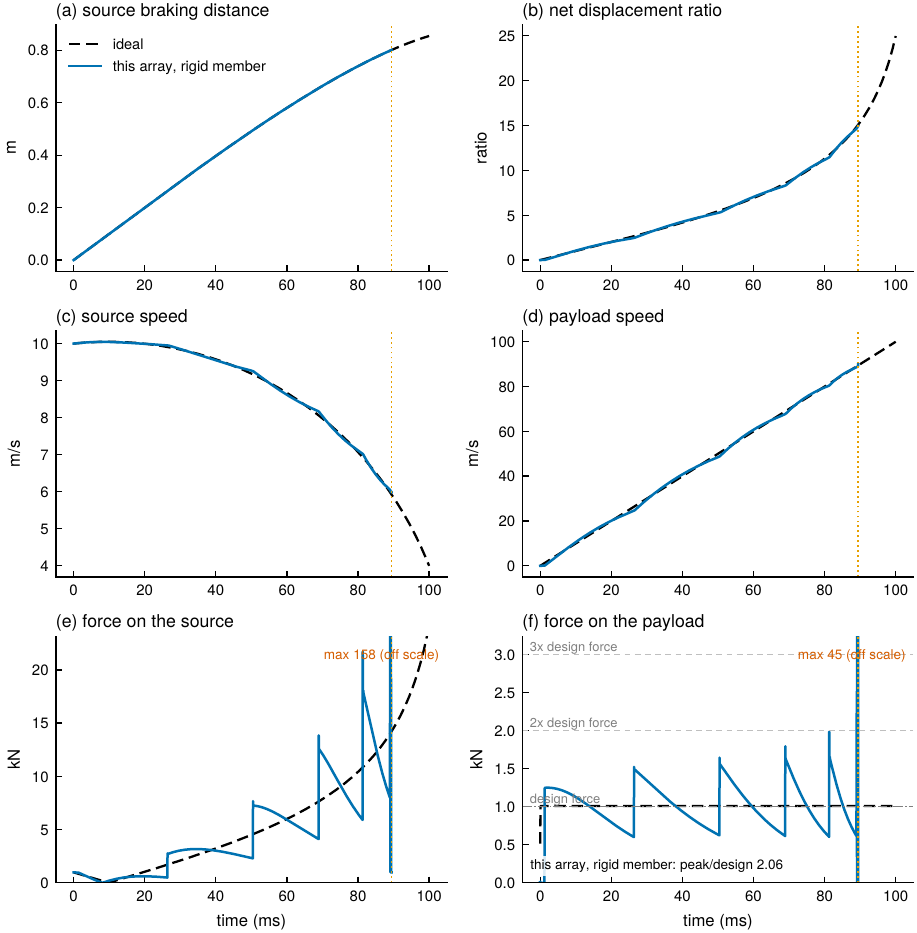}
\caption{Rigid-body dynamics, 100:1: exit velocity 89.8~m/s against an ideal 99.9~m/s; \ptm{} 2.05 over the release window, rising to roughly 45 times the mean if the terminal traction-loss divergence is included, a multiple that is not time-step converged.}
\end{figure}

\begin{figure}[!htbp]
\centering
\includegraphics[width=\linewidth]{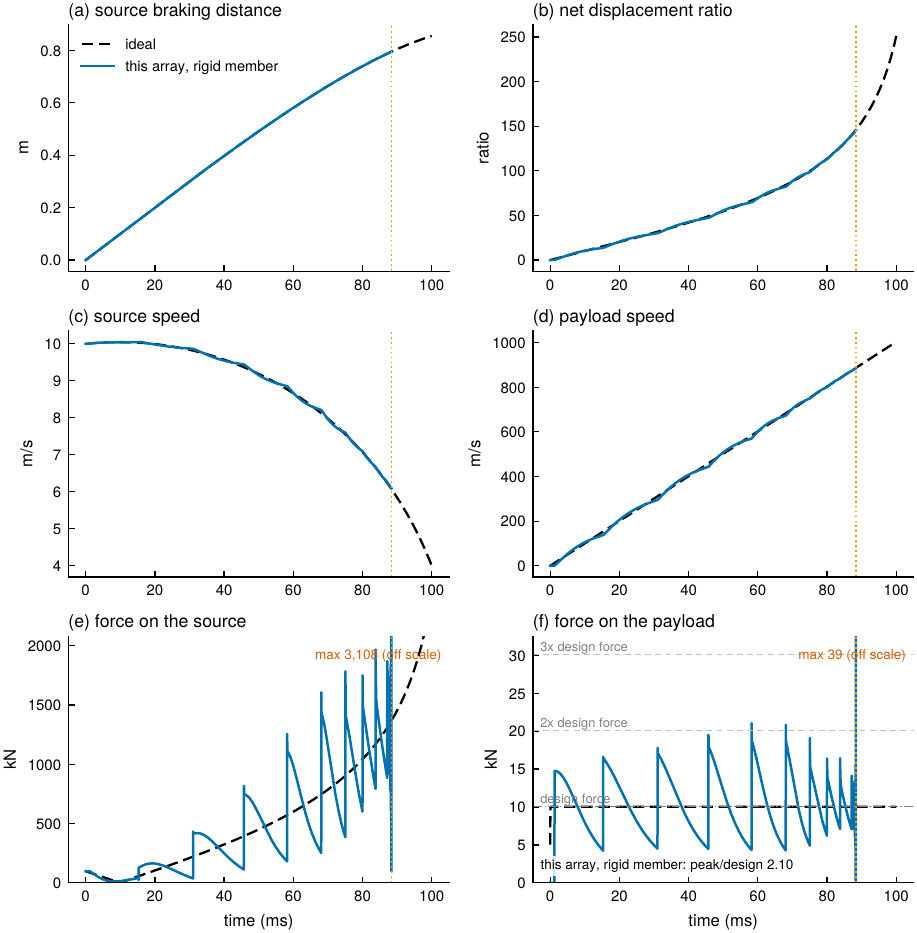}
\caption{Rigid-body dynamics, 10,000:1: exit velocity 885.9~m/s against an ideal 1,003.5~m/s; \ptm{} 2.10 over the release window, comparable to the 100:1 case rather than markedly better, since once $N$ and $k$ are selected jointly mass ratio no longer orders severity.}
\end{figure}

\begin{figure}[!htbp]
\centering
\includegraphics[width=0.8\linewidth]{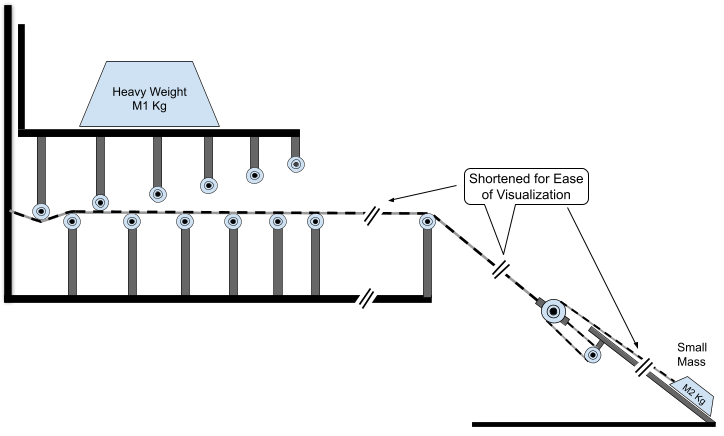}
\caption{A single array with a 3:1 stage arranged so that the output member passes over one fewer sheave at payload speed than the arrangement of Figure~\ref{fig:hero} of the main text, taking $\neff$ from about 2 to about 1 and recovering about 4\% of exit velocity (Table~\ref{tab:si-sheave} and Section~\ref{sec:sheave} of the main text). The stage ratio is unchanged; only the number of direction changes imposed on the output member differs. Breaks in the rope runs are drawn for compactness and do not represent joints.}
\end{figure}

\begin{table}[!htbp]
\centering
\caption{Exit velocity of the single-array references corrected for the rotational inertia of sheaves in the fast path, which acts as added payload mass by \eqref{eq:sheave} of the main text; $\neff = \sum_i (v_i/v)^2$ counts each sheave by the square of its rim speed.}
\label{tab:si-sheave}
\begin{tabular}{lcccc}
\toprule
& lossless model & $\neff = 3$ & $\neff = 2$ & $\neff = 1$ \\
\midrule
100:1    & 100.2 m/s  & 87.9  & 91.5  & 95.5 \\
1,000:1  & 318.2 m/s  & 279.1 & 290.5 & 303.4 \\
10,000:1 & 1,004.8 m/s & 881.3 & 917.2 & 958.0 \\
\bottomrule
\end{tabular}
\end{table}

\FloatBarrier
\section{Guide moment, bearing couple and guide friction}
\label{si:guide}

The statics behind Sections~\ref{sec:reactions}, \ref{sec:configs} and~\ref{sec:moment} of the main text. Let the carriage be guided by two bearings at spacing $h$ along the guide. Each engagement member pushes on the carriage along the guide with $f_i = n_j T_j g_i(d)$, $g_i = 2D_i/\sqrt{R_i^2 + D_i^2}$, the transverse components of its two rope segments cancelling by the symmetry of the fold about the member; the array's reaction $F_{\mathrm{carriage}} = \sum_i f_i$ is therefore axial and acts at the centre of pressure $\bar{x}(d) = \sum_i x_i f_i / \sum_i f_i$, $x_i$ the member's position along the array. A single array on one side of the guide loads it with the moment $\mathcal{M}(d) = F_{\mathrm{carriage}}\,\bar{x}$, reacted by the two bearings as a couple, each carrying $\mathcal{M}/h$ across the guide, and with a friction coefficient $\mu$ at each bearing the friction force opposing the carriage is $2\mu \mathcal{M}/h$. The work done against it over the stroke is
\begin{equation}
E_{\mathrm{guide}} = \frac{2\mu}{h} \int_0^{\dmax} \mathcal{M}\,dd = 2\mu\,\frac{\bar{e}}{h}\,E_{\mathrm{transferred}}, \qquad \bar{e} = \frac{\int_0^{\dmax} \mathcal{M}\,dd}{\int_0^{\dmax} F_{\mathrm{carriage}}\,dd},
\label{eq:si-guide}
\end{equation}
since in the lossless model the work done by the reaction on the carriage is the energy transferred to the payload; $\bar{e}$ is the force-weighted mean of the centre of pressure. The fraction of the transferred energy lost in the guide is $2\mu\bar{e}/h$, and it depends on neither the mass ratio nor the fit: it is set by the guide geometry, the bearings and where the members sit. For the 1,000:1 single-array reference of main-text Table~\ref{tab:single}, with the spans in engagement order outward from the guide, $\bar{x}$ runs from 0.25~m at first engagement to 1.58~m at the end of the stroke, $\bar{e} = 0.89$~m, the peak moment is 360~kN\,m and $\int \mathcal{M}\,dd = 45.1$~kJ\,m against $\int F_{\mathrm{carriage}}\,dd = 50.4$~kJ, the work the reaction does on the carriage over the stroke. Table~\ref{tab:si-guide} evaluates \eqref{eq:si-guide} over the range of $\mu$ and $\bar{e}/h$ an embodiment may take, rolling linear guides at the low end of $\mu$ and plain sliding bearings at the high end, a compact carriage with the array close to the guide at the low end of $\bar{e}/h$ and one with the array a full bearing spacing outboard at the high end. Placing the driving mass at an offset $x_{\mathrm{cg}}$ from the guide axis subtracts $F_{\mathrm{carriage}}\,x_{\mathrm{cg}}$ from the moment, the inertia force of the decelerating mass acting at its centre, and no fixed offset cancels a sweeping $\bar{x}$: at $x_{\mathrm{cg}} = 1.2$~m the net moment runs from $-35$ to $+86$~kN\,m and reverses sign mid-stroke, and at the terminal centre of pressure, 1.58~m, it reaches $-63$~kN\,m earlier in the stroke.

\begin{table}[!htbp]
\centering
\caption{Guide friction for a single-sided array at 1,000:1 (50.4~kJ of work done by the reaction over the stroke), from \eqref{eq:si-guide}: the fraction of the transferred energy lost in the guide. $\bar{e}$ is the force-weighted centre of pressure, 0.89~m for the reference design in the drawing placement; the peak bearing couple is $360$~kN per metre of bearing spacing, independent of $\mu$.}
\label{tab:si-guide}
\begin{tabular}{lccc}
\toprule
& $\bar{e}/h = 0.25$ & $\bar{e}/h = 0.5$ & $\bar{e}/h = 1$ \\
\midrule
$\mu = 0.005$ & 0.25\% & 0.5\% & 1.0\% \\
$\mu = 0.02$  & 1.0\%  & 2.0\% & 4.0\% \\
$\mu = 0.1$   & 5.0\%  & 10\%  & 20\% \\
\bottomrule
\end{tabular}
\end{table}

For the dual array the two reactions oppose across the guide and the moment is main-text equation~\eqref{eq:moment}, $\mathcal{M} = \mathcal{M}_2 - \mathcal{M}_1$ with $\mathcal{M}_j = \sum_i x_i f_i$ over array $j$. The bearings react it by the same couple, $|\mathcal{M}|/h$ each, the friction is $2\mu|\mathcal{M}|/h$, and the energy lost is $(2\mu/h)\int_0^{\dmax}|\mathcal{M}|\,dd$, so the reduction against the single array is $\int \mathcal{M}_{\mathrm{single}}\,dd / \int |\mathcal{M}_2 - \mathcal{M}_1|\,dd$, a property of the two geometries alone. The reaction imbalance $\Delta F = |F_1 - F_2|$ that the synthesis balances is a force; the moment equals $\bar{e}\,\Delta F$ only where the two centres of pressure coincide, and since the ratio is a sum over members the order of the spans along each array is free and sets the centres of pressure without touching the fit. Table~\ref{tab:si-moment} gives, for the five designs of main-text Table~\ref{tab:counts}, the peak moment, its stroke integral and the reduction factors in the drawing placement and at the span order minimising the peak moment, found by exhaustive search over both arrays' orders at $k=9$ and by pairwise-swap descent from the drawing order and thirty random orders at $k=7$ (\texttt{moment\_statics.py}, \texttt{moment\_statics.json}); the orders are listed beneath. Minimising the peak is not the same as minimising the integral, which is why $\cfg{5{:}4}{9}$ gains on one and loses on the other. Under the friction model of Section~\ref{si:friction} the order of the members along the tension member changes the reactions at second order, members nearer the anchor carrying the higher tension, and nothing else about the design.

\begin{table}[!htbp]
\centering
\caption{The guide moment of the configured designs at 1,000:1, from the frozen coordinates: peak of $|\mathcal{M}_2 - \mathcal{M}_1|$, its stroke integral, and the reductions against the single-array reference (360~kN\,m, 45.1~kJ\,m), in the drawing placement and at the searched span order. The peak bearing couple is the peak moment per metre of bearing spacing.}
\label{tab:si-moment}
\footnotesize
\begin{adjustbox}{max width=\linewidth}
\begin{tabular}{@{}lcccccc@{}}
\toprule
& \multicolumn{3}{c}{drawing placement} & \multicolumn{3}{c}{searched placement} \\
\cmidrule(lr){2-4}\cmidrule(lr){5-7}
& peak (kN\,m) & $\int|\mathcal{M}|\,dd$ (kJ\,m) & reduction, peak / integral & peak (kN\,m) & $\int|\mathcal{M}|\,dd$ (kJ\,m) & reduction, peak / integral \\
\midrule
$\cfg{7}{7}$ & 21.3 & 3.60 & 16.9 / 12.6 & 7.5 & 1.99 & 48.0 / 22.7 \\
$\cfg{5}{9}$ & 32.0 & 7.15 & 11.2 / 6.3 & 5.0 & 2.11 & 72.2 / 21.4 \\
$\cfg{8{:}6}{7}$ & 18.7 & 3.65 & 19.2 / 12.4 & 3.8 & 1.06 & 94.6 / 42.5 \\
$\cfg{5{:}4}{9}$ & 11.2 & 2.82 & 32.2 / 16.0 & 8.6 & 3.33 & 42.0 / 13.6 \\
$\cfg{6{:}5}{9}$ & 21.5 & 5.43 & 16.7 / 8.3 & 6.8 & 3.15 & 53.0 / 14.4 \\
\bottomrule
\end{tabular}
\end{adjustbox}
\end{table}

\noindent Searched span orders, members numbered in engagement order and listed from the guide outward, array 1 then array 2: $\cfg{7}{7}$: 1,2,3,4,5,7,6 and 6,1,7,3,2,4,5; $\cfg{5}{9}$: 3,4,2,1,5 and 2,5,3,4,1; $\cfg{8{:}6}{7}$: 8,3,4,2,5,6,7,1 and 3,2,6,4,5,1; $\cfg{5{:}4}{9}$: 3,1,2,4,5 and 2,1,3,4; $\cfg{6{:}5}{9}$: 3,2,4,6,1,5 and 2,3,5,1,4. In every case the late, floor-width members sit at or near the guide and the two arrays' centres of pressure track one another through the stroke.

\paragraph{Rope passed through the contacts.} The searched orders are bounds on what placement can do for the moment, because the span order also sets how much rope passes over each contact. Rope enters each array at its outboard end and flows toward the dead-end anchor at the guide, so the rope sliding over member $i$ during the stroke is half its own take-up, $Y_i = 2(\sqrt{R_i^2 + D_i^2} - R_i)$, plus the take-up of every fold anchor-side of it, and each stationary support passes the take-up of every fold anchor-side of it. Weighting the rope passed by the wrap angle, $2\arctan(D_i/R_i)$ at a member and the sum of the two adjacent half-angles at a support, and integrating over the stroke gives a proxy for the friction work and wear at the contacts, in radian-metres, that the order-blind per-contact model of Section~\ref{si:friction} does not resolve. Table~\ref{tab:si-ropeflow} evaluates it for the two $k=7$ designs in three orders: the order of the drawings, engagement order from the anchor outward with the deepest fold at the dead end (main-text Section~\ref{sec:spanorder}); its reverse; and the searched order of Table~\ref{tab:si-moment}. The drawing order is the minimum over all orders of every configured array, checked exhaustively (\texttt{rope\_flow.py}, \texttt{rope\_flow.json}); the total moves by only 4 to 7\% between orders, but the rope passed over the deepest member doubles when it moves to the entry, and the searched orders of $\cfg{8{:}6}{7}$ do exactly that in both arrays.

\begin{table}[!htbp]
\centering
\caption{Rope passed through the contacts, weighted by wrap angle and integrated over the stroke (radian-metres), for three span orders of each array at 1,000:1: total over the array's members and supports, and the worst single contact.}
\label{tab:si-ropeflow}
\footnotesize
\begin{tabular}{@{}llcccccc@{}}
\toprule
& & \multicolumn{2}{c}{drawing order} & \multicolumn{2}{c}{reversed} & \multicolumn{2}{c}{searched (Table~\ref{tab:si-moment})} \\
\cmidrule(lr){3-4}\cmidrule(lr){5-6}\cmidrule(lr){7-8}
& & total & worst & total & worst & total & worst \\
\midrule
$\cfg{7}{7}$ & array 1 & 15.84 & 2.27 & 16.48 & 4.22 & 15.84 & 2.27 \\
             & array 2 & 9.10 & 1.46 & 9.58 & 2.92 & 9.16 & 1.87 \\
$\cfg{8{:}6}{7}$ & array 1 & 16.00 & 2.26 & 16.65 & 4.22 & 16.43 & 4.22 \\
                 & array 2 & 9.04 & 1.46 & 9.49 & 2.92 & 9.42 & 2.92 \\
\bottomrule
\end{tabular}
\end{table}

Two cautions apply. The estimate treats the bearings as Coulomb, and rolling-element guides at 10~m/s have velocity-dependent and lubricant-dependent losses that a single $\mu$ does not capture. And it is an estimate of the friction cost only; the more consequential penalty of a single-sided reaction at 228~kN may be the bearing rating and the structure needed to react a 360~kN\,m moment rather than the energy lost, and those are matters of embodiment outside this paper.

\FloatBarrier
\section{Selection of the source final speed}
\label{si:knee}

\begin{figure}[!htbp]
\centering
\includegraphics[width=\linewidth]{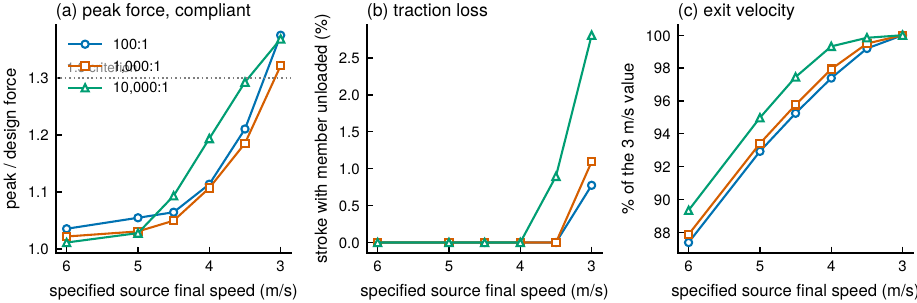}
\caption{Why the source is stopped at 4~m/s. Left: compliant peak payload force. Centre: fraction of the stroke during which the tension member carries no load. Right: exit velocity relative to the 3~m/s value. Harder deceleration is to the right in each panel.}
\label{fig:si-knee}
\end{figure}

The final speed of 4~m/s in the specification of Table~\ref{tab:spec} of the main text is a design input chosen at the knee of the trade in Table~\ref{tab:si-decel} and Figure~\ref{fig:si-knee}, which sweep the specified final speed with the member count and stage ratio of each reference design held fixed and the geometry refitted at each target with a reduced restart budget. Three things happen together as the deceleration is made more severe. Peak payload force rises monotonically, and by a final speed of 3.0~m/s all three cases have crossed the 1.3 bound. The tension member begins to unload during the stroke: at 3.0~m/s it carries no load for 0.8\%, 1.1\% and 2.8\% of the stroke at 100:1, 1,000:1 and 10,000:1, and the 10,000:1 case also shows slack at 3.5~m/s. And exit velocity stops reaching the ideal the target itself defines. Easing the deceleration in the other direction costs velocity and array width without buying much: going from 4 to 5~m/s at 1,000:1 lowers the ideal exit velocity from 317 to 303~m/s and more than doubles the array, from 2.4~m to 5.5~m, for 0.08 of peak force. The knee is a joint property of the target and the member count, since it is the discreteness of engagement that drives the unloading; an earlier sweep at seven members rather than nine for the 1,000:1 case showed slack at 3.5~m/s as well. What is robust across every configuration examined is that 4~m/s is clear of it while 3.0~m/s is not.

\begin{table}[!htbp]
\centering
\caption{Peak payload force (multiples of the design force, compliant model), fraction of the stroke during which the tension member is unloaded, and exit velocity relative to each case's ideal, against the specified source final speed.}
\label{tab:si-decel}
\begin{tabular}{lcccccc}
\toprule
final speed (m/s) & 3.0 & 3.5 & \textbf{4.0} & 4.5 & 5.0 & 6.0 \\
\midrule
peak force, 100:1    & 1.38 & 1.21 & \textbf{1.11} & 1.06 & 1.05 & 1.04 \\
peak force, 1,000:1  & 1.32 & 1.19 & \textbf{1.11} & 1.05 & 1.03 & 1.02 \\
peak force, 10,000:1 & 1.37 & 1.29 & \textbf{1.19} & 1.09 & 1.03 & 1.01 \\
slack fraction, 100:1    & 0.8\% & 0.0\% & \textbf{0.0\%} & 0.0\% & 0.0\% & 0.0\% \\
slack fraction, 1,000:1  & 1.1\% & 0.0\% & \textbf{0.0\%} & 0.0\% & 0.0\% & 0.0\% \\
slack fraction, 10,000:1 & 2.8\% & 0.9\% & \textbf{0.0\%} & 0.0\% & 0.0\% & 0.0\% \\
exit, \% of that case's ideal & 98--100 & 99--100 & \textbf{100} & 100 & 100 & 100 \\
\bottomrule
\end{tabular}
\end{table}

\FloatBarrier
\section{Fitting against search}
\label{si:search}

\begin{figure}[!htbp]
\centering
\includegraphics[width=0.62\linewidth]{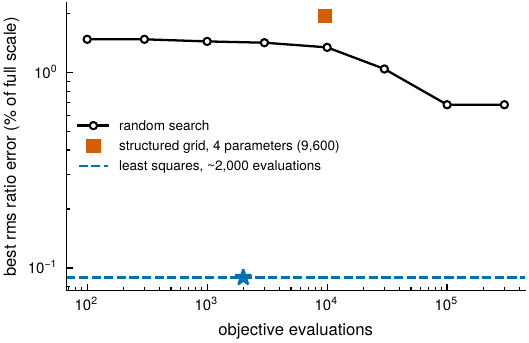}
\caption{Best residual against objective evaluations for a 1,000:1 target with eight members. Least squares reaches 0.18\% of full scale in roughly two thousand evaluations; random search remains above 1.3\% after three hundred thousand; a structured grid over four intuitive parameters is worse than random sampling, the optimum lying outside that family.}
\label{fig:si-search}
\end{figure}

Table~\ref{tab:si-search} and Figure~\ref{fig:si-search} compare least-squares fitting with direct search on a 1,000:1 target with eight members, sixteen free parameters and a full-scale ratio range of 49.1. Least squares improves on 300,000 random samples by a factor of 7.7 using 150 times fewer evaluations. The structured grid parametrises the array as a geometric progression of span widths and a power-law stagger, four parameters and 9,600 combinations, and performs worse than uniform random sampling because the optimum lies outside that family.

\begin{table}[!htbp]
\centering
\caption{Least-squares fitting against direct search.}
\label{tab:si-search}
\begin{tabular}{llll}
\toprule
method & evaluations & best rms & \% of full scale \\
\midrule
random search & 300 & 1.474 & 3.00\% \\
random search & 30,000 & 1.040 & 2.12\% \\
random search & 300,000 & 0.683 & 1.39\% \\
structured grid, 4 parameters & 9,600 & 1.937 & 3.94\% \\
least squares, 10 starts & $\sim$2,000 & \textbf{0.089} & \textbf{0.18\%} \\
\bottomrule
\end{tabular}
\end{table}

\FloatBarrier
\section{Additional figures}
\label{si:figs}

\begin{figure}[!htbp]
\centering
\includegraphics[width=\linewidth]{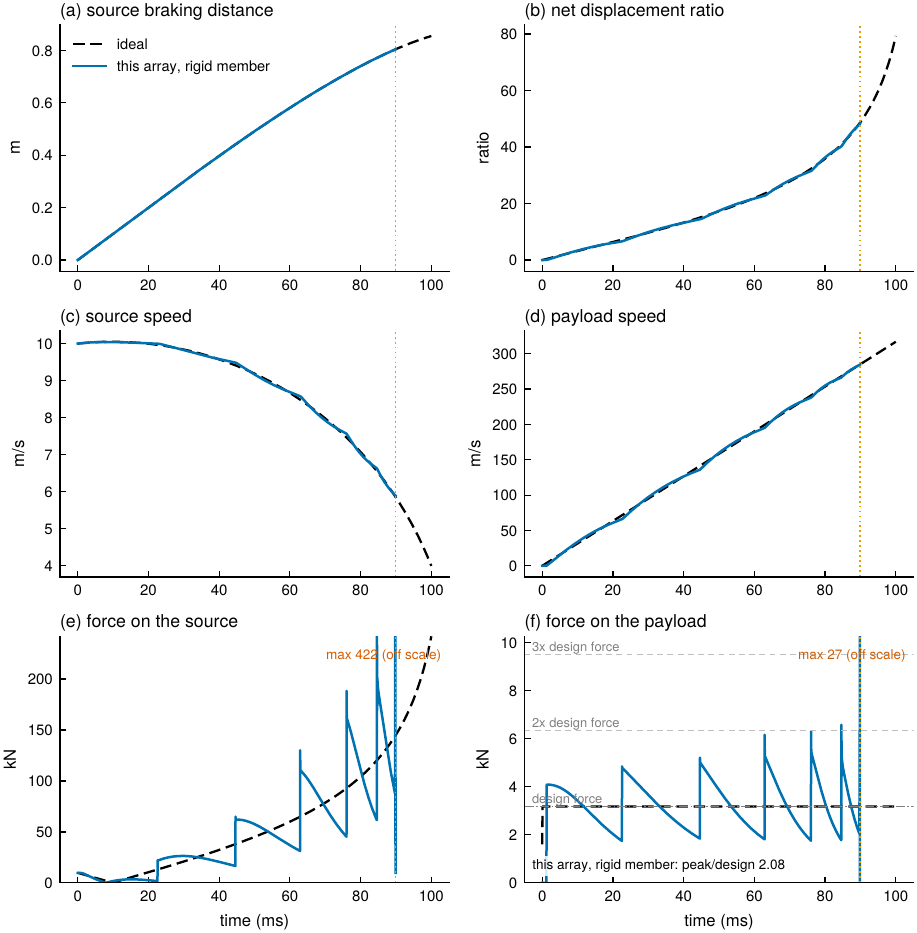}
\caption{Rigid-body dynamics of the 1,000:1 single-array reference. Dashed curves are the ideal run, a uniform 3,170~N on the payload, computed by the same integration loop with the ratio solved for at each step. Panel (f) shows one bounded sawtooth in payload force per engagement.}
\label{fig:si-rigid-single}
\end{figure}

\begin{figure}[!htbp]
\centering
\includegraphics[width=0.85\linewidth]{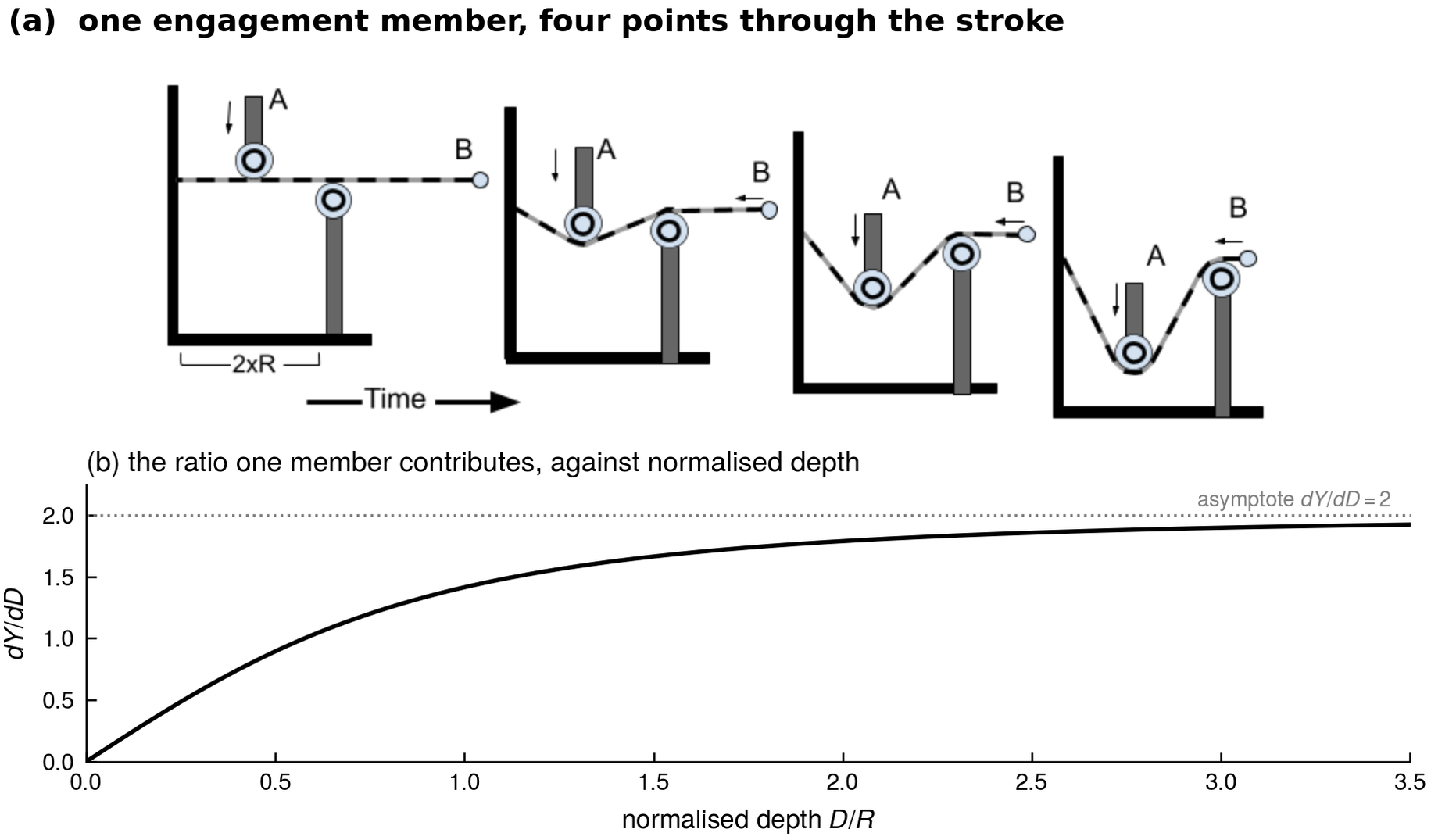}
\caption{A single engagement member deflecting the tension member into a span of half-width $R$, and the resulting displacement ratio $dY/dD$ against normalised depth $D/R$. The ratio rises from zero toward an asymptote of 2; a wider span rises more slowly.}
\label{fig:si-member}
\end{figure}

\begin{figure}[!htbp]
\centering
\includegraphics[width=\linewidth]{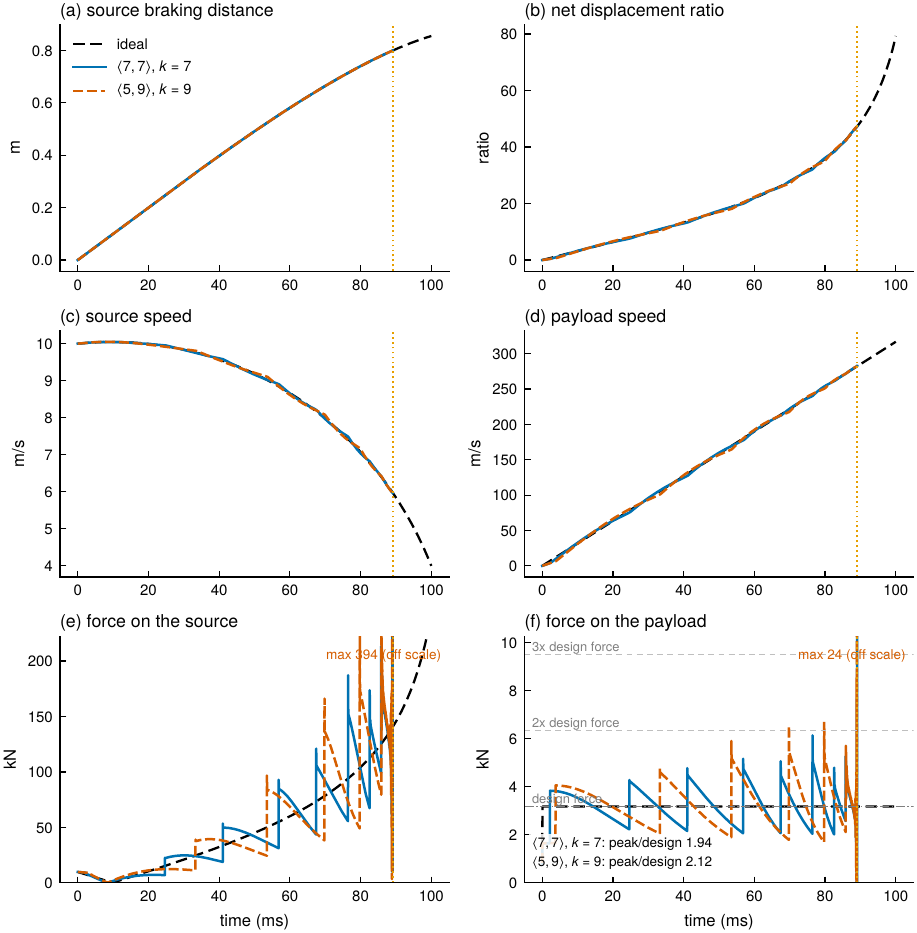}
\caption{Rigid-body dynamics for both equal-count dual designs against the ideal profile. Dashed black is the ideal, computed by the same integration loop with the ratio solved for at each step. Panel (f) shows one bounded sawtooth in payload force per engagement.}
\label{fig:si-runsrigid}
\end{figure}

\begin{figure}[!htbp]
\centering
\includegraphics[width=\linewidth]{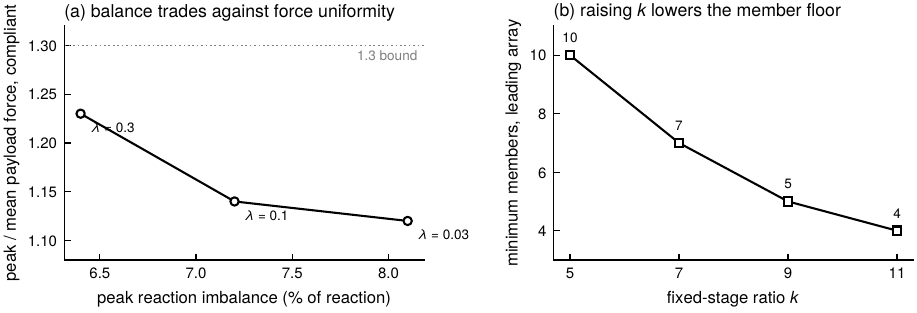}
\caption{(a) The trade between reaction-imbalance cancellation and force uniformity traced by $\lambda$ for the $\cfg{7}{7}$ design of Section~\ref{sec:duals} of the main text, over the sweep of Table~\ref{tab:lamsweep} there. (b) Minimum members on the leading array against stage ratio, the array-1 column of Table~\ref{tab:floors} of the main text.}
\label{fig:si-trade}
\end{figure}

Figure~\ref{fig:si-member} shows the single-member ratio of equation~\eqref{eq:member} of the main text; Figure~\ref{fig:si-rigid-single} the rigid dynamics of the 1,000:1 reference array, whose compliant dynamics are Figure~\ref{fig:single-compliant} of the main text; Figure~\ref{fig:si-runsrigid} the rigid dynamics of the two equal-count dual designs; and Figure~\ref{fig:si-trade} the accuracy--balance trade of the $\cfg{7}{7}$ design together with the member floor against stage ratio.

\FloatBarrier
\section{Verification of the scaling relations}
\label{si:scaling}

Table~\ref{tab:si-scaling} evaluates the endpoint relations of Section~\ref{sec:scaling} of the main text at $\nu = 0.4$ and $\gamma = 2g\dmax/v_0^2 = 0.167$ ($g = 9.8$~m/s$^2$, $\dmax = 0.854$~m, $v_0 = 10$~m/s) and sets them beside the values obtained by the simulator's root find and first-order integration. The endpoint relations follow from energy conservation alone and are exact for the continuous problem; the residual is the integrator's, which reaches the specified final speed to within 0.002~m/s but carries a one-step lag in the reaction force. Payload travels are compared against the values implied by the release-tolerance figures of Section~\ref{sec:instab} of the main text, where 5\% of the payload's run is 0.25, 0.79 and 2.52~m. The script \texttt{scaling\_check.py} reproduces the table and also integrates \eqref{eq:si-dimless} numerically from the frozen $(F, \dmax)$ of each case, returning stroke durations of 99.4, 99.8 and 100.0~ms against the specified 100~ms, and confirms the closed form of Section~\ref{sec:closed} at $\gamma = 0$ to machine precision.
\begin{equation}
\frac{d\hat y}{d\delta} = \Phi\,\sqrt{\frac{\hat y}{1 + \gamma\delta - \hat y}}
\label{eq:si-dimless}
\end{equation}

\begin{table}[!htbp]
\centering
\caption{Endpoint relations of the dimensionless target against the simulated values of Table~\ref{tab:spec} of the main text, at $\nu = 0.4$, $\gamma = 0.167$.}
\label{tab:si-scaling}
\footnotesize
\begin{tabular}{@{}lcccccc@{}}
\toprule
& \multicolumn{2}{c}{100:1} & \multicolumn{2}{c}{1,000:1} & \multicolumn{2}{c}{10,000:1} \\
\cmidrule(lr){2-3}\cmidrule(lr){4-5}\cmidrule(lr){6-7}
& analytic & simulated & analytic & simulated & analytic & simulated \\
\midrule
terminal ratio & 25.1 & 25.0 & 79.3 & 79.3 & 250.9 & 251.0 \\
exit velocity (m/s) & 100.4 & 99.9 & 317.4 & 317.0 & 1,003.7 & 1,003.5 \\
payload travel (m) & 4.9 & 5.0 & 15.8 & 15.8 & 50.1 & 50.4 \\
force number $\Phi$ & --- & 1.707 & --- & 1.712 & --- & 1.714 \\
\bottomrule
\end{tabular}
\end{table}

\FloatBarrier
\section{Per-contact friction model and full results}
\label{si:friction}

The integrators of Sections~\ref{si:rigid} and~\ref{si:compliant} are extended by a second function of carriage displacement, the reaction ratio $\rho(d)$, used in place of the kinematic ratio in the force on the source: the reaction becomes $\rho(d)\,F_{\mathrm{payload}}(t)$ while the rope tip still advances by $\Gnet(d)\,dd$. For a lossless machine $\rho = \Gnet$ and the loops reduce exactly to the originals, which was verified by reproducing every entry of \texttt{designs\_16471.json} (peak-to-mean rigid and compliant, exit velocity, reaction imbalance) and the matched single-array fits at $\eta_c = 1$.

With a per-contact efficiency $\eta_c$ the tension steps by $1/\eta_c$ across every sheave in the direction opposite to the flow of power. Along the output member, numbered from the payload end, the segments $s_0, s_1, \dots, s_p$ carry $F\eta_c^{-r}$; the sheaves $r = 1 \dots p$ alternate between the two counter-moving blocks starting with block~2, as drawn in Figure~\ref{fig:arch}(b) of the main text, and the dead end goes to the block that keeps $n_1 = \lfloor k/2 \rfloor$ and $n_2 = \lceil k/2 \rceil$. The pull on block $j$ is $\tau_j F$ with $\tau_j$ the sum of $\eta_c^{-r}$ over the segments touching it; for the one-line stage, a single block with $k$ falls, $\tau = \sum_{r=0}^{k-1}\eta_c^{-r}$. Along tension member $j$, from its block back to its anchor on the guide, the line passes the entry sheave and then, per engagement member, a fixed support and the member itself; members are numbered in engagement order with member~1 nearest the anchor, as in the design drawings, so the block-side segment at member $i$ sits $2(N-i)+2$ contacts from the block and its anchor-side segment one more:
\begin{equation}
T_i^{b} = \tau_j F\,\eta_c^{-(2(N-i)+2)}, \qquad T_i^{a} = T_i^{b}/\eta_c, \qquad
\rho_j(d) = \frac{1}{F}\sum_{i=1}^{N} \cos\theta_i(d)\,\bigl(T_i^{b} + T_i^{a}\bigr),
\label{eq:si-rho}
\end{equation}
with $\cos\theta_i = D_i/\sqrt{R_i^2 + D_i^2}$ the member's geometric term, so that $\rho_j = n_j G_j$ at $\eta_c = 1$. The reaction imbalance with friction is $\max_d |\rho_1 - \rho_2| / \max_d(\rho_1 + \rho_2)$ over the design stroke. The geometry is not refitted for the losses; a design intended for a lossy machine would carry $\rho_j$ in the objective of Algorithm~1 in place of $n_j G_j$. Table~\ref{tab:si-friction} gives the full sweep behind Table~\ref{tab:friction} of the main text. \texttt{friction\_sim.py} and \texttt{friction\_study.py} in \texttt{paper3/sim/} reproduce it.

\begin{table}[!htbp]
\centering
\caption{Per-contact friction on the frozen 1,000:1 designs. Energy retained is relative to the same design at $\eta_c = 1$; $v_h$ is the source velocity at release; $\tau_1/\tau_2$ is the ratio of the two block pulls, 0.75 at $k=7$ and 0.8 at $k=9$ without losses. Compliant model at 16,471~N/m, released at $\dmax$; rigid \ptm{} over the release window.}
\label{tab:si-friction}
\scriptsize
\begin{tabular}{@{}lcccccccc@{}}
\toprule
$\eta_c$ & energy & exit (m/s) & $v_h$ release & p2m compl. & p2m rigid & imbalance & reaction (kN) & $\tau_1/\tau_2$ \\
\midrule
\multicolumn{9}{l}{\emph{$\cfg{9}{5}$ one-line single}}\\
1 & 100.0\% & 318.2 & 3.94 & 1.099 & 2.07 & --- & 228 & 1.0000 \\
0.995 & 92.6\% & 306.1 & 3.79 & 1.097 & 2.00 & --- & 244 & 1.0000 \\
0.99 & 85.5\% & 294.1 & 3.66 & 1.095 & 1.91 & --- & 261 & 1.0000 \\
0.98 & 72.2\% & 270.4 & 3.40 & 1.089 & 1.78 & --- & 300 & 1.0000 \\
\multicolumn{9}{l}{\emph{$\cfg{9}{5}$ two-line single / mirror}}\\
1 & 100.0\% & 318.2 & 3.94 & 1.099 & 2.07 & 20.00\% & 228 & 0.6667 \\
0.995 & 92.9\% & 306.7 & 3.80 & 1.097 & 1.98 & 19.88\% & 243 & 0.6683 \\
0.99 & 86.1\% & 295.2 & 3.67 & 1.095 & 1.93 & 19.76\% & 259 & 0.6700 \\
0.98 & 73.4\% & 272.5 & 3.43 & 1.090 & 1.81 & 19.52\% & 295 & 0.6734 \\
\multicolumn{9}{l}{\emph{$\cfg{7}{7}$ one-line single}}\\
1 & 100.0\% & 321.0 & 3.72 & 1.402 & 2.44 & --- & 235 & 1.0000 \\
0.995 & 92.7\% & 309.0 & 3.70 & 1.373 & 2.09 & --- & 250 & 1.0000 \\
0.99 & 85.3\% & 296.5 & 3.75 & 1.344 & 2.07 & --- & 266 & 1.0000 \\
0.98 & 72.0\% & 272.4 & 3.79 & 1.311 & 2.05 & --- & 302 & 1.0000 \\
\multicolumn{9}{l}{\emph{$\cfg{7}{7}$ two-line single / mirror}}\\
1 & 100.0\% & 321.0 & 3.72 & 1.402 & 2.44 & 14.29\% & 235 & 0.7500 \\
0.995 & 93.3\% & 310.0 & 3.70 & 1.375 & 2.28 & 14.16\% & 248 & 0.7519 \\
0.99 & 86.5\% & 298.5 & 3.74 & 1.348 & 2.32 & 14.04\% & 262 & 0.7538 \\
0.98 & 73.9\% & 276.0 & 3.80 & 1.313 & 2.00 & 13.80\% & 294 & 0.7575 \\
\multicolumn{9}{l}{\emph{$\cfg{7}{7}$ unequal pair}}\\
1 & 100.0\% & 318.7 & 3.88 & 1.123 & 1.93 & 8.09\% & 236 & 0.7500 \\
0.995 & 93.9\% & 308.9 & 3.77 & 1.098 & 1.84 & 8.04\% & 249 & 0.7519 \\
0.99 & 88.0\% & 299.0 & 3.67 & 1.095 & 1.76 & 7.99\% & 264 & 0.7538 \\
0.98 & 76.6\% & 278.8 & 3.53 & 1.088 & 1.62 & 7.90\% & 296 & 0.7575 \\
\multicolumn{9}{l}{\emph{$\cfg{8{:}6}{7}$ count-balanced}}\\
1 & 100.0\% & 318.8 & 3.88 & 1.133 & 1.90 & 2.36\% & 235 & 0.7500 \\
0.995 & 93.9\% & 309.0 & 3.76 & 1.098 & 1.90 & 2.43\% & 248 & 0.7519 \\
0.99 & 87.9\% & 299.0 & 3.67 & 1.095 & 1.77 & 2.67\% & 262 & 0.7538 \\
0.98 & 76.5\% & 278.8 & 3.53 & 1.088 & 1.75 & 3.44\% & 294 & 0.7575 \\
\multicolumn{9}{l}{\emph{$\cfg{5}{9}$ one-line single}}\\
1 & 100.0\% & 328.1 & 3.08 & 2.191 & 2.88 & --- & 224 & 1.0000 \\
0.995 & 93.6\% & 317.4 & 3.09 & 2.103 & 2.74 & --- & 236 & 1.0000 \\
0.99 & 88.0\% & 307.7 & 3.02 & 2.086 & 2.70 & --- & 249 & 1.0000 \\
0.98 & 77.9\% & 289.6 & 2.74 & 2.105 & 2.18 & --- & 279 & 1.0000 \\
\multicolumn{9}{l}{\emph{$\cfg{5}{9}$ two-line single / mirror}}\\
1 & 100.0\% & 328.1 & 3.08 & 2.191 & 2.88 & 11.11\% & 224 & 0.8000 \\
0.995 & 94.4\% & 318.8 & 3.09 & 2.111 & 2.78 & 10.99\% & 234 & 0.8020 \\
0.99 & 89.5\% & 310.3 & 3.04 & 2.090 & 2.77 & 10.86\% & 245 & 0.8040 \\
0.98 & 80.6\% & 294.5 & 2.82 & 2.105 & 2.62 & 10.62\% & 269 & 0.8080 \\
\multicolumn{9}{l}{\emph{$\cfg{5}{9}$ unequal pair}}\\
1 & 100.0\% & 321.2 & 3.67 & 1.480 & 2.11 & 7.04\% & 227 & 0.8000 \\
0.995 & 95.6\% & 314.0 & 3.53 & 1.385 & 2.02 & 6.97\% & 238 & 0.8020 \\
0.99 & 90.8\% & 306.1 & 3.45 & 1.281 & 1.94 & 6.91\% & 249 & 0.8040 \\
0.98 & 80.6\% & 288.4 & 3.47 & 1.230 & 1.92 & 6.79\% & 274 & 0.8080 \\
\multicolumn{9}{l}{\emph{$\cfg{5{:}4}{9}$ count-balanced}}\\
1 & 100.0\% & 314.2 & 4.20 & 1.447 & 2.56 & 7.19\% & 217 & 0.8000 \\
0.995 & 96.4\% & 308.4 & 3.99 & 1.415 & 2.42 & 7.36\% & 226 & 0.8020 \\
0.99 & 93.0\% & 303.0 & 3.76 & 1.383 & 2.36 & 7.52\% & 236 & 0.8040 \\
0.98 & 86.7\% & 292.5 & 3.24 & 1.401 & 2.21 & 7.86\% & 258 & 0.8080 \\
\multicolumn{9}{l}{\emph{$\cfg{6{:}5}{9}$ comparison row}}\\
1 & 100.0\% & 321.6 & 3.62 & 1.484 & 1.89 & 3.93\% & 235 & 0.8000 \\
0.995 & 95.2\% & 313.7 & 3.48 & 1.366 & 1.79 & 4.10\% & 247 & 0.8020 \\
0.99 & 89.9\% & 305.0 & 3.41 & 1.304 & 1.77 & 4.28\% & 260 & 0.8040 \\
0.98 & 78.4\% & 284.8 & 3.53 & 1.245 & 1.71 & 4.65\% & 287 & 0.8080 \\
\bottomrule
\end{tabular}
\end{table}

\FloatBarrier
\section{The balance trade to its end, and a third reference design}
\label{si:frontier}

Table~\ref{tab:si-frontier} gives the full sweeps behind Table~\ref{tab:frontier} of the main text: seeded fits of Algorithm~1 (\texttt{fit\_semisym\_staged}, one solve per point from the single-array seed, no restarts) at each $\lambda$, simulated on both integrators. Because each point is a single solve the table is an upper bound on the attainable frontier; a restart budget would move points toward lower residual at the same $\lambda$. The $\lambda = 0.03$ point of the first family is the re-fit of the configured $\cfg{7}{7}$ design and differs from the frozen design of \texttt{designs\_16471.json} by the solver's path (imbalance 8.2\% against 8.1\%, rigid \ptm{} 2.07 against 1.93, compliant 1.122 against 1.123); the main text quotes the frozen values. \texttt{balanced\_study.py}, \texttt{balanced\_study2.py} and \texttt{balanced\_study3.py} reproduce the table.

\begin{table}[!htbp]
\centering
\caption{Balance sweeps at 1,000:1 and 16,471~N/m. Reaction imbalance is geometric over the design stroke; kN on the corresponding peak reaction; "terminal" is the net ratio reached at $\dmax$ as a fraction of the target 79.3; "min tension" is the minimum tension in the compliant trace over the first 99.5\% of the stroke, as a fraction of the design force, zero meaning the member went slack.}
\label{tab:si-frontier}
\scriptsize
\begin{tabular}{@{}lcccccccc@{}}
\toprule
$\lambda$ & imbalance & (kN) & rms & terminal & \ptm{} rigid & \ptm{} compliant & min tension & exit (m/s) \\
\midrule
\multicolumn{9}{l}{\emph{$\cfg{7}{7}$, width budget 2.11 m per array (the configured design's budget)}}\\
0.03 & 8.2\% & 19.4 & 0.378 & 94\% & 2.07 & 1.12 & 0.90 & 318.7 \\
0.3 & 6.4\% & 14.9 & 0.431 & 93\% & 2.08 & 1.23 & 0.80 & 319.9 \\
1 & 5.0\% & 11.5 & 0.558 & 92\% & 2.13 & 1.59 & 0.48 & 323.3 \\
3 & 3.0\% & 6.8 & 0.796 & 91\% & 1.91 & 1.94 & 0.14 & 326.2 \\
10 & 1.2\% & 2.6 & 0.948 & 90\% & 2.14 & 1.97 & 0.05 & 324.8 \\
30 & 0.9\% & 2.0 & 1.108 & 87\% & 2.53 & 1.90 & 0.00 & 320.3 \\
\multicolumn{9}{l}{\emph{$\cfg{5}{9}$, width budget 1.96 m per array}}\\
0.03 & 7.0\% & 16.0 & 0.640 & 91\% & 2.11 & 1.48 & 0.66 & 321.2 \\
0.3 & 6.0\% & 13.5 & 0.699 & 90\% & 2.00 & 1.82 & 0.38 & 324.6 \\
1 & 4.8\% & 10.8 & 0.873 & 89\% & 1.85 & 2.28 & 0.05 & 328.1 \\
3 & 3.4\% & 7.5 & 1.085 & 89\% & 2.29 & 2.47 & 0.00 & 316.6 \\
10 & 2.1\% & 4.6 & 1.181 & 88\% & 2.59 & 2.49 & 0.00 & 308.0 \\
30 & 1.4\% & 3.0 & 1.292 & 87\% & 2.54 & 2.38 & 0.00 & 295.8 \\
\multicolumn{9}{l}{\emph{$\cfg{9}{7}$, width budget 2.11 m per array}}\\
0.03 & 8.7\% & 21.6 & 0.364 & 98\% & 2.00 & 1.31 & 0.72 & 314.0 \\
1 & 2.3\% & 5.8 & 0.371 & 98\% & 2.21 & 1.72 & 0.38 & 324.3 \\
3 & 1.4\% & 3.4 & 0.707 & 97\% & 2.79 & 2.14 & 0.00 & 327.9 \\
10 & 0.8\% & 2.0 & 0.861 & 97\% & 3.42 & 2.21 & 0.00 & 328.2 \\
30 & 1.0\% & 2.3 & 0.980 & 96\% & 2.88 & 1.97 & 0.00 & 325.0 \\
\multicolumn{9}{l}{\emph{$\cfg{7}{7}$, width budget 2.7 m per array}}\\
0.03 & 8.1\% & 19.1 & 0.382 & 94\% & 1.90 & 1.11 & 0.91 & 318.5 \\
1 & 4.3\% & 9.9 & 0.569 & 92\% & 1.61 & 1.26 & 0.69 & 320.0 \\
3 & 2.9\% & 6.7 & 0.720 & 91\% & 1.93 & 1.37 & 0.53 & 321.0 \\
10 & 1.3\% & 2.9 & 0.823 & 90\% & 2.07 & 1.39 & 0.57 & 320.6 \\
\multicolumn{9}{l}{\emph{$\cfg{9}{7}$, width budget 2.7 m per array}}\\
0.03 & 4.3\% & 10.6 & 0.134 & 99\% & 1.74 & 1.07 & 0.93 & 317.9 \\
1 & 2.0\% & 5.0 & 0.267 & 98\% & 1.88 & 1.35 & 0.70 & 321.0 \\
3 & 1.1\% & 2.8 & 0.458 & 97\% & 2.31 & 1.47 & 0.56 & 321.5 \\
10 & 0.6\% & 1.4 & 0.553 & 97\% & 2.39 & 1.52 & 0.49 & 319.2 \\
\multicolumn{9}{l}{\emph{$\cfg{7}{7}$, width budget 3.3 m per array}}\\
0.03 & 8.1\% & 19.1 & 0.382 & 94\% & 1.78 & 1.11 & 0.91 & 318.6 \\
1 & 3.9\% & 9.0 & 0.626 & 91\% & 1.95 & 1.19 & 0.76 & 318.6 \\
3 & 2.9\% & 6.6 & 0.689 & 91\% & 1.98 & 1.22 & 0.72 & 318.4 \\
10 & 1.5\% & 3.4 & 0.778 & 89\% & 1.94 & 1.23 & 0.67 & 317.9 \\
\multicolumn{9}{l}{\emph{$\cfg{9}{7}$, width budget 3.3 m per array}}\\
0.03 & 3.5\% & 8.7 & 0.115 & 99\% & 1.47 & 1.03 & 0.97 & 317.5 \\
1 & 2.0\% & 4.9 & 0.205 & 98\% & 1.55 & 1.17 & 0.83 & 318.9 \\
3 & 1.2\% & 3.0 & 0.365 & 97\% & 1.96 & 1.23 & 0.76 & 318.4 \\
10 & 0.5\% & 1.2 & 0.439 & 97\% & 1.77 & 1.29 & 0.68 & 319.1 \\
\bottomrule
\end{tabular}
\end{table}

The pattern is the same in every family: $\lambda$ buys balance and costs force uniformity, and what sets how much force uniformity is available to spend is the width budget. At 2.11~m per array the $\cfg{7}{7}$ family crosses the compliant bound of 1.3 between $\lambda = 0.3$ and $1$; the $\cfg{9}{7}$ family at the same budget is worse at every $\lambda$, its spans having narrowed; at 2.7~m nine members reach 0.56\% at 1.52; at 3.3~m nine members reach 0.49\% at 1.29 and seven members reach 1.5\% at 1.23. Table~\ref{tab:si-bf} gives the near-balanced designs under path-loss asymmetry, and Table~\ref{tab:si-k97} the coordinates of the $\cfg{9}{7}$, 3.3~m, $\lambda = 10$ design, which the main text carries as a third reference design: rms 0.439, terminal ratio 97.3\% of target, array widths 1.99 and 3.30~m, reaction imbalance 0.49\% (1.2~kN), compliant \ptm{} 1.29, rigid 1.77, exit 319.1~m/s. The first members of the two arrays engage together at 26~mm and the last member of array~2 engages at $\dmax$ itself, so it contributes nothing to the stroke and could be omitted.

\paragraph{Why the compliant figure rises.} \texttt{balance\_diagnosis.py} evaluates every design of the sweeps above and of the two objective variants below (56 designs) for the minimum tension in its compliant trace, the fraction of the stroke spent slack, and the minimum over 15 to 97\% of the stroke of the fitted profile's slope relative to the target's, both averaged over 3~cm windows. The compliant \ptm{} force correlates with the minimum tension at $-0.88$ and with the minimum slope ratio at $-0.66$. The eighteen admissible designs (\ptm{} $\le 1.3$) have minimum tensions of 0.58 to 0.97 and slope ratios of 0.53 to 0.81; every design whose tension falls below half the design force has a \ptm{} above 1.39, every design above 1.5 has a minimum tension below 0.49, and every design that goes slack has a \ptm{} of 1.90 or more. As a control, the ideal profile itself perturbed by smooth 1\% errors (uniform, half-sine, full-sine, late ramp) gives compliant \ptm{} figures of 1.02 to 1.07, and a 2\% full-sine error 1.13, so the residuals of Table~\ref{tab:si-frontier} are harmless as long as they are smooth; the peaks are produced by the local flats. Two reweightings of the objective were tried to test whether the flats are an artefact of the value residual of Algorithm~1 (\texttt{balanced\_relative.py}, \texttt{balanced\_slope.py}). A relative residual $(n_1 G_1 + n_2 G_2 - \Gstar)/(\Gstar + 0.1\max\Gstar)$ removes the early-stroke under-tracking of the large-$\lambda$ fits (net ratio over target at 10\% of stroke 1.03 to 1.05 against 0.55 at $\lambda = 10$) but leaves the frontier where it was: for $\cfg{7}{7}$ at 2.11~m, imbalance 8.6, 7.3, 6.2, 3.9, 2.9 and 1.3\% at $\lambda = 0.03$ to 30 with compliant \ptm{} 1.07, 1.20, 1.54, 2.05, 2.07 and 3.47 and minimum tensions 0.92, 0.72, 0.23 and then zero; at 2.7~m the same family gives 1.05 to 1.72 with minimum tensions 0.93 to 0.18. Adding a slope residual $\sqrt{\beta}\,(G_{\mathrm{net}}' - \Gstar{}')/\max\Gstar{}'$ with weight $\beta$ is worse on every axis, because the target's slope is dominated by its near-singular terminal rise and the term pulls the fit toward it: at $\beta = 1$ the 2.11~m family reaches only 4.7\% reaction imbalance at $\lambda = 10$ with \ptm{} 1.72, and at $\beta = 10$ only 7.0\% at 2.71. The frontier is therefore a property of the geometry at the given width budget rather than of the weighting.

\begin{table}[!htbp]
\centering
\caption{Reaction imbalance of the near-balanced designs and of the configured $\cfg{7}{7}$ design under per-contact friction, from \eqref{eq:si-rho}.}
\label{tab:si-bf}
\begin{tabular}{lccc}
\toprule
design & $\eta_c = 1$ & $\eta_c = 0.99$ & $\eta_c = 0.98$ \\
\midrule
$\cfg{9}{7}$, 3.3 m, $\lambda=10$ & 0.49\% (1.2 kN) & 0.64\% (1.8 kN) & 0.78\% (2.6 kN) \\
$\cfg{9}{7}$, 3.3 m, $\lambda=3$ & 1.23\% (3.0 kN) & 1.64\% (4.6 kN) & 2.02\% (6.5 kN) \\
$\cfg{7}{7}$, 2.11 m, $\lambda=0.03$ (configured) & 8.09\% (19.1 kN) & 7.99\% (21.1 kN) & 7.90\% (23.4 kN) \\
\bottomrule
\end{tabular}
\end{table}

\begin{table}[!htbp]
\centering
\caption{Coordinates (m) of the near-balanced $\cfg{9}{7}$ design, 3.3~m width budget, $\lambda = 10$, from \texttt{balanced\_study3.json}. Array 1 leads and serves 3 falls, array 2 serves 4.}
\label{tab:si-k97}
\begin{tabular}{lcccc}
\toprule
member & $R_{1,i}$ & $s_{1,i}$ & $R_{2,i}$ & $s_{2,i}$ \\
\midrule
1 & 0.2566 & 0.0262 & 0.3835 & 0.0262 \\
2 & 0.2426 & 0.3174 & 0.4753 & 0.3287 \\
3 & 0.1545 & 0.5542 & 0.3045 & 0.5613 \\
4 & 0.0890 & 0.6933 & 0.1740 & 0.6985 \\
5 & 0.0518 & 0.7682 & 0.0959 & 0.7721 \\
6 & 0.0500 & 0.8060 & 0.0545 & 0.8133 \\
7 & 0.0500 & 0.8207 & 0.0592 & 0.8337 \\
8 & 0.0500 & 0.8337 & 0.0548 & 0.8394 \\
9 & 0.0500 & 0.8394 & 0.0500 & 0.8542 \\
\bottomrule
\end{tabular}
\end{table}

\FloatBarrier
\section{Count-balanced designs}
\label{si:counts}

This section gives the two designs of main-text Section~\ref{sec:counts}, whose member counts satisfy
\begin{equation}
n_1 N_1 = n_2 N_2 ,
\label{eq:counts-si}
\end{equation}
with their coordinates, balance-weight ladders and the solver checks referred to there. All numbers are from \texttt{designs\_counts\_16471.json}, produced by \texttt{freeze\_counts.py} on the setup of \texttt{designs\_16471.json}: 1,000:1, 16,471~N/m, width budgets 2.11~m at $k=7$ and 1.96~m at $k=9$, span floor 0.05~m, width penalty weight 50. The frozen equal-count designs re-evaluated through the same functions reproduce their published figures exactly (\ptm{} 1.123 and 1.480 compliant, 1.930 and 2.106 rigid, rms 0.380 and 0.640, reaction imbalance 8.09 and 7.04\%).

\paragraph{Seeding and restarts.} Each design is seeded by one array of $N_1 + N_2$ members fitted at the ratio $k/2$ (width cap three times the budget) and dealt in engagement order alternately to the two arrays, the leading array taking the first member and the excess. Because the single-array fit converges to one design from any start (thirty seeds spread its coordinates by 3~mm), restarting it supplies no diversity; the restarts are instead thirty perturbations of the dealt seed, with log-normal factors of standard deviation 0.3 on every span width and on every gap between successive engagements, each followed by the staged solve, and the best is kept by the joint objective. The plain seed ranked 17th and 18th of the 31 starts for $\cfg{8{:}6}{7}$ at $\lambda = 0.03$ and $0.02$ and 19th and 3rd for $\cfg{5{:}4}{9}$ at $\lambda = 0.03$ and $0.02$, in objective 1.1, 35, 2.4 and 0.0\% above the best; 15 to 20 of the 31 starts finish within 1\% of the best objective. Two other seeds corroborate the optimum: an $N_1$-member single array at $k$ with the second array on its midpoints, and thirty random starts of the joint objective, land within 0.6\% of it in every case, once 0.2\% below it. The seed as dealt, before any optimisation, already tracks the target to rms 0.634 at $\cfg{8{:}6}{7}$ (peak reaction imbalance 6.6\%, terminal ratio 92.9\%) and to rms 1.069 at $\cfg{5{:}4}{9}$ (8.3\%, 85.9\%).

\paragraph{Engagement order.} The parametrisation of Algorithm~1 deals the merged offset sequence alternately, which makes the order a constraint. Released from it, with each array's offsets built from its own positive increments so that members may cross, and started from the constrained optimum, the fit does not move: the objective changes by less than 0.02\% and no coordinate by more than 0.7~mm, the only visible effect being that tied final engagements exchange places. The perturbed starts of the free solve reach basins with two consecutive engagements of one array, and every one is worse: at $\cfg{5{:}4}{9}$ and $\lambda = 0.02$, LRLRLRR[LL] at $+1.7\%$ in objective and 6.9\% reaction imbalance with 1.5\% at the end, LRLRLLRRL at $+5.5\%$ and 11.4\%; at $\lambda = 0.01$, LRLRLRRLL at $+1.0\%$ and 8.1\% and LRLRLLRRL at $+3.1\%$ and 11.8\%, against 8.25\% for the alternating optimum (\texttt{unequal\_free\_order.py}). Where the members the leading array has in excess are dealt also matters, Table~\ref{tab:si-placement}: dealt first they raise the peak reaction imbalance to 14\%, dealt in the middle to 4 to 11\%, so they are dealt last, where the terminal balance of \eqref{eq:counts-si} needs them.

\begin{table}[!htbp]
\centering
\caption{Placement of the members the leading array has in excess, best of 31 starts at each placement. Peak reaction imbalance, its value at the end of the stroke, and the joint objective.}
\label{tab:si-placement}
\begin{tabular}{llccc}
\toprule
design & excess dealt & peak reaction imbalance & at the end & objective \\
\midrule
$\cfg{8{:}6}{7}$, $\lambda = 0.03$ & last & \textbf{2.3\%} & 0.2\% & \textbf{0.01202} \\
 & first & 14.4\% & 0.8\% & 0.12106 \\
 & middle & 4.1\% & 1.7\% & 0.01381 \\
$\cfg{8{:}6}{7}$, $\lambda = 0.02$ & last & \textbf{2.7\%} & 0.4\% & \textbf{0.01162} \\
 & first & 14.4\% & 1.3\% & 0.09081 \\
 & middle & 4.8\% & 2.3\% & 0.01290 \\
$\cfg{5{:}4}{9}$, $\lambda = 0.02$ & last & \textbf{7.2\%} & 0.4\% & \textbf{0.06796} \\
 & first & 13.9\% & 0.2\% & 0.11467 \\
 & middle & 11.3\% & 2.6\% & 0.07185 \\
$\cfg{5{:}4}{9}$, $\lambda = 0.03$ & last & \textbf{5.5\%} & 0.1\% & \textbf{0.07003} \\
 & first & 13.2\% & 0.2\% & 0.13127 \\
 & middle & 10.7\% & 1.7\% & 0.07510 \\
\bottomrule
\end{tabular}
\end{table}

\paragraph{Balance-weight ladders.} Table~\ref{tab:si-ladder} traces both designs over $\lambda$. At $k=7$ the ladder is flat: the count has removed the terminal residual, the peak reaction imbalance falls from 4.1 to 2.0\% across a fivefold change in $\lambda$, and the force profile barely moves. At $k=9$ the ladder crosses a basin boundary between $\lambda = 0.02$ and $0.03$: the narrow design (arrays of 1.0 to 1.1~m, the trailing array's first member engaging at 0.18 to 0.21~m, imbalance peak of 7 to 8\% there) gives way to a wide one (1.6~m, first trailing engagement at 0.06~m, imbalance peak 5\% at mid-stroke) whose 0.88~m leading span rings the payload mode, compliant \ptm{} 1.78 against 1.45. The two basins are different machines rather than two tunings of one, which is why the main text reports the $\lambda = 0.02$ design and mentions the other.

\begin{table}[!htbp]
\centering
\caption{Balance-weight ladders of the count-balanced designs, best of 31 starts at each $\lambda$, at 16,471~N/m.}
\label{tab:si-ladder}
\footnotesize
\begin{adjustbox}{max width=\linewidth}
\begin{tabular}{lccccccccc}
\toprule
$\lambda$ & imbalance peak & mean & end & rms & terminal & \ptm{} rigid & \ptm{} compl. & exit (m/s) & widths (m) \\
\midrule
\multicolumn{10}{l}{\emph{$\cfg{8{:}6}{7}$}}\\
0.01 & 4.1\% & 1.0\% & 1.3\% & 0.406 & 93\% & 1.83 & 1.10 & 318.4 & 2.10 / 1.87 \\
0.02 & 2.7\% & 0.9\% & 0.4\% & 0.411 & 93\% & 1.85 & 1.11 & 318.6 & 2.11 / 1.87 \\
0.03 & 2.3\% & 0.9\% & 0.2\% & 0.413 & 93\% & 1.90 & 1.13 & 318.8 & 2.05 / 1.83 \\
0.05 & 2.0\% & 0.8\% & 0.2\% & 0.419 & 93\% & 2.01 & 1.13 & 318.9 & 2.10 / 1.89 \\
\multicolumn{10}{l}{\emph{$\cfg{5{:}4}{9}$}}\\
0.01 & 8.3\% & 3.6\% & 0.5\% & 0.949 & 86\% & 2.56 & 1.26 & 315.6 & 1.02 / 0.94 \\
0.02 & 7.2\% & 3.2\% & 0.4\% & 0.965 & 86\% & 2.56 & 1.45 & 314.2 & 1.13 / 0.98 \\
0.03 & 5.5\% & 1.6\% & 0.1\% & 1.019 & 86\% & 2.57 & 1.78 & 321.4 & 1.60 / 1.22 \\
0.05 & 4.9\% & 1.6\% & 0.2\% & 1.024 & 85\% & 2.48 & 1.80 & 321.8 & 1.68 / 1.25 \\
\bottomrule
\end{tabular}
\end{adjustbox}
\end{table}

\paragraph{The comparison row.} $\cfg{6{:}5}{9}$ at $\lambda = 0.03$ is the nearest count above the floors of 5 and 4. It misses the count condition by one part in 49, $4 \times 6$ against $5 \times 5$, a saturation residual of 2.0\%, and gives a peak reaction imbalance of 3.9\% (9.2~kN, 96.1\% of the single array's one-sided reaction cancelled) with 1.2\% at the end of the stroke, where the near-miss of the count shows; its compliant \ptm{} is 1.48 at 321.6~m/s, one member more than $\cfg{5}{9}$ for a peak 44\% lower, and it retains 78.4\% of its payload energy at 2\% loss per contact (Table~\ref{tab:si-friction}). Its coordinates are Table~\ref{tab:si-k9d}.

\begin{figure}[!htbp]
\centering
\includegraphics[width=\linewidth,height=0.86\textheight,keepaspectratio]{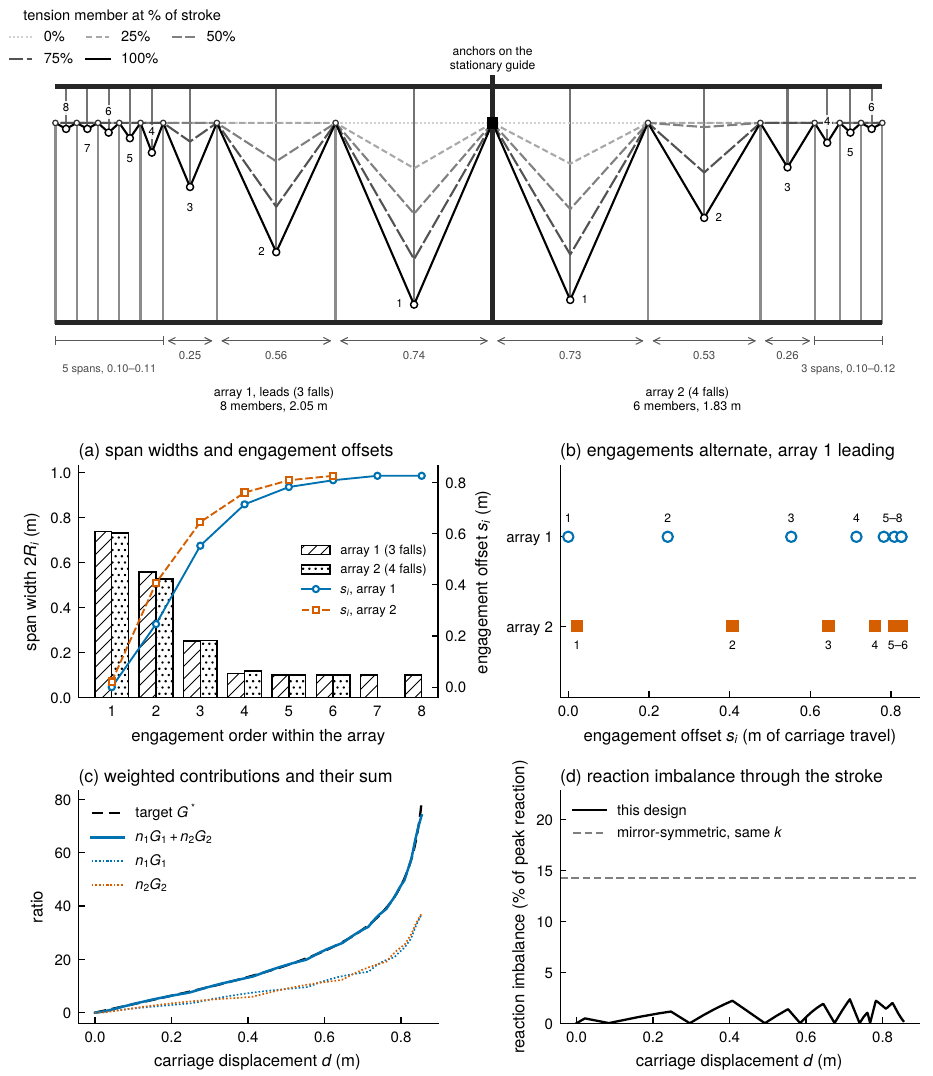}
\caption{The $\cfg{8{:}6}{7}$ design, $\lambda = 0.03$, panels as main-text Figure~\ref{fig:k7}.}
\label{fig:si-k7c}
\end{figure}

\begin{figure}[!htbp]
\centering
\includegraphics[width=\linewidth,height=0.86\textheight,keepaspectratio]{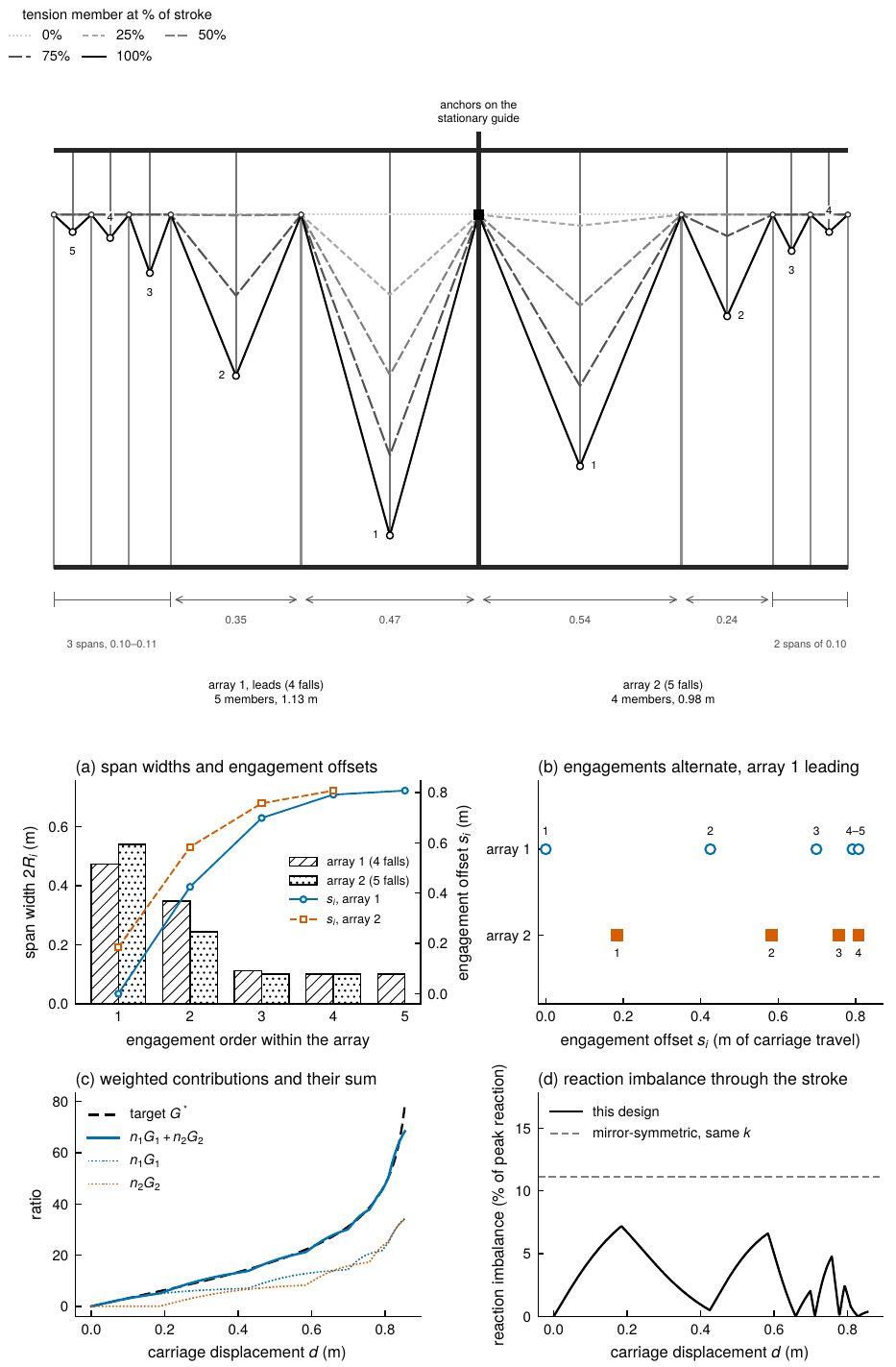}
\caption{The $\cfg{5{:}4}{9}$ design, $\lambda = 0.02$, panels as main-text Figure~\ref{fig:k7}.}
\label{fig:si-k9c}
\end{figure}

\paragraph{The designs.} Figures~\ref{fig:si-k7c} and~\ref{fig:si-k9c} draw the two designs in the layout of main-text Figures~\ref{fig:k7} and~\ref{fig:k9}, and Tables~\ref{tab:si-k7c} to~\ref{tab:si-k9d} give their coordinates and those of the comparison row $\cfg{6{:}5}{9}$. In $\cfg{8{:}6}{7}$ the last two members of the leading array coincide, at the span floor and 28~mm before the end of the stroke, with the last member of the trailing array beside them: the three engage together, and the excess member is the one whose transfer from the trailing array balances the end of the stroke (Section~\ref{sec:counts}).

\begin{table}[!htbp]\centering\caption{Full-precision coordinates (m) of the $\cfg{8{:}6}{7}$ design, $\lambda=0.03$, from \texttt{designs\_counts\_16471.json}. Array 1 leads and serves 3 falls, array 2 serves 4. Fitted rms 0.4127; reaction imbalance 2.31\%; terminal ratio 74.05 (93.4\% of target); array widths 2.055 and 1.833 m.}\label{tab:si-k7c}
\begin{tabular}{lcccc}\toprule
member & $R_{1,i}$ & $s_{1,i}$ & $R_{2,i}$ & $s_{2,i}$ \\ \midrule
1 & 0.3688 & 0.0000 & 0.3655 & 0.0218 \\
2 & 0.2792 & 0.2465 & 0.2642 & 0.4074 \\
3 & 0.1260 & 0.5528 & 0.1275 & 0.6450 \\
4 & 0.0534 & 0.7145 & 0.0591 & 0.7606 \\
5 & 0.0500 & 0.7823 & 0.0500 & 0.8087 \\
6 & 0.0500 & 0.8087 & 0.0500 & 0.8261 \\
7 & 0.0500 & 0.8261 & --- & --- \\
8 & 0.0500 & 0.8261 & --- & --- \\
\bottomrule\end{tabular}\end{table}
\begin{table}[!htbp]\centering\caption{Full-precision coordinates (m) of the $\cfg{5{:}4}{9}$ design, $\lambda=0.02$, from \texttt{designs\_counts\_16471.json}. Array 1 leads and serves 4 falls, array 2 serves 5. Fitted rms 0.9647; reaction imbalance 7.19\%; terminal ratio 68.40 (86.3\% of target); array widths 1.131 and 0.984 m.}\label{tab:si-k9c}
\begin{tabular}{lcccc}\toprule
member & $R_{1,i}$ & $s_{1,i}$ & $R_{2,i}$ & $s_{2,i}$ \\ \midrule
1 & 0.2363 & 0.0000 & 0.2701 & 0.1843 \\
2 & 0.1736 & 0.4249 & 0.1217 & 0.5835 \\
3 & 0.0558 & 0.6988 & 0.0500 & 0.7574 \\
4 & 0.0500 & 0.7920 & 0.0500 & 0.8079 \\
5 & 0.0500 & 0.8079 & --- & --- \\
\bottomrule\end{tabular}\end{table}
\begin{table}[!htbp]\centering\caption{Full-precision coordinates (m) of the comparison row $\cfg{6{:}5}{9}$, $\lambda=0.03$, from \texttt{designs\_counts\_16471.json}. Array 1 leads and serves 4 falls, array 2 serves 5. Fitted rms 0.4463; reaction imbalance 3.91\%; terminal ratio 74.21 (93.6\% of target); array widths 1.960 and 1.681 m.}\label{tab:si-k9d}
\begin{tabular}{lcccc}\toprule
member & $R_{1,i}$ & $s_{1,i}$ & $R_{2,i}$ & $s_{2,i}$ \\ \midrule
1 & 0.4467 & 0.0000 & 0.4332 & 0.0431 \\
2 & 0.2816 & 0.3354 & 0.2222 & 0.5111 \\
3 & 0.1019 & 0.6500 & 0.0852 & 0.7252 \\
4 & 0.0500 & 0.7735 & 0.0500 & 0.8022 \\
5 & 0.0500 & 0.8172 & 0.0500 & 0.8269 \\
6 & 0.0500 & 0.8269 & --- & --- \\
\bottomrule\end{tabular}\end{table}

\FloatBarrier
\section{Operation in reverse: arrest}
\label{si:arrest}

The transmission of the main text has a rising ratio when the carriage deflects the tension members into the spans. Run in reverse, with the carriage held low on stops and the folds deep, it has a falling ratio: a body that engages the output member draws the tension members out of the spans and lifts the carriage against a resisting element, and the ratio between the body's travel and the carriage lift falls as the folds straighten. This section configures one such arrest and simulates it. Everything in it is separate from the launch machinery of the main text: the fit is a direct least-squares in a relative residual (not Algorithm~1), and the integrator carries a damped output member (not the undamped model of Section~\ref{si:compliant}). The section shows what the geometry does in this mode; it is not a validation at the level of the launch results.

\paragraph{Arrangement and kinematics.} A braking member of half-span $R_b$, anchored at one end and continuing at the other as the output member, lies across the path of the body. The carriage is pre-lowered so that the net ratio is at its greatest, and a hold-down mass $M$ resists its lift through a tackle of ratio $j$, so that the carriage sees an effective mass $M_e = j^2 M$ and an effective weight $jMg$. The body engages the braking member at mid-span and deflects it by $x$: the braking member is itself a first fold, drawing in $Y_b(x) = 2(\sqrt{R_b^2 + x^2} - R_b)$ of the output member at the ratio $g_b(x) = 2x/\sqrt{R_b^2 + x^2}$, which rises from zero at engagement toward 2, the form of main-text equation~\eqref{eq:member}. The output member drawn in pulls the blocks of the stage together, the blocks draw the tension members out of the spans, and the carriage rises by $u$. With $t = S - u$ the remaining depth, $S$ the working lift, each member is deflected by $\max(0, t - s_i)$ and the net ratio of the arrays is $C(u) = n_1 G_1(S - u) + n_2 G_2(S - u)$. While the output member is taut, $g_b\,dx = C\,du$, the force on the body is $F = T g_b$ and the carriage reaction $T C$, so by power balance $F = F_{\mathrm{carriage}}\, g_b / C$. Because $g_b$ starts at zero, the carriage speed per unit body speed, $g_b/C$, starts at zero: the hold-down mass is picked up from rest without a step in its speed, whatever the comb's initial ratio, the mirror of the snatch-free engagement of the launch.

\paragraph{Target.} For a lossless, taut system the body's speed and the carriage lift obey
\begin{equation}
v(x) = \sqrt{v_0^2 - 2E(x)/m}, \qquad \frac{du}{dx} = \frac{1}{v(x)}\sqrt{\frac{2\,[E(x) - jMg\,u]}{M_e}}, \qquad C^*(x) = \frac{g_b(x)}{du/dx},
\label{eq:si-arrest}
\end{equation}
with $E(x) = \int_0^x F\,dx'$ the energy removed from the body, which is main-text equation~\eqref{eq:ode} with the two sides exchanged: the element accelerated from rest, here the carriage, starts at zero speed. A uniform force from the first instant would need unbounded tension while $g_b$ is still near zero, so the target has two phases: a ramp, in which the comb is held at its greatest ratio $C_0$ and the force rises with $g_b$ as the mass is picked up, and a uniform phase at the force $F_u$ that brings the body to its specified exit speed at the specified distance, in which \eqref{eq:si-arrest} gives $C^*$. $C_0$ is set near the greatest value the uniform phase would demand, so that $C^*$ falls monotonically after the ramp.

\begin{figure}[!htbp]
\centering
\includegraphics[width=\linewidth]{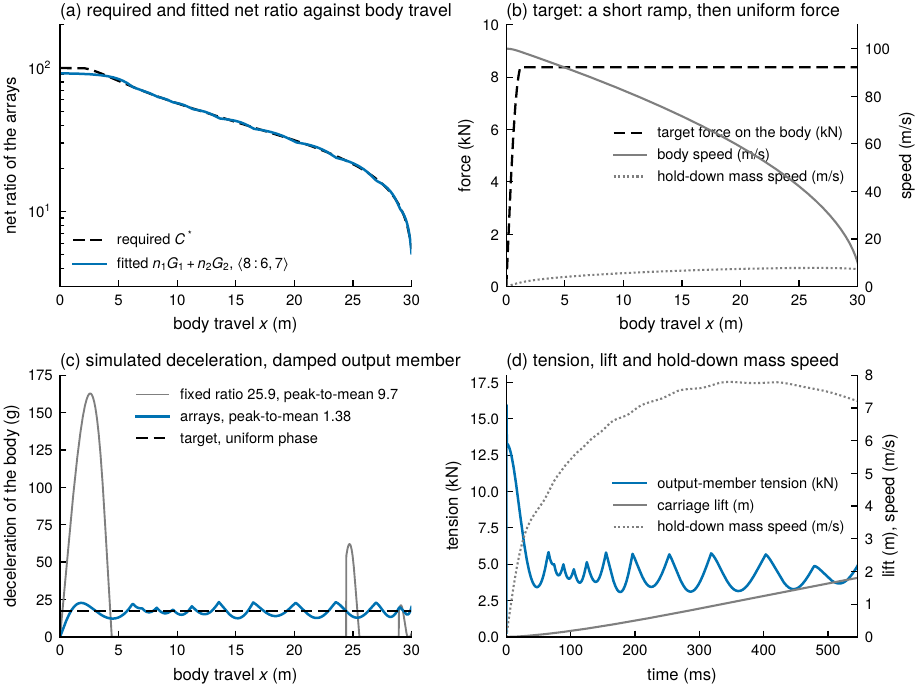}
\caption{The transmission run in reverse: arrest of a 50~kg body from 100 to 10~m/s in 30~m into a 4~t hold-down mass through a 2:1 tackle. (a) The required net ratio $C^*$ against body travel and the fitted $\langle 8{:}6, 7\rangle$ arrays; (b) the target, a ramp of 1.19~m at $C_0 = 100$ then a uniform 8.38~kN, with the body's and the hold-down mass's speeds; (c) simulated deceleration with the damped output member, and the same hold-down at a fixed ratio of 25.9; (d) output-member tension, carriage lift and hold-down mass speed against time.}
\label{fig:si-arrest}
\end{figure}

\paragraph{Worked case.} A 50~kg body enters at 100~m/s and is to leave at 10~m/s after 30~m, 99\% of its kinetic energy, 247.6~kJ, absorbed; the hold-down mass is 4,000~kg through a 2:1 tackle, $M_e = 16{,}000$~kg and an effective weight of 78.5~kN; $R_b = 2.5$~m; the stage is $k = 7$ with $n_1 = 3$ and $n_2 = 4$, and the arrays carry eight and six members, the counts of main-text equation~\eqref{eq:counts}. With $C_0 = 100$ the ramp ends at $x = 1.19$~m, the uniform force is 8.38~kN (17.1~$g$), the working lift is 1.80~m and $C^*$ falls from 100 to 5.5. The arrays were fitted to $C^*$ as a function of lift by least squares in the residual $(C - C^*)/(C^* + 0.05\max C^*)$, sampled uniformly along the body's travel because the force is inversely proportional to the ratio, with the balance term of main-text equation~\eqref{eq:dual-obj} at $\lambda = 0.03$, a width budget of 2.5~m per array and a 50~mm span floor, from 40 random starts (\texttt{sim/arrest/fit.py}). The fit (Table~\ref{tab:si-arrest}) has widths of 2.30 and 2.18~m, a relative residual of 2.7\%, a fitted ratio falling from 92 to 5.1 and, at uniform tension, a peak reaction imbalance of 6.1\% of the peak reaction against 14.3\% for a mirror pair at $k = 7$. The narrow members flatten first, while the body is fastest; the widest, with negative offsets, remain engaged throughout and supply the low ratio at the end. Figure~\ref{fig:si-arrest}(a) and (b) show the target and the fit.

\begin{table}[!htbp]
\centering
\caption{The arrest arrays, $\langle 8{:}6, 7\rangle$, in metres. Members are numbered in the order in which they flatten during the arrest; offsets are in the depth coordinate $t$, with $S = 1.7995$~m, and a negative offset denotes a member that remains engaged throughout the working lift.}
\label{tab:si-arrest}
\footnotesize
\begin{tabular}{ccccc}
\toprule
member & $R_{1,i}$ & $s_{1,i}$ & $R_{2,i}$ & $s_{2,i}$ \\
\midrule
1 & 0.0500 & 1.7048 & 0.0500 & 1.7030 \\
2 & 0.0500 & 1.6831 & 0.0500 & 1.6471 \\
3 & 0.0500 & 1.6037 & 0.0784 & 1.4541 \\
4 & 0.0500 & 1.5462 & 0.1979 & 1.1117 \\
5 & 0.0838 & 1.3166 & 0.3874 & 0.5363 \\
6 & 0.2032 & 0.8656 & 0.3270 & $-$0.0942 \\
7 & 0.3401 & 0.2458 & --- & --- \\
8 & 0.3243 & $-$0.1791 & --- & --- \\
\bottomrule
\end{tabular}
\end{table}

\paragraph{Simulation.} The integrator (\texttt{sim/arrest/sim.py}, semi-implicit Euler at 20~$\mu$s) carries the body, the carriage with its effective mass, the braking fold and the comb, coupled by an output member of stiffness 29~kN/m, about 2.3~MN of axial stiffness over 80~m of routed length, pre-tensioned to 300~N, with viscous damping set by a damping ratio of 0.3 on the body mass reflected through $g_b$; the member goes slack at zero tension. Per-contact friction and the release are not modelled. Figure~\ref{fig:si-arrest}(c) and (d) and \texttt{arrest\_results.json} give the run. The arrest takes 0.546~s and the body leaves the 30~m at 9.81~m/s, 99.0\% of its energy absorbed. The force on the body averages 8.25~kN (16.8~$g$) and peaks at 11.4~kN (23.3~$g$): a \ptm{} of 1.38 over the whole arrest, including the ramp, and 1.37 after it. It reaches 90\% of its uniform value within 0.8~m and does not fall below 6.0~kN thereafter; the output member never goes slack. Its tension peaks at 16.0~kN during the pick-up of the hold-down mass, stays below 8.9~kN after the first 20~ms and averages 4.4~kN in the uniform phase, at a speed of up to 166~m/s. The carriage rises 1.80~m; the hold-down mass rises 3.6~m and leaves the working stroke at 7.2~m/s, its kinetic and potential energy accounting for 245 of the 248~kJ absorbed. The carriage reaction averages 184~kN over the lift but reaches 1.46~MN in the first 10~ms while the mass is picked up, scaling roughly as $2 M_e v_0^2 / (R_b C_0)$; the stops, guide and frame are sized for that transient, and a longer braking span, a lower tackle ratio or a mass picked up in stages reduce it.

The same hold-down mass, braking member and output member at a fixed ratio, 25.9, chosen by bisection so that the same 30~m ends at 10~m/s, pick the mass up as an impact: a peak of 163~$g$, a \ptm{} of 9.7, the output member slack for 81\% of the stroke (Figure~\ref{fig:si-arrest}c). Conventional arresting gear shapes its force with programmed valves or active control; here the geometry does it.

\paragraph{Damping, and the scaled family.} Unlike the launch, the arrest needs a damped output member. Undamped at 29~kN/m the line alternates between slack and taut after the pick-up, slack for 64\% of the stroke, and the \ptm{} is 4.9 (3.6 and 5.3 at 10 and 580~kN/m); at a damping ratio of 0.1, 0.2, 0.3, 0.4 and 0.6 it is 1.80, 1.46, 1.38, 1.46 and 1.58 at 29~kN/m, and at 10~kN/m and a ratio of 0.4 it is 1.29 with the body leaving at 11.6~m/s, a softer line trading a little stopping distance for smoothness. The launch member of Section~\ref{si:compliant} needs no damping because the rising ratio keeps a pre-tensioned member loaded; here the mass is picked up impulsively and the member is unloaded by the falling ratio. With the hold-down mass, stiffness and pre-tension scaled with the body's mass, the same comb arrests 20, 50 and 100~kg bodies (1.6, 4 and 8~t) at the same \ptm{} of 1.38 and peak of 23.3~$g$, at mean forces of 3.3, 8.3 and 16.5~kN and peak tensions of 6.4, 16.0 and 31.9~kN: for a given stopping distance, braking span and tackle ratio the geometry depends on the ratio of hold-down to body mass, 80:1 here, and on the entry and exit speeds, not on scale.

\paragraph{Balance and the single array.} The balance condition, the count condition and the span order along the arrays carry over unchanged, because they depend only on the forces the arrays apply to the carriage, which have the same form in either direction; the engagement order is simply traversed in reverse, the members that engage last in a launch flattening first in an arrest. Where the guide can carry the moment of a one-sided reaction, a small or light body, a low deceleration, a short lift, a single array with one tension member serves, coupled directly ($k = 1$, net ratio $G(t)$), through a stage with one movable element ($kG$), or with two lines through one array driving the counter-moving stage; its member floor is $C_{\max}/(2k)$, about 46 to 50 members for direct coupling at the ratio of this case and 7 or 8 at $k = 7$.

\end{document}